\documentclass{agujournal2019}
\usepackage{url} 
\usepackage{lineno}
\usepackage[inline]{trackchanges} 
\usepackage{soul}
\usepackage{float, epsfig, appendix}
\usepackage{color, amssymb, amsmath, amsthm, verbatim, wasysym, mathrsfs}
\usepackage{siunitx}
\usepackage{multirow}
\usepackage{booktabs}
\usepackage{cancel}

\newcommand{\pd}[2]{\frac{\partial #1}{\partial #2}}

\newcommand{\dd}{\text{d}}

\journalname{Journal of Advances in Modeling Earth Systems (JAMES)}

\usepackage{hyperref}  
\hypersetup{
    colorlinks=true,        
    linkcolor=[rgb]{0,0,0.5},
    citecolor=[rgb]{0,0,0.5},
    urlcolor=[rgb]{0.5,0,0},
    filecolor=[rgb]{0.7,0.35,0},      
}

\begin{document}
\justifying

%
%


\title{NORi: An ML-Augmented Ocean Boundary Layer Parameterization}

%
%




\authors{Xin Kai Lee\affil{1,2, 3}, Ali Ramadhan\affil{1,2}\thanks{Current affiliation: atdepth}, Andre Souza\affil{1}, Gregory LeClaire Wagner\affil{1}\thanks{Current affiliation: Aeolus Labs}, Simone Silvestri\affil{1}\thanks{Current affiliation: Department of Environment, Land and Infrastructure Engineering, Politecnico di Torino}, John Marshall\affil{1,2} and Raffaele Ferrari\affil{1,2}}


\affiliation{1}{Department of Earth, Atmospheric and Planetary Sciences, Massachusetts Institute of Technology}
\affiliation{2}{Center for Computational Science and Engineering, Massachusetts Institute of Technology}
\affiliation{3}{Department of Physics, Imperial College London}




\correspondingauthor{Xin Kai Lee}{me@xinkailee.com}



\begin{keypoints}
    \item NORi uses neural networks to augment a Richardson number-based closure of turbulent vertical mixing in the ocean boundary layer.
    \item \textit{A posteriori} calibration, where loss depends on variable trajectories rather than instantaneous fluxes, ensures robust numerical stability.
    \item NORi outperforms baselines on idealized simulations, closely tracks column observations, and is numerically stable in an idealized 3D setup.
\end{keypoints}

%
%

%
%


\begin{abstract}
    
NORi is a machine learning (ML) parameterization of ocean boundary layer turbulence that is physics-based and augmented with neural networks.
NORi stands for neural ordinary differential equations (NODEs) Richardson number (Ri) closure. 
The physical parameterization is controlled by a Richardson number-dependent diffusivity and viscosity. 
The neural ODEs are trained to capture the entrainment through the base of the boundary layer, which cannot be represented with a local diffusive closure.
The parameterization is trained using large-eddy simulations in an \textit{a posteriori} fashion, where parameters are calibrated with a loss function that explicitly depends on the actual time-integrated variables of interest rather than the instantaneous subgrid fluxes, which are inherently noisy.
NORi conserves tracers by design, uses realistic nonlinear thermodynamics, and demonstrates excellent prediction and generalization capabilities in capturing entrainment dynamics under different convective strengths, background stratifications, rotation, and wind forcings. 
NORi is shown to simulate the seasonal evolution of the boundary layer at Ocean Weather Station Papa with similar performance to the state-of-the-art two-equation $k$-$\epsilon$ closure. 
When implemented in a double-gyre simulation, it is numerically stable for at least 100 years, despite only being trained on two-day horizons, and can be run with time steps as long as one hour. 
Combining highly expressive neural networks with a physically grounded base closure proves to be a robust paradigm for designing parameterizations for climate models: data required and training cost are drastically reduced, inference performance can be directly optimized as a primary objective, and numerical stability is implicitly promoted through training.

\end{abstract}

\section*{Plain Language Summary}
Climate models cannot represent small-scale ocean turbulence at scales of 1--100 meters because the immense computational cost means they can, at best, resolve 10-kilometer features.
These mixing processes, driven by winds, evaporation, and surface cooling, affect how heat, salinity, and other properties are exchanged between the atmosphere and ocean interior, influencing climate predictions.
We introduce NORi, a new approach that combines traditional physics-based equations with modern machine learning to better represent these processes.
Traditional methods often sacrifice accuracy for speed or vice versa.
NORi achieves both by using neural networks to enhance simpler physics-based models where they fall short.
What makes NORi effective is our training approach: using high-resolution simulations as ``ground truth'' and focusing on predicting evolution over time rather than matching instantaneous snapshots.
This produced a model that remains stable when run for 100 years, despite being trained on two-day-long datasets, and compares favorably to observations from a long-term monitoring site in the Northeast Pacific.
NORi has a lower computational cost than traditional models, making it practical for long-term climate simulations.
By bridging physics and machine learning in this way, NORi represents a new paradigm for fast and accurate models that require less training.


%
%

%


%
%
%
%

\section{Introduction}
State-of-the-art global ocean models used in climate studies solve the equations that govern ocean dynamics and thermodynamics on grids with resolutions of 10~km or coarser~\cite{hewitt_resolving_2020,silvestri_gpubased_2025}. 
At this resolution, many subgrid-scale processes cannot be explicitly resolved. 
In the upper ocean boundary layers (BLs), these include turbulent mixing driven by cooling, evaporation, and wind stresses on scales from a few hundred meters down to centimeters. 
Despite occurring at small spatial scales, this mixing has a fundamental impact on the large-scale structure of the ocean by regulating the exchange of heat, carbon, and other climatically important tracers between the atmosphere and the ocean interior. 
Subgrid-scale parameterizations are therefore introduced in climate models to represent the impact of these processes on large-scale variables. 

\begin{figure}
    \centering
    \includegraphics[width=0.5\linewidth]{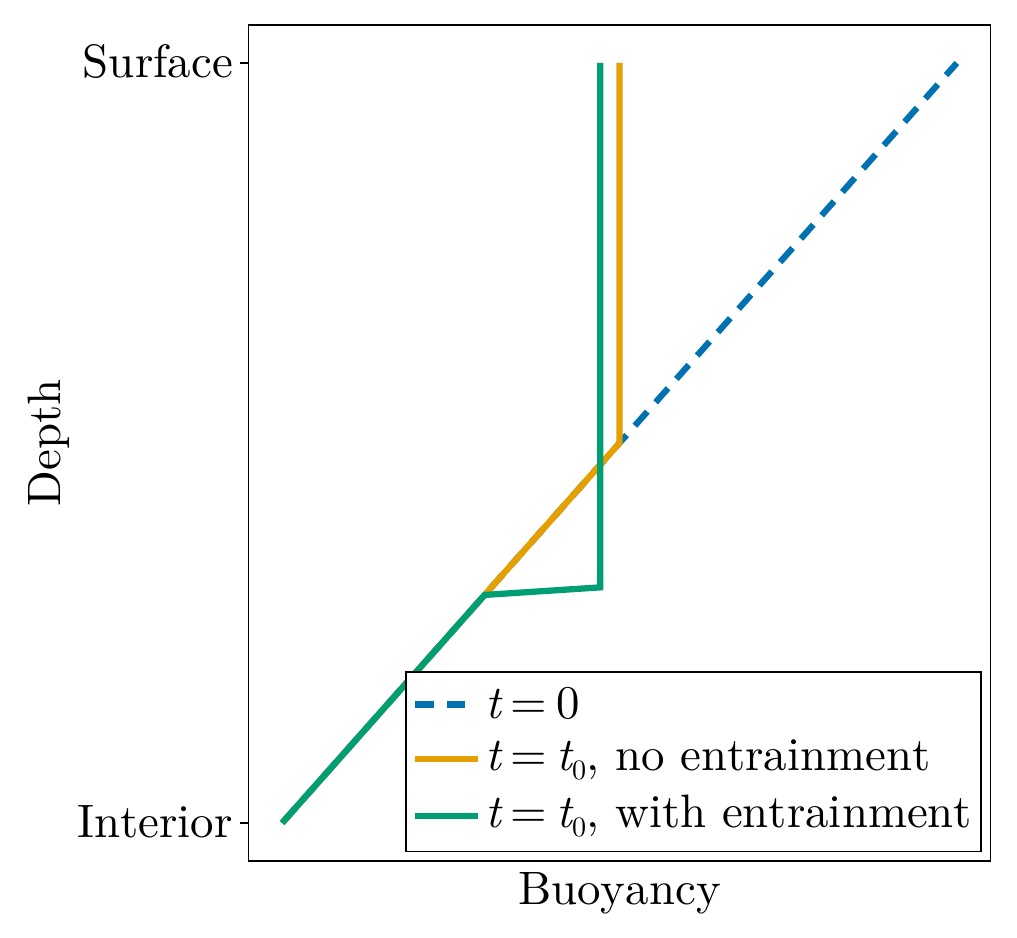}
    \caption{Schematic of the vertical profile of buoyancy as a function of depth below the ocean surface---buoyancy decreases linearly with the density anomaly generated by a change in temperature and salinity.
    At time $t=0$, the buoyancy decreases linearly with depth.
    After some time, in response to buoyancy loss at the surface, a well-mixed layer develops in the upper ocean.
    In the absence of entrainment through the base of the mixed layer, the buoyancy profile smoothly connects to the stratified interior.
    Entrainment results in a further deepening of the boundary layer and the development of a sharp buoyancy jump at its base.}
    \label{fig:entrainment}
\end{figure}

Traditional parameterizations of turbulent mixing in the upper ocean rely on known physics, typically encoded in scaling laws with a number of free parameters that are determined with empirical data. 
The most parsimonious parameterizations are zero-equation first-order diffusive closures, where vertical diffusivities are increased when surface winds, cooling, and/or evaporation are strong enough to trigger turbulent mixing.
The Pacanowski--Philander parameterization~\cite{pacanowski_parameterization_1981} is one such example where the vertical diffusivity is a function of local stratification and vertical shear via the Richardson number. 
Although this diffusive closure smooths local gradients of momentum, temperature, and salinity, it cannot accurately represent the entrainment at the base of the BL. 
Entrainment is driven by convective plumes formed by surface cooling and/or evaporation that penetrate beneath the mixed layer and draw denser water from the thermocline into it. 
This process sharpens the gradient at the BL base---an inherently anti-diffusive, nonlocal mechanism, as illustrated in Figure~\ref{fig:entrainment}. 
To represent processes that are not described purely by downgradient diffusive first-order closures, the widely used K-Profile Parameterization (KPP, \shortciteNP{large_oceanic_1994}) adds a nonlocal flux term to the diffusive flux. 
This nonlocal term does not depend on local gradients and can capture nonlocal entrainment fluxes at the BL base~\cite{noh_improvement_2003}.
However, bulk parameterizations, originally proposed by \citeA{niiler_one-dimensional_1977} and recently adopted for use in large-scale ocean models (ePBL, \shortciteNP{reichl_simplified_2018}), are formulated only in terms of vertically averaged properties of the BL through bulk energetic arguments.
This can be thought of as a fully nonlocal model, since the behavior at the BL base depends explicitly on the bulk properties of the entire mixed layer column above it, including surface fluxes.
Two-equation first-order models, such as the celebrated $k$-$\epsilon$ parameterizations and related generic length-scale closures~\cite{umlauf_generic_2003}, address the nonlocality of turbulent mixing by formulating equations that predict the evolution and transport of quantities such as turbulent kinetic energy (TKE) and energy dissipation. 
In these models, the tracer and momentum fluxes are based on local eddy diffusivities while the TKE and dissipation equations include nonlocal transport terms.
Although more accurate than zero-equation closure models, they are not widely used in global climate models, as they require shorter time steps than the equations for large-scale ocean dynamics and therefore add substantial computational cost~\cite{reffray_modelling_2015, reichl_simplified_2018}. 
Most recently, the Convective Adjustment TKE parameterization (CATKE, \shortciteNP{wagner_formulation_2025}) has been introduced to leverage the benefits of two-equation models while reducing their computational cost. 
CATKE is a 1.5-equation closure that uses a prognostic equation for TKE to compute the diffusivity. 
Its skill compares favorably with the $k$-$\epsilon$ parameterization at a computational cost comparable to that of KPP.

Despite much progress in the formulation of BL parameterizations, substantial biases persist in the simulation of the upper ocean in large-scale ocean models, especially in the Southern Ocean~\cite{duvivier_argo_2018, sallee_assessment_2013, treguier_mixed-layer_2023} and the tropics~\cite{li_tropical_2014}.
In particular, the BL depth bias is often of the same order as the depth itself~\cite{duvivier_argo_2018, treguier_mixed-layer_2023}.

Machine learning methods offer a potential route to reduce these biases and are being increasingly used in the parameterization of climate models for subgrid-scale processes thanks to advances in computing hardware and machine learning software~\cite{bolton_applications_2019,zanna_framework_2025}.
For upper ocean BL turbulence, a Bayesian approach has been used to tune and uncover structural deficiencies of KPP~\cite{souza_uncertainty_2020}.
Neural networks have been used to predict the evolution of temperature and salinity in the Northeast Pacific Ocean~\cite{liang_exploring_2022}, improve the ePBL parameterization~\cite{sane_parameterizing_2023}, and calibrate the vertical diffusivity profile in KPP~\cite{yuan_kprofile_2024}.
Equation-discovery approaches have been investigated as a means to enhance the ePBL parameterization while preserving its physical interpretability~\cite{sane_machine_2026}.
While all these approaches have reduced some biases, none has been completely successful.
For example, \citeA{sane_parameterizing_2023} and \citeA{sane_machine_2026} show that machine learning-enhanced ocean BL parameterizations in Ocean Model Intercomparison Project (OMIP) configurations reduce mixed layer depth biases in the summer but not in the winter.

One of the most important challenges facing neural network-based parameterizations applied to climate modeling is the loss of numerical stability and physical consistency over long integration times~\cite{chattopadhyay_challenges_2023}. 
The causes of such numerical instabilities are manifold, including the spectral bias of neural networks, which prioritize learning low-frequency dynamics at the expense of accurate high-frequency dynamics, essential to represent turbulence and chaotic processes~\cite{chattopadhyay_challenges_2023}, as well as the use of \textit{a priori} calibration to train the neural network~\cite{frezat_posteriori_2022}. 
In \textit{a priori} calibration, the neural network parameterization is trained offline to match the diagnosed, unresolved subgrid-scale fluxes as a function of large-scale variables.
Since no parameterization is perfect, once implemented in the dynamical equations of the climate model, it will introduce some random errors in addition to the desired fluxes.
On top of that, there are unavoidable numerical errors due to the spatial and temporal discretization of the PDEs.
Over time, these errors accumulate and amplify, leading to the climate variables diverging from the correct trajectories, not unlike how trajectories of chaotic systems diverge in response to infinitesimal perturbations. 
This challenge has been well documented in data-driven neural network-based weather forecast models, which can be considered the state-of-the-art in the application of machine learning tools to atmospheric models, including FourCastNet~\cite{pathak_fourcastnet_2022}, GraphCast~\cite{lam_graphcast_2023}, and Pangu~\cite{bi_accurate_2023}. 
These models can develop instabilities or accumulate errors when run autoregressively over long times~\cite{chattopadhyay_challenges_2023}.

A recently suggested alternative to obtain stable solutions with neural networks is \textit{a posteriori} training, albeit with an increase in training cost.
In \textit{a posteriori} training, the parameterization is trained to match the temporal evolution of coarse-grained variables rather than the instantaneous subgrid-scale fluxes. 
The parameterized fluxes are not required to exactly match those in the training dataset, as the training data are noisy and the fluxes required to produce the correct tendencies are not unique. 
Instead, the parameterization is trained end-to-end by evolving the coarse-grained variables over time and including the entire solution's evolution in the loss function. 
A key advantage of this approach is that it aligns the training data distribution with the conditions expected during inference, unlike \textit{a priori} training, where the data distribution at inference might differ from that during training due to drifts in prognostic variables. 
In addition, \textit{a posteriori} training naturally accounts for numerical discretization errors, as models are trained on trajectories of coarse-grained variables over time. 
Examples of parameterizations calibrated \textit{a posteriori} include closures for 2D turbulence~\cite{kochkov_machine_2021}, quasi-geostrophic turbulence~\cite{frezat_posteriori_2022}, 3D turbulence~\cite{sirignano_dpm_2020, stachenfeld_learned_2021}, the viscous Burgers' equation~\cite{kang_learning_2023}, as well as for BL turbulence~\cite{wagner_formulation_2025}. 
Another notable example of \textit{a posteriori} calibration in earth system modeling is NeuralGCM~\cite{kochkov_neural_2024}, where the model variable tendencies were used to train the neural networks, rather than the more traditional approach of training \textit{a priori} on eddy fluxes.

A challenge of \textit{a posteriori} calibration is that it requires the forward model to be fully automatic-differentiable and is serial in time, resulting in a high computational and memory cost. 
For physically based models, this can be mitigated with ensemble-based methods such as Ensemble Kalman Inversion (EKI)~\cite{dunbar_ensemblekalmanprocessesjl_2022, iglesias_ensemble_2013} that do not require differentiability, but these methods perform well only when parameter counts are small and priors are informative~\cite{gjini_ensemble_2025}. 
Neural network models have large numbers of free parameters and informative priors are difficult to find because the parameters do not represent interpretable physical processes. 
The differentiability bottleneck has been addressed through modern languages that offer libraries for automatic differentiation, such as JAX~\cite{frostig_compiling_2018} in Python and Lux.jl~\cite{pal_lux_2023} and Enzyme.jl~\cite{moses_instead_2020} in Julia.

We present NORi (pronounced noh-ree, a dried edible seaweed in Japanese): an ocean BL parameterization that combines the strengths of physics-based and machine-learned parameterizations.
The philosophy is to start with as simple a physics-based parameterization as possible and then augment it with a data-trained neural network.
The physics-based model is a first-order diffusive closure similar to the Pacanowski--Philander model~\cite{pacanowski_parameterization_1981} where the eddy diffusivity is a simple function of the local gradient Richardson number.
To capture the additional mixing physics of entrainment, the parameterization is augmented with neural networks trained \textit{a posteriori} with high-resolution simulations of upper ocean turbulence.
The \textit{a posteriori} calibration uses a neural ordinary differential equation framework~\cite{chen_neural_2019, rackauckas_universal_2021} where the parameterization is integrated forward in time within fully differentiable ODE solvers.
NORi considers the full nonlinear equation of state~\cite{roquet_accurate_2015} and provides a closure for the complete suite of prognostic variables for the dynamics of the ocean BL: momentum, temperature, and salinity.
We show that incorporating known physics into the parameterization reduces the amount of data required for training, allows the use of a small neural network, and improves the model's generalization ability and numerical stability. 
We find that NORi compares favorably in accuracy and computational performance with the $k$-$\epsilon$ parameterization~\cite{umlauf_second-order_2005}, considered a gold standard for high-resolution BL simulations, and the CATKE parameterization~\cite{wagner_formulation_2025}, a recently developed closure for use in global ocean simulations. 
Like all BL vertical mixing schemes, NORi parameterizes processes that deepen the BL, assuming that the large-scale model resolves lateral mesoscale and submesoscale instabilities that restratify the BL. 
The novelty of NORi lies in the design paradigm as well as in the training strategy, which are applicable for the design of neural network-based parameterizations in general.
Instead of using a large neural network to learn the entire parameterization scheme or restricting a neural network to calibrate specific parameters in a complex physics-based model, we design a simple scheme that captures the bulk of the physical process to be parameterized and then leverage the expressivity of neural networks to learn the residual physics that the base scheme does not capture.

\begin{figure}
    \centering
    \includegraphics[trim={0 12cm 0 7.5cm}, width=\linewidth]{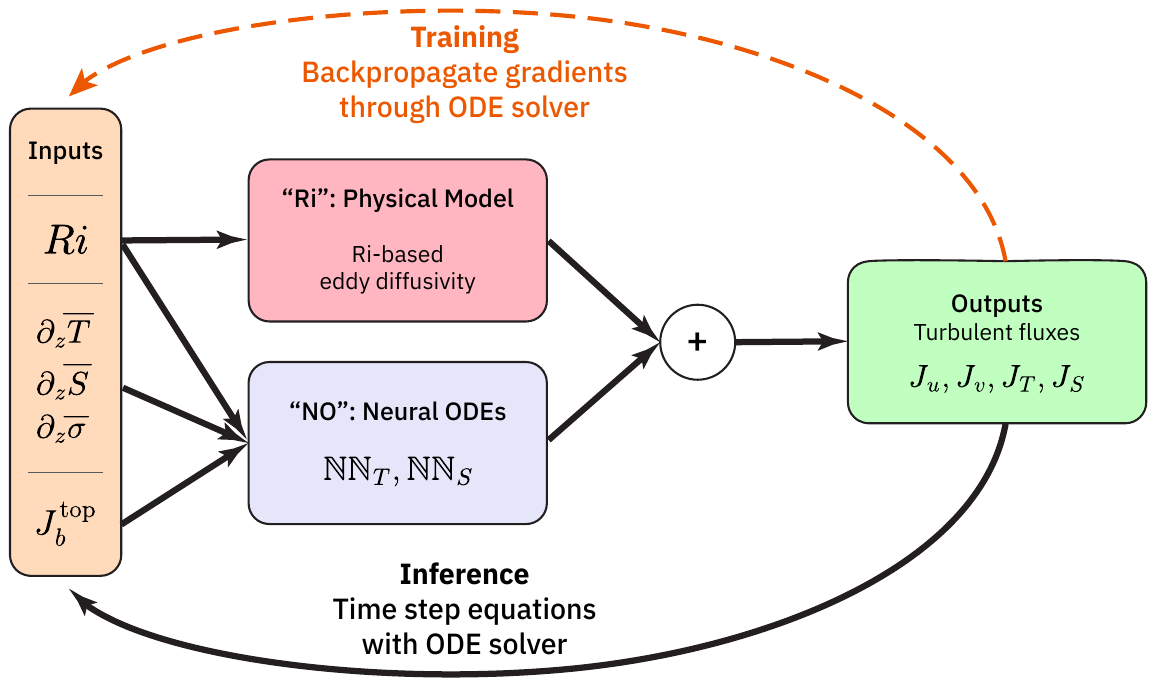}
    \caption{Schematic summary of NORi's design and training framework.
    NORi consists of a physics-based eddy-diffusivity closure based on the gradient Richardson number (``Ri'') augmented by neural ODEs (``NO'').
    The physics-based closure takes local gradient Richardson number as input and captures local mixing, while the neural networks take Richardson number, temperature, salinity, and potential density gradients, and the surface buoyancy flux as inputs and capture nonlocal entrainment fluxes at the base of the mixed layer.
    The base-closure outputs are summed with the neural-network entrainment fluxes for temperature and salinity (the base closure alone provides the momentum flux), producing the turbulent fluxes used to step the column-model equations forward in time.
    Calibration is done over the full time trajectory by backpropagating gradients through the ODE solve to update neural network parameters.
    }
    \label{figure nori summary}
\end{figure}

Section~\ref{section LES to 1D model} introduces the high-resolution, high-fidelity, large-eddy simulations (LES) that we use to train and validate NORi.
Section~\ref{section 1D model formulation} describes the formulation of the NORi column model.
Section~\ref{section local base closure} describes the local eddy-diffusivity closure which encodes the known physics.
Section~\ref{section neural network} describes the neural network formulation and the \textit{a posteriori} training paradigm used in NORi.
We then assess NORi's performance in single-column configurations in Section~\ref{section results column} and in a large-scale long-time double-gyre ocean simulation in Section~\ref{section results double gyre}.
Finally, Section~\ref{section conclusion} presents our conclusions. 
Table~\ref{table nori summary} and the schematic in Figure~\ref{figure nori summary} summarize the key components and design choices of NORi.

\begin{table}[htbp]

\centering
\small
\renewcommand{\arraystretch}{1.1}
\caption{Overview of the NORi framework and training approach.}
\label{table nori summary}
\begin{tabular}{p{0.3\linewidth} p{0.7\linewidth}}
\toprule
\textbf{Aspect} & \textbf{Description} \\
\midrule
Overall approach & Augments a simple physics-based model with neural networks to capture missing physics \\
Physics-based component & \textit{Local} gradient Richardson number-based eddy-diffusivity closure to capture local mixing \\
Neural network component & \textit{Nonlocal} entrainment fluxes of temperature and salinity at the base of the mixed layer due to convective plumes \\
Neural network design & Small feedforward networks (3 hidden layers $\times$ 128 ReLU units) shared across vertical grid points \\
Neural network inputs & Local gradients of temperature, salinity, and potential density, the local Richardson number, and the nonlocal surface buoyancy flux\\
Training data & Large-eddy simulations (LES) spanning convective and shear turbulence regimes, with varying rotation, background stratification, and mean temperatures and salinities\\
Training strategy & Trained on the full temporal evolution of temperature and salinity rather than on instantaneous turbulent fluxes \\
Validation and testing & Held-out LES scenarios and the Ocean Weather Station Papa observations for accuracy; idealized 3D double-gyre simulation for numerical stability \\
\bottomrule
\end{tabular}
\end{table}

\section{High-resolution dataset for boundary layer turbulence} \label{section LES to 1D model}
A parameterization must, first and foremost, accurately represent the physics of the processes it parameterizes.
Therefore, a well-designed suite of ``ground-truth'' data is required for the training, validation, and testing of NORi.
To that end, we run a wide range of high-resolution large-eddy simulations (LES) of upper ocean turbulence, which aim to cover the conditions under which BL turbulence occurs in the real ocean.

\subsection{Large-eddy simulations}

LES are simulations that resolve the largest turbulent eddies in the BL and use a subgrid-scale model to represent the smaller, more universal 3D eddies.
LES are generated with Oceananigans.jl~\cite{ramadhan_oceananigansjl_2020, wagner_high-level_2025}, a finite volume ocean model written in Julia~\cite{bezanson_julia_2017} and optimized for GPUs~\cite{silvestri_gpubased_2025}.
The LES data suite focuses on shear-driven and convection-driven deepening of the BL against a stable background stratification.
Mathematically, LES solve the incompressible Boussinesq form of the Navier--Stokes equations given by
\begin{align}
    \label{equation LES u} \pd{\boldsymbol{u}}{t} &= -\nabla \cdot (\boldsymbol{u} \otimes \boldsymbol{u}) - f \hat{\boldsymbol{k}} \times \boldsymbol{u} - \nabla p + b \hat{\boldsymbol{k}}, \\
    \label{equation LES T} \pd{T}{t} &= -\nabla \cdot (\boldsymbol{u}T), \\
    \label{equation LES S} \pd{S}{t} &= -\nabla \cdot (\boldsymbol{u}S), \\
    \label{equation LES incompressibility} \nabla \cdot \boldsymbol{u} &= 0, \\
    \label{equation LES b} b &= -g \frac{\sigma - \sigma_0}{\sigma_0}, \\
    \label{equation LES sigma} \sigma &= \sigma_\text{TEOS-10}(T, S, z = 0),
\end{align}
where $\boldsymbol{u}$ are the velocities in the $x$, $y$, and $z$ directions, $f$ is the Coriolis parameter, $\hat{\boldsymbol{k}}$ is the unit vector in the vertical direction (normal to the fluid surface), $p$ is the kinematic pressure, $b$ is the buoyancy of the fluid, $T$ and $S$ are the temperature and salinity of the fluid, $\sigma$ is the potential density, $g$ is the gravitational acceleration, and $\sigma_0 = \SI{1020}{kg.m^{-3}}$ is the reference density.
The LES are run with laterally doubly periodic boundary conditions, prescribed advective fluxes of $T$, $S$, and $u$ at the top given by $J_T^\text{top}$, $J_S^\text{top}$, and $J_u^\text{top}$, with $J_v^\text{top}=0$ in all simulations, and linear gradient boundary conditions for $T$ and $S$ and free-slip boundary conditions for $u$ and $v$ at the bottom.
Additionally, we assume that the vertical velocities at the top and bottom surfaces are zero, imposing the no-penetration boundary condition given by
\begin{equation} \label{equation LES w no penetration}
    w(z = 0) = w(z=-L_z) = 0.
\end{equation}

To model a realistic ocean, we use the TEOS-10 equation of state~\cite{roquet_accurate_2015}, which takes into account the nonlinear dependence of buoyancy on temperature and salinity.
This nonlinear dependence can lead to effects such as cabbeling, which requires that temperature and salinity be modeled separately.
The prognostic variables in \eqref{equation LES u}-\eqref{equation LES incompressibility} are evolved using a third-order Runge--Kutta scheme with a pressure projection (or predictor-corrector) method that involves the solution of a three-dimensional Poisson equation for pressure~\cite{schumann_fast_1988}.

The simulations do not use an explicit LES closure; instead, they employ a 9th-order weighted essentially non-oscillatory (WENO, \shortciteNP{balsara_monotonicity_2000}) advection scheme. 
This dissipative scheme acts as an implicit viscous closure, capturing kinetic energy dissipation while maintaining high resolution through gradient-preserving properties~\cite{silvestri_new_2024}. 
The implicit LES (ILES) formulation~\cite{grinstein_implicit_2007} has been shown to outperform explicit closures in ocean and atmospheric simulations where turbulence coexists with internal waves~\cite{chen_evaluating_2025, pressel_numerics_2017, silvestri_new_2024, smolarkiewicz_implicit_2007, wagner_formulation_2025}. 
The ILES prevents oscillatory errors that lead to spuriously large diffusivities in explicit closures at sharp density interfaces, such as BL bases, where internal wave shears can be large without generating significant mixing. 
In particular, ILES with prescribed surface fluxes used in the Coastal and Regional Ocean COmmunity model (CROCO) has been shown to give results comparable to the NCAR-LES~\cite{fan_comparison_2024} and the commonly used explicit Smagorinsky closure to simulate BL turbulence in convective and shear turbulent regimes with vertical resolutions below a few meters~\cite{chen_evaluating_2025}.

\begin{figure}[htbp]
    \centering
    \includegraphics[width=\textwidth]{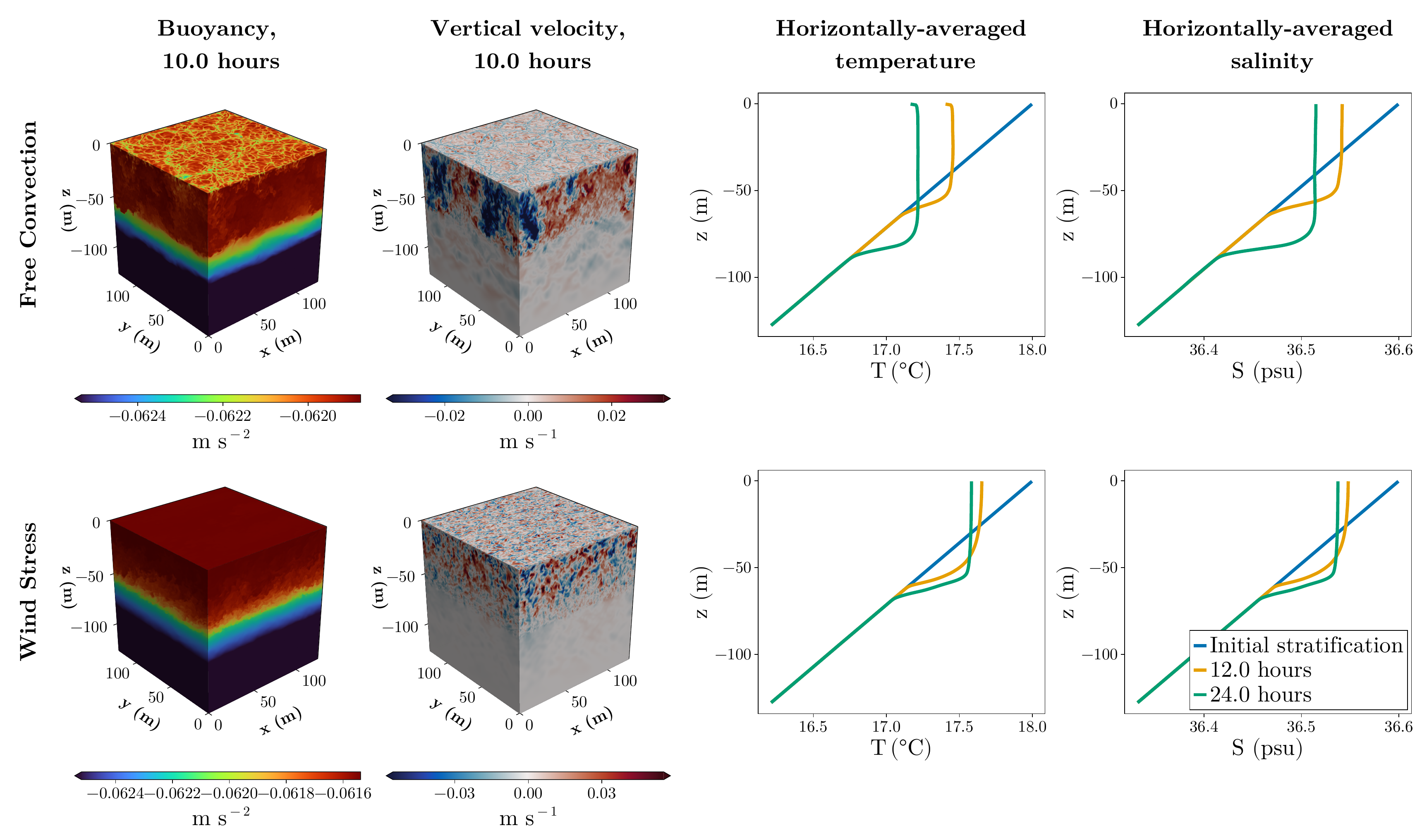}
    \caption{Large-eddy simulations (LES) for free convection and pure wind stress scenarios in a horizontally doubly periodic domain of size $(L_x, L_y, L_z) = (128, 128, 128) \, \si{m}$ with a grid resolution of $\SI{0.5}{m}$ and a Coriolis parameter of $f = \SI{8d-5}{s^{-1}}$.
    The top row shows a convective turbulence LES driven by surface cooling, while the bottom row shows a shear turbulence LES driven by surface wind stress.
    The first and second columns show snapshots of the buoyancy and vertical velocity fields.
    The third and fourth columns show the time evolution of the horizontally averaged temperature and salinity profiles characterized by a deepening mixed layer in response to surface forcing.}
    \label{figure LES 3D field}
\end{figure}

The snapshots in Figure~\ref{figure LES 3D field} show the simulated buoyancy and vertical velocity fields of two representative examples of upper ocean turbulence: free convection and shear turbulence acting on a fluid initialized with constant vertical stratification.
These simulations, used for illustration purposes in Figure~\ref{figure LES 3D field}, are run with an isotropic resolution of $\SI{0.5}{m}$ and a Coriolis parameter of $f = \SI{8d-5}{s^{-1}}$.
Turbulence in the free convection simulation is generated by buoyancy loss at the ocean surface, leading to static instability, and in the shear simulation by a surface wind stress.
Over time, convection- and wind-driven turbulence create a well-mixed layer at the top of the fluid, which deepens over time.
Between the mixed layer and the stratified interior, a thermocline with sharp gradients is formed.
This vertical structure can be seen in the horizontally averaged temperature and salinity fields in the third and fourth columns of Figure~\ref{figure LES 3D field} for different times.

The two simulations illustrate the fundamental differences between the two scenarios of BL turbulence. 
In free convection, the case shown in the upper row of Figure~\ref{figure LES 3D field}, coherent plumes with strong downward velocities plunge from the surface towards and through the thermocline.
As the plumes reach the base of the BL, they overshoot into the stratified ocean interior due to inertia and entrain denser fluid from the thermocline into the mixed layer, increasing the mixed layer's mean density.
Entrainment mixing is nonlocal as it occurs at the BL base, driven by surface air-sea fluxes.
However, in the shear-driven case in the bottom row of Figure~\ref{figure LES 3D field}, surface wind stress creates shear instability, which generates local eddies that decay away from the surface.
The turbulent eddies generated by wind stress are much smaller in scale than the convective plumes created by buoyancy loss.
Thus, from a parameterization perspective, the shear instability can be regarded as a local process.
We will therefore develop NORi, starting with a diffusive closure that is known to represent well the local mixing induced by both convective and shear turbulence.
To address the nonlocal entrainment that has proven very challenging to parameterize, we will add a machine-learned neural network component to NORi.

\subsection{Training and validation data generation}

An LES training and validation suite is run to calibrate NORi.
All LES are initialized with uniform temperature and salinity profiles in the horizontal direction and with a linear gradient in the vertical direction.
The simulations are run under a variety of surface forcing scenarios: winds only, cooling or evaporation only, cooling + evaporation, winds + cooling and/or evaporation, and winds + heating + precipitation.
All of these forcings produce destabilizing fluxes that mix away the ocean stratification and result in a deepening mixed layer in the upper ocean.

A convection-resolving simulation requires a resolution of $\mathcal{O}(\SI{1}{cm})$ to capture the homogenization of turbulent filaments at molecular scales. 
LES allow us to simulate the same process at a resolution of $\mathcal{O}(\SI{1}{m})$. 
Through experimentation, we found that our LES converged at an isotropic resolution of $\SI{2}{m}$ in domains of $(L_x, L_y, L_z) = (512, 512, 256) \, \si{m}$. 
On an NVIDIA V100 GPU we achieved about 12 simulated days per day. 
This is the setup used for all LES in the NORi training suite.

The initial temperature and salinity stratifications are based on linear fits to summertime profiles from the ISAS reanalysis product of Argo profiles from 2002 to 2020~\cite{gaillard_situbased_2016} spanning a representative range of oceanic stratification and equation of state regimes---we initialize with summertime profiles because we are interested in simulating the deepening of the mixed layer in fall and winter.

By referencing Argo profiles, we ensure that we cover a representative and realistic range of oceanic stratifications that we expect at inference time, including regimes with nonlinear thermodynamics.
To that end, we consider profiles from three latitude bands: midlatitude Atlantic (strong temperature and salinity gradient), equatorial Pacific (strong temperature but weak salinity gradient), and Southern Ocean (salinity-dominated stratification, nonlinear thermodynamics).
We complement these observationally inspired initial profiles with additional ones designed to span a broader parameter space, which is key to ensure generalizability and avoid overfitting.
In particular, some training cases lack temperature or salinity gradients, others have inverse (stabilizing) salinity stratification, and others operate in the more thermodynamically nonlinear regime at lower temperatures.
The actual values of the physical parameters used in the LES suite can be found in \ref{section appendix LES table}.

The prescribed air-sea fluxes are constant in time and space and chosen to cover the range of forcing regimes encountered in the ocean from free convection (no wind stress) to pure wind scenarios (no buoyancy flux) to mixed conditions.
While the training is based on simulations forced with constant fluxes, we demonstrate in Section~\ref{subsection 1D long integration} that NORi remains accurate even under time-varying fluxes.
The range of flux values is chosen to be broad so that NORi is as unlikely as possible to run outside the training distribution at inference time.
In terms of magnitudes, cooling rates range from $\SI{0}{W.m^{-2}}$ to $\SI{2500}{W.m^{-2}}$, evaporation rates range from $\SI{0}{m.yr^{-1}}$ to around $\SI{45}{m.yr^{-1}}$, and wind stresses range from $\SI{0}{N.m^{-2}}$ to $\SI{0.5}{N.m^{-2}}$.
For context, the mean monthly cooling rate of the Gulf Stream can reach $\SI{600}{W.m^{-2}}$~\cite{the_climode_group_climode_2009}, latent heat fluxes during strong wind events such as tropical cyclones can reach $\SI{500}{W.m^{-2}}$~\cite{vreugdenhil_ocean_2021}, the winter average evaporation rate over the Gulf Stream is around $\SI{3}{m.yr^{-1}}$~\cite{yu_global_2007}, and typical monthly averaged wind stress is around $\SI{0.2}{N.m^{-2}}$~\cite{copernicus_climate_change_service_era5_2019}.
The main training suite consists of 62 simulations, and the LES parameters can be found in Tables~\ref{table wind driven LES suite}, \ref{table free convection LES training suite}, \ref{table wind + convection LES training suite}, and \ref{table misc LES training suite}.
The parameters for the 32 simulations used in the validation and testing suite are given in Tables~\ref{table wind driven LES suite} and \ref{table LES validation suite}. 
Since heating and precipitation are stabilizing fluxes that do not contribute to the deepening of the BL, we did not include them in the training suite.
However, we tested NORi's performance under a winds + heating + precipitation scenario to ascertain whether it is able to capture mixing accurately under stabilizing fluxes, which we have shown in the fourth column in the bottom panel of Figure~\ref{figure local base closure results} (the parameters for this case are shown in row 7 of Table~\ref{table wind driven LES suite}).
This suite of datasets is publicly available for easy download and access \href{https://doi.org/10.5281/zenodo.16278000}{on Zenodo}~\cite{lee_soblles_2025}.

All LES are run for two days, a compromise between simulating long enough to capture the physics of turbulent mixing under different scenarios and keeping the memory footprint of the \textit{a posteriori} calibration of the 1D column model manageable.
The backward propagation of gradients requires storing intermediate model states to compute parameter sensitivities over the entire simulation period. 
During the forward--backward automatic differentiation, we do not employ checkpointing; instead, we store the model state at every time step rather than only at a limited number of checkpoints. 
Introducing checkpointing would reduce the memory footprint but increase computational cost, because model states at specific times would need to be recomputed from their nearest checkpoint during backpropagation.

\section{Modeling unresolved fluxes in a coarse-grained column model} \label{section 1D model formulation}

Global oceanic simulations cannot be run at LES resolutions of $\mathcal{O}(\SI{1}{m})$.
Our goal is thus to develop a parameterization that predicts the deepening of the BL in response to forcing without the need to resolve the small-scale horizontal motions.
In practice, we want to find a closed-form 1D equation that predicts the evolution of the LES area-averaged profiles of velocity, temperature, and salinity as a function of initial conditions and air-sea fluxes. 

Let us start by defining the area averaging operator of some LES variable $\phi$ in a doubly periodic horizontal domain $\Omega$,
\begin{align}
    \phi(x, y, z, t) &= \overline{\phi}(z, t) + \phi^\prime (x, y, z, t), \\
    \overline{\phi} &\equiv \frac{1}{L_x L_y} \iint_\Omega \phi \dd x \dd y, \\
    \overline{\phi^\prime} &\equiv 0,
\end{align}
where $\overline{\phi}$ is the horizontal area-average of $\phi$, $L_x$ and $L_y$ are the dimensions of the lateral domain, and $\phi^\prime$ is its deviation from the horizontal average.
Applying this Reynolds-like averaging as well as the no-penetration boundary conditions at the top and bottom of the LES domain according to Equation~\eqref{equation LES w no penetration}, the momentum, temperature, and salinity Equations~\eqref{equation LES u}, \eqref{equation LES T}, and \eqref{equation LES S} reduce to a set of 1D equations (see complete derivation in \ref{section appendix column model}),
\begin{align}
    \label{equation u 1D} \pd{\overline{u}}{t} &= -\pd{}{z}\overline{u^\prime w^\prime} + f \overline{v}, \\
    \label{equation v 1D} \pd{\overline{v}}{t} &= -\pd{}{z}\overline{v^\prime w^\prime} - f \overline{u}, \\
    \label{equation T 1D} \pd{\overline{T}}{t} &= -\pd{}{z}\overline{w^\prime T^\prime},\\
    \label{equation S 1D} \pd{\overline{S}}{t} &= -\pd{}{z}\overline{w^\prime S^\prime},
\end{align}
where $\overline{u^\prime w^\prime}$, $\overline{v^\prime w^\prime}$, $\overline{w^\prime T^\prime}$, and $\overline{w^\prime S^\prime}$ are the total unresolved momentum, temperature, and salinity fluxes due to subgrid-scale turbulence.
Their surface boundary conditions are given by
\begin{align}
    \overline{u^\prime w^\prime}(z=0) &= J_u^\text{top},\\
    \overline{v^\prime w^\prime}(z=0) &= J_v^\text{top}, \\
    \overline{w^\prime T^\prime}(z=0) &= J_T^\text{top},\\
    \overline{w^\prime S^\prime}(z=0) &= J_S^\text{top}.
\end{align}
At the bottom $z=-L_z$ we impose free-slip boundary conditions on momentum and a gradient condition on temperature and salinity matched to the background stratification,
\begin{align}
  \overline{u^\prime w^\prime} (z = -L_z) &= \overline{v^\prime w^\prime} (z = -L_z) = 0, \\
  \partial_z \overline{T} \big|_{z=-L_z} &= \pd{T_0}{z}, \\
  \partial_z \overline{S} \big|_{z=-L_z} &= \pd{S_0}{z},
\end{align}
so that the initial mean gradients $\pd{T_0}{z}$ and $\pd{S_0}{z}$ are preserved at the lower boundary.
We write equations for temperature and salinity separately because we use the full nonlinear equation of state of seawater, and we cannot combine the two equations into a single one for buoyancy.

\section{``Ri'': Eddy-diffusivity closure based on local gradient Richardson number} \label{section local base closure}

Following \citeA{kochkov_neural_2024}, we first attempted to develop a parameterization of the BL using a purely data-driven approach by modeling the full \textit{fluxes} of $\overline{u^\prime w^\prime}$, $\overline{v^\prime w^\prime}$, $\overline{w^\prime T^\prime}$, and $\overline{w^\prime S^\prime}$ with neural networks. 
(Note that \citeA{kochkov_neural_2024} modeled the \textit{tendencies} instead of the fluxes).
However, we found that neural networks were not able to learn that vertical profiles of temperature and salinity must be statically stable in simulations that parameterize rather than resolve turbulence, i.e., $N^2 =  \pd{b}{z} \ge 0$ where $N^2$ is the buoyancy frequency.
Instead, neural networks produced a BL that was nearly well-mixed near the surface, but with small, random fluctuations in $z$ resulting in negative $N^2$ patches. 
In nonhydrostatic formulations, this stratification inversion would trigger convective instabilities that would rapidly remove the negative $N^2$. 
However, large-scale ocean models use the hydrostatic approximation to lower computational costs, which is inconsistent with the physics of density overturns. 
This deficiency could likely be addressed by increasing the size of the training data or the expressivity of the neural networks~\cite{subel_data-driven_2021}, but both incur significant computational and memory costs given the serial-in-time nature of \textit{a posteriori} calibration.

\begin{figure}[htbp]
    \centering
    \includegraphics[width=0.8\linewidth]{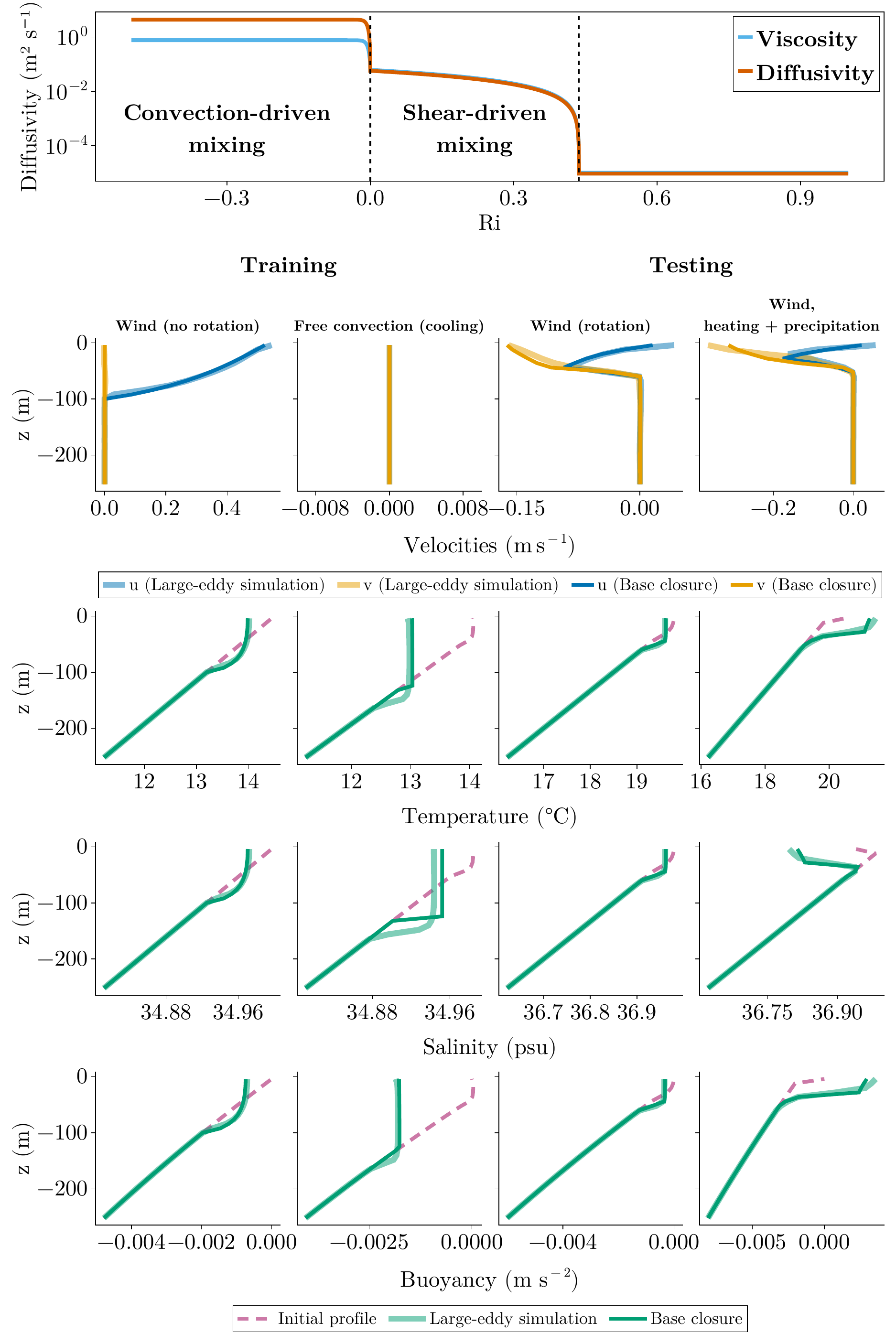}
    \caption{The top panel shows the calibrated diffusivity and viscosity values of the base closure as a function of the local gradient Richardson number $Ri$.
    In the convective range, the viscosity and diffusivity are constant, while they decrease linearly (note the logarithmic scale on the $y$-axis) to a background value across the shear range, with a continuous transition between regimes.
    The lower panels show vertical profiles of momentum, temperature, salinity, and buoyancy.
    Solid colored lines with a lighter shade are LES results forced with four different air-sea fluxes; the corresponding predictions from the base closure are overlaid with a darker shade, while the dashed lines represent the initial profiles. 
    From left to right: a wind-forced example with no rotation and a free convection example with rotation used to train the base closure; a wind-forced example with rotation and a wind-forced example with heating and precipitation used for testing.}
    \label{figure local base closure results}
\end{figure}

In view of these challenges, we instead start by formulating a simple physics-based parameterization that captures most of the BL physics and then augment it with a neural network.
We will refer to this physics-based component as the \textit{base closure}.
Training a neural network to learn the physics missing in the base closure rather than all BL physics reduces the data requirement and model size, dramatically improves parameterization performance, and reduces numerical instability during training and validation.
Mathematically, the base closure represents all subgrid-scale fluxes as downgradient fluxes of momentum, temperature, and salinity,
\begin{align}
    \label{u local closure} \overline{u^\prime w^\prime} = J_{u,\text{ local}} &= -\nu \pd{\overline{u}}{z}, \\
    \label{v local closure} \overline{v^\prime w^\prime} = J_{v,\text{ local}} &= -\nu \pd{\overline{v}}{z}, \\
    \label{T local closure} \overline{w^\prime T^\prime} = J_{T,\text{ local}} &= -\kappa \pd{\overline{T}}{z}, \\
    \label{S local closure} \overline{w^\prime S^\prime} = J_{S,\text{ local}} &= -\kappa \pd{\overline{S}}{z},
\end{align}
where $\nu$ and $\kappa$ are the (positive) eddy viscosity and diffusivity, which depend on the local gradient Richardson number $Ri$, given by 
\begin{equation}
    Ri = -\frac{g}{\sigma_0} \frac{\pd{\overline{\sigma}}{z}}{\left( \pd{\overline{u}}{z} \right)^2 + \left( \pd{\overline{v}}{z} \right)^2}.
\end{equation}
The $Ri$-dependence is well supported by theory~\cite{drazin_hydrodynamic_2004} and ocean observations~\cite<e.g.>[]{price_diurnal_1986}, and it forms the basis of many ocean BL parameterizations~\cite<e.g.>[]{pacanowski_parameterization_1981,large_validation_1999,wagner_formulation_2025}.
When $Ri < 0$, the BL stratification is convectively unstable.
When the BL is statically stable, but the shear is large enough so that $Ri$ is below some critical value $Ri^c$, the BL is shear unstable.
The specific functional dependence of diffusivity and viscosity on $Ri$ is illustrated in the upper row of Figure~\ref{figure local base closure results} and takes the form,
\begin{align}
  \nu &=
  \begin{cases} \label{local base closure viscosity}
    (\nu_{\text{shear}} - \nu_{\text{conv}}) \tanh{\frac{Ri}{\Delta Ri}} + \nu_{\text{shear}}, & \text{for } Ri < 0 \\
    (\nu_0 - \nu_{\text{shear}}) \frac{Ri}{Ri^c} + \nu_{\text{shear}}, & \text{for } 0 \leq Ri < Ri^c \\
    \nu_0, & \text{for } Ri \geq Ri^c
  \end{cases}\\
    \kappa &=
    \begin{cases} \label{local base closure diffusivity}
    (\kappa_{\text{shear}} - \kappa_{\text{conv}}) \tanh{\frac{Ri}{\Delta Ri}} + \kappa_{\text{shear}}, & \text{for } Ri < 0 \\
    (\kappa_0 - \kappa_{\text{shear}}) \frac{Ri}{Ri^c} + \kappa_{\text{shear}}, & \text{for } 0 \leq Ri < Ri^c \\
    \kappa_0, & \text{for } Ri \geq Ri^c
  \end{cases}\\
  \label{local base closure diffusivity shear} \kappa_\text{shear} &= \frac{\nu_\text{shear}}{Pr_\text{shear}}, \\
  \label{local base closure diffusivity convection} \kappa_\text{conv} &= \frac{\nu_\text{conv}}{Pr_\text{conv}},
\end{align}
where $\nu_{\text{conv}}$, $\nu_{\text{shear}}$, $Ri^c$, $\Delta Ri$, $Pr_\text{conv}$, and $Pr_\text{shear}$ are parameters that are trained with the LES suite.
$\nu_{\text{conv}}$ and $\nu_{\text{shear}}$ represent the viscosity under convective and shear turbulence, while $Pr_\text{conv}$ and $Pr_\text{shear}$ are their respective turbulent Prandtl numbers.
$Ri^c$ is the critical Richardson number above which the flow is stable, and $\Delta Ri$ sets the width of the transition between the convective and shear regimes.
The other parameters are instead fixed: $\sigma_0 = \SI{1020}{kg.m^{-3}}$ is the reference potential density, $g = \SI{9.80665}{m.s^{-2}}$ is the gravitational acceleration, $\nu_0 = \SI{1d-5}{m^{2}.s^{-1}}$ is the background viscosity whose value is chosen to be small enough to have no impact on the training and validation tests, while $\kappa_0$ is calculated using the shear Prandtl number $\kappa_0 = \frac{\nu_0}{Pr_\text{shear}}$.

Physically, when $Ri$ drops below its critical value $Ri^c$ but remains positive (i.e., $0 \leq Ri < Ri^c$), the flow becomes shear unstable (but convectively stable), and the viscosity and diffusivity increase from their background values $\nu_0$ and $\kappa_0$ linearly to their shear values $\nu_{\text{shear}}$ and $\kappa_{\text{shear}}$ as $Ri$ decreases from $Ri^c$ to $0$. 
When $Ri$ drops below zero the flow becomes convectively unstable and the viscosity and diffusivity increase to their convective values $\nu_{\text{conv}}$ and $\kappa_{\text{conv}}$ over a transition width set by $\Delta Ri$. 

This formulation is inspired by the Pacanowski--Philander model~\cite{pacanowski_parameterization_1981}, but differs from it by using different functional forms and accounting for different mixing rates in the convective and shear regimes. 
Regardless, a purely $Ri$-based parameterization cannot fully capture the entire range of small-scale mixing in ocean BLs, as it does not account for nonlocal dynamics such as entrainment~\cite{zaron_new_2009}. 
Indeed, the Pacanowski--Philander model is not used in modern coarse-resolution ocean models.
However, in NORi the base closure is only the first step and nonlocal mixing processes will be captured with neural network components.

The base closure implies a change in both the magnitude of viscosity and the Prandtl numbers between the convective, shear, and stable regimes.
This is supported by the LES solution and mixing length arguments.
The eddy diffusivity/viscosity can be expressed as the product of a turbulent velocity and a mixing length scale.
Although the turbulent velocity spans similar ranges in all regimes, the mixing length scale is significantly longer in the convective regime, spanning the entire turbulent BL and resulting in stronger mixing of tracers.
In the convective regime, the turbulent Prandtl number is smaller than one because rotation slows momentum mixing by establishing a thermal wind when the vertical mixing time scale is longer than the Coriolis period~\cite{young_subinertial_1994}.
This is not relevant in the shear regime, which is associated with faster mixing time scales. 

The free parameters in the base closure are calibrated in an \textit{a posteriori} fashion by minimizing the mean squared difference between the column profiles and vertical gradients predicted by the closure and those diagnosed from the LES, for all scenarios and times.
The loss function $\mathcal{L}_\text{base}$ is defined as
\begin{align}
    &\label{loss function base closure} \mathcal{L}_\text{base}(\overline{u}, \overline{v}, \overline{T}, \overline{S}; \boldsymbol{\theta}_\text{base})
    = \frac{1}{N_\text{sim}} \sum_{a=1}^{N_\text{sim}} \left( \mathcal{L}_u^a + \mathcal{L}_v^a + \mathcal{L}_T^a + \mathcal{L}_S^a + \mathcal{L}_\sigma^a\right)\\
   &\hbox{with}\quad \mathcal{L}_\phi^a =  \frac{A_\phi^a}{N_tN_z} \sum_{i=1}^{N_z} \sum_{j=1}^{N_t} \left| \overline{\phi}^{a,i,j} - \overline{{\phi}}^{a,i,j}_{LES} \right|^2 + \frac{A_{\partial_z \phi}^a}{N_tN_z} \sum_{i=1}^{N_z} \sum_{j=1}^{N_t} \left| {\partial_z \overline{\phi}}^{a,i,j} - {\partial_z \overline{\phi}}^{a,i,j}_{LES} \right|^2,
   \label{eq:lossfunction}
\end{align}    
where $\overline{\phi}$ and $\overline{\phi}_{LES}$ are the profiles of one of the variables (velocity, temperature, salinity) as a function of vertical level $i$, time $j$, and scenario $a$---the overbars indicate that we are only parameterizing the area-averaged profiles.
Normalization factors $A_\phi^a$ are adaptively calculated based on initial conditions and losses in iteration 0 before training to ensure balanced learning in all variables and loss components.
This is done through a two-stage process.
Firstly, the magnitudes of the temperature and salinity losses are reweighted to be inversely proportional to their contributions to density variations, preventing either variable from dominating the loss. 
The density loss function is then weighted such that
\begin{equation}
    \frac{\mathcal{L}_\sigma^a}{\mathcal{L}_T^a + \mathcal{L}_S^a} = \frac{1}{9}.
\end{equation}
The $1/9$ ratio is chosen to ensure that the density contribution to the loss function acts as a weak constraint compared to temperature and salinity ones.
This lets density guide the optimization process without overwhelming the temperature and salinity signals.
The value of $1/9$ is chosen through experimentation, and the calibration is insensitive to it as long as the density weighting is small enough for the density contribution to remain a weak constraint relative to the temperature and salinity losses.
Momentum losses are normalized to match the sum of density, temperature, and salinity losses.
The same weights are also applied to their respective gradient loss counterparts.
Secondly, after these physical weightings are applied, the total profile losses and gradient losses are rescaled to be equal.
The mathematical details of how normalization factors are computed are provided in \ref{section appendix loss scalings}.

The free parameters $\boldsymbol{\theta}_\text{base} = \left[ \nu_\text{conv}, \nu_\text{shear}, Ri^c, \Delta Ri, Pr_\text{conv}, Pr_\text{shear} \right]^\top$ in the base closure are calibrated with Ensemble Kalman Inversion (EKI) using the EnsembleKalmanProcesses.jl package~\cite{iglesias_ensemble_2013,dunbar_ensemblekalmanprocessesjl_2022}, a gradient-free optimization algorithm well-suited for problems with a small number of parameters.
Since there are only six tunable parameters in the base closure, we can afford a large ensemble (200 members) to explore the parameter space more effectively than gradient-based methods and take advantage of EKI's robust performance when dealing with noisy loss landscapes~\cite{gjini_ensemble_2025}.

Free parameters in the base closure are calibrated in two stages.
Firstly, we calibrate only the shear-relevant parameters $\nu_\text{shear}$, $Ri^c$, and $Pr_\text{shear}$ using LES that are purely wind-driven as shown in Table~\ref{table wind driven LES suite}, freezing the convection-relevant parameters that are never activated in these scenarios since $Ri \geq 0$.
Secondly, we freeze the shear-relevant parameters and calibrate the convection-relevant parameters $\nu_\text{conv}$, $\Delta Ri$, and $Pr_\text{conv}$ using LES that contain convective forcing as shown in Tables~\ref{table free convection LES training suite}, \ref{table wind + convection LES training suite}, and \ref{table misc LES training suite}.
Even though the base closure is unable to capture nonlocal entrainment, free convection cases are still useful to calibrate the convection-relevant parameters which quantify the strength of local convective mixing.
With this two-stage approach, we aim to minimize error compensation during the calibration, ensuring that we are capturing the physics in a structurally and parametrically correct way.
The calibrated parameters of the base closure are $\nu_\text{conv} = \SI{0.761}{m^{2}.s^{-1}}$, $\nu_\text{shear} = \SI{6.16d-2}{m^{2}.s^{-1}}$, $Ri^c = 0.437$, $\Delta Ri = 9.70 \times 10^{-3}$, $Pr_\text{conv} = 0.175$, and $Pr_\text{shear} = 1.08$.

The base closure is capable of capturing the evolution of the vertical profile of the BLs under training and testing scenarios where instabilities are driven purely by surface wind stresses.
As expected from a purely local parameterization, the base closure underestimates the mixed layer depth when convection is present, since entrainment is a nonlocal process (as discussed in Section~\ref{section LES to 1D model}). 
The bottom panels of Figure~\ref{figure local base closure results} show examples of vertical profiles of temperature, salinity, and potential density from LES forced with four different air-sea fluxes and the corresponding predictions from the base closure. 
The base closure is unable to fully capture the deepening of the BL in the free convection training case (second from left), which could be attributed to the missing nonlocal entrainment physics.
The other three training and validation cases shown in Figure~\ref{figure local base closure results} include purely wind-driven mixing in rotational and irrotational cases, as well as (destabilizing) wind stress combined with (stabilizing) heating and precipitation.
In all three cases, mixing is purely local and the base closure is able to accurately predict the depth of the mixed layer.
Compared with traditional physical closures, the base closure is comparable with $k$-$\epsilon$ (first column of Figure~\ref{figure vs k epsilon results}) and CATKE (first column of Figure~\ref{figure vs CATKE results}).
Since the base closure performs comparably to CATKE, which has been shown to outperform KPP~\cite{wagner_formulation_2025}, the base closure also outperforms KPP in scenarios without convection.

\section{``NO'': Using neural networks to capture nonlocal entrainment} \label{section neural network}
We now use neural networks as a residual model to capture the nonlocal entrainment physics missing from the base closure.

\subsection{Enforcing physical knowledge and constraints on neural networks with architecture design} \label{subsection neural network architecture}
In NORi, the neural networks are used to predict the missing entrainment fluxes of temperature and salinity at the base of the BL.
Initially, we included additional networks for $u$ and $v$, but they did not improve skill despite a substantial increase in training cost.
This is likely because the momentum fluxes are dominated by inertia-gravity wave signals rather than turbulent fluctuations, obscuring the entrainment signal that contributes to vertical mixing.
Through our experimentation, we also found that entrainment fluxes in $T$ and $S$ still lead to enhanced momentum mixing at the base of the BL, because the entrainment-modified stratification lowers $Ri$, which raises the eddy viscosity of the base closure.

Mathematically, we add nonlocal terms $J_{\mathbb{NN}_T}$ and $J_{\mathbb{NN}_S}$ which are computed using neural networks $\mathbb{NN}_T$ and $\mathbb{NN}_S$ to Equations~\eqref{T local closure} and \eqref{S local closure},
\begin{align}
    \overline{w^\prime T^\prime} = J_{T,\text{ local}} + J_{T,\text{ nonlocal}}  &= -\kappa \pd{\overline{T}}{z} + J_{\mathbb{NN}_T} \\
    \overline{w^\prime S^\prime} = J_{S,\text{ local}} + J_{S,\text{ nonlocal}}  &= -\kappa \pd{\overline{S}}{z} + J_{\mathbb{NN}_S}.
\end{align}
By construction, the surface and bottom fluxes of $J_{\mathbb{NN}_T}$ and $J_{\mathbb{NN}_S}$ are set to zero, ensuring that the neural network only redistributes tracers within the water column without adding or removing them from the domain.
The output of each neural network $\mathbb{NN}_T$ and $\mathbb{NN}_S$ consists of only a scalar value that represents the nonlocal flux at the target grid point.
To infer the missing entrainment fluxes along the column, the same neural networks are ``convolved'' along the vertical dimension, i.e., the weights of the neural networks are shared among all vertical grid points.
The inputs to $\mathbb{NN}_T$ and $\mathbb{NN}_S$ are the temperature gradient, salinity gradient, potential density gradient, and gradient Richardson number of five grid points within the neighborhood of the grid point intended for inference, as well as the surface buoyancy flux.
The information of local shear is encoded in the local Richardson number $Ri$ while information regarding nonlinear seawater thermodynamics from TEOS-10~\cite{roquet_accurate_2015} is encoded in the potential density gradient inputs.
Location-dependent inputs (neighborhood tracer gradients and $Ri$) change accordingly as the neural networks are vertically ``convolved''.
This design makes the effect of entrainment local to the BL base, so only local information about the flow is required to predict the effects of entrainment-driven mixing. 
However, nonlocal information of the surface buoyancy flux is provided as inputs to predict the strength of the convective plumes, consistent with evidence that the entrainment fluxes scale nonlocally with air-sea fluxes~\cite{deardorff_laboratory_1980,van_roekel_kpp_2018}.
More technical details on the neural network architecture are provided in \ref{section appendix neural network architecture}.

Under this formulation, the neural networks' behavior is independent of depth, implicitly assuming that entrainment depends only on the local flow structure and the surface buoyancy flux, not on the absolute depth of the BL base.
This assumption is valid for BLs with depths up to roughly a kilometer.
In very deep BLs, rotational effects partly suppress entrainment as the time scale for convective plumes to reach the BL base becomes comparable to the Coriolis time scale~\cite{marshall_openocean_1999}.
Such deep BLs are only found in regions of dense water formation at high latitudes and require a dedicated parameterization that includes both vertical mixing by convection under the influence of rotation and lateral restratification by lateral baroclinic instabilities~\cite{boccaletti_mixed_2007}.
NORi, like most ocean BL parameterization schemes, does not explicitly account for deep convection physics.

Taking advantage of the fact that entrainment creates upgradient fluxes near the base of the BL, the neural networks are active for inference within five grid points below and ten grid points above the base of the BL, diagnosed as the shallowest point at which the gradient Richardson number exceeds the critical value $Ri^c$.
Five grid points below the BL base, equivalent to $\SI{40}{m}$, proved sufficient to capture plume-driven entrainment in all LES spanning the wide range of forcings considered.
While the upgradient fluxes are located near the base of the BL, they bring tracers into the mixed layer where they are rapidly redistributed across the entire BL by the BL-filling convective plumes.
Ten grid points above the BL base proved necessary to keep tracers well mixed over the BL.
In effect, the neural networks have implicitly learned to detect the base of the BL through the Richardson number criterion and to apply entrainment fluxes in its vicinity.
We experimented with using fewer and more grid points above and below the base of the BL and found that this configuration gave us the best results based on the LES we have run for the training suite.

\subsection{Neural ODEs: online calibration to promote accuracy, generalizability, and stability}

The 1D column model consists of a system of ODEs for two velocity components, temperature, and salinity, with neural networks embedded in the temperature and salinity equations.
Two separate neural networks that represent the nonlocal turbulent fluxes of temperature and salinity are embedded in their respective equations.
The neural ODEs are calibrated \textit{a posteriori} to promote numerical stability and generalizability of the solutions, as explained in the introduction.
This means that the loss $\mathcal{L}$ is a function of the full time history of the evolving variables rather than the subgrid fluxes.
Mathematically, for some integration time from $t_0$ to $t_1$ where the variables are time-marched forward,
\begin{equation}
    \mathcal{L}(\overline{u}, \overline{v}, \overline{T}, \overline{S}; \boldsymbol{\theta}_T, \boldsymbol{\theta}_S) = \mathcal{L} \left( {\overline{u}}(t),  {\overline{v}}(t),  {\overline{T}}(t),  {\overline{S}}(t),  {\overline{\sigma}}(t) \right),
\end{equation}
where $\boldsymbol{\theta}_T$ and $\boldsymbol{\theta}_S$ are the trainable weights of the neural networks, and the variables in $\mathcal{L}$ are the full time series from $t_0$ to $t_1$.
The potential density $\sigma$ is included in the loss because it is a nontrivial function of temperature and salinity due to nonlinearities in the equation of state. 

The neural networks are trained with \textit{a posteriori} calibration.
We utilize automatic differentiation (AD) to compute how loss changes with respect to the weights of the neural networks.
AD propagates derivatives through the ODE solver, enabling gradient-based updates to parameters.
In the context of the parameterization problem addressed in this study, the climate model acts as the forward model, implying that online calibration requires the climate model to be entirely differentiable.
However, contemporary climate models are not differentiable.
Additionally, due to the serial-in-time nature of online calibration, training requires substantial computational power and memory.
Below, we describe how we were able to overcome both challenges.
First, however, we explain why we did not resort to \textit{a priori} calibration, which is a common strategy to simplify the calibration task.  

\textit{A priori} (or offline) calibration trains the neural networks to match the turbulent fluxes diagnosed from the LES as a function of the inputs with no information about the temporal evolution of the system.
Mathematically, offline training for our problem would take the form
\begin{equation}
     \mathcal{L}_\text{offline}(\overline{u}, \overline{v}, \overline{T}, \overline{S}; \boldsymbol{\theta}_T, \boldsymbol{\theta}_S) = \mathcal{L} \left( \overline{u^\prime w^\prime}, \overline{v^\prime w^\prime}, \overline{w^\prime T^\prime}, \overline{w^\prime S^\prime} \right).
\end{equation}
This approach led to numerical instabilities within a few time steps when applied to our parameterization problem. 
The reason is that the diagnosed horizontally averaged LES turbulent fluxes contain irrelevant signal, like internal waves, in addition to that relevant to entrainment: linear internal waves generate fluctuations in all variables but have no impact on entrainment, and discretization errors are unavoidable. 
Although regularization techniques can be used to reduce the impact of noise contamination in training data, this is not appropriate for our purposes because the target entrainment flux signal is characterized by sharp gradients at the BL base. 
Any smoothing or further coarse-graining of its vertical profile would affect the stratification at the base of the BL, which is a crucial variable in setting the deepening of the BL itself.
As a result, \textit{a priori} calibration results in parameterizations of fluxes with a small amount of grid-scale noise.
This noise inevitably accumulates as the neural ODEs are integrated forward in time, eventually leading to a finite-time blowup.
We speculate that this is due to the iterative rollout of neural ODEs and the lack of explicit variance-dissipation mechanisms in Equations~\eqref{equation u 1D} through \eqref{equation S 1D}. 
As the neural networks receive noisy inputs (the fluxes), they generate noisy outputs.
Lacking strong damping, this noise accumulates, leading to blowup.
However, this approach might succeed with substantially more training data, but at considerably greater cost.

The loss function used for \textit{a posteriori} calibration of the neural networks has the same form as the one used to train the base closure and given in Equation~\eqref{eq:lossfunction} except for the omission of the momentum fields, $\mathcal{L}_u^a$ and $\mathcal{L}_v^a$.
Including the momentum terms in the loss function significantly deteriorates the quality of the training, because the velocity fields are dominated by inertial oscillations, which have no impact on entrainment.
In an attempt to capture them, neural networks deteriorate the quality of $T$ and $S$ predictions.
However, flow shear information is implicitly encoded in the local Richardson number $Ri$, which is invariant to inertial oscillations and Galilean transformations.
We also found that including vertical derivatives of tracers in the loss function during optimization promotes smooth solutions free of grid-scale oscillations about the training data, acting as a form of regularization.
Training is carried out over $N_\text{sim} = 62$ simulations outlined in Tables~\ref{table wind driven LES suite}, \ref{table free convection LES training suite}, \ref{table wind + convection LES training suite}, and \ref{table misc LES training suite}.
The details of how normalization factors are calculated are discussed in Section~\ref{section local base closure} and \ref{section appendix loss scalings}.

At the time of NORi's development, there were no fully differentiable ocean models that provided the functionality and interface required for our \textit{a posteriori} calibration.
Therefore, we built a standalone implementation of the 1D BL model described in Equations~\eqref{equation u 1D} through \eqref{equation S 1D} in the Julia programming language~\cite{bezanson_julia_2017}.
At the time of writing, this training is likely to become feasible directly in Oceananigans.jl in the near future, due to recent developments in its automatic differentiation capabilities~\cite{moses_dj4earth_2025}.

The neural ODEs~\eqref{equation u 1D} through \eqref{equation S 1D} are quite stiff~\cite{kim_stiff_2021} due to the multiscale nature of the BL physics, where the diffusion terms, Coriolis terms, and neural network entrainment terms evolve on three significantly different length and time scales.
Explicit time stepping schemes are thus prohibitively expensive for both the forward and backward passes.
Instead, we implemented a split implicit-explicit algorithm, where the diffusion term is time stepped with the implicit Euler method and the nonlocal flux is time stepped with the explicit Euler method.
The neural ODEs are time stepped by iterating over the nondimensional, discretized equations with a constant $\Delta t_\text{train} = \SI{10}{minutes}$ during training and $\Delta t_\text{validation} = \SI{5}{minutes}$ during validation.
(The discretized equations are provided in \ref{section appendix discretization} and the nondimensionalization is derived in \ref{section appendix normalization}.)
As we will also show in Section~\ref{section results column}, NORi is insensitive to time step size up to $\SI{1}{hour}$, so with a training time step of $\SI{10}{minutes}$, explicit higher-order time stepping schemes are not required.

The neural networks are built employing Lux.jl~\cite{pal_lux_2023}, with the standard Glorot weight initialization technique~\cite{glorot_understanding_2010}.
During training, both input and output data are normalized to a mean of zero and a variance of one across all training simulations.
The backpropagation of the weight gradients of the neural network with respect to the loss function is accomplished through the Enzyme.jl AD tool~\cite{moses_instead_2020} in the Julia language; see Equations~\eqref{u IMEX} through \eqref{S IMEX}.
The weights of the neural network are updated using the Adam optimizer~\cite{kingma_adam_2017}.
Through a grid hyperparameter sweep over different activation functions, neural network layers, and the number of units in the hidden layers, we found that 3 hidden layers with 128 units each and rectified linear units~\cite<ReLU,>[]{glorot_deep_2011} provided the best performance as computed following the loss function introduced above.
Each neural network in the final configuration contains 35,969 parameters.
To avoid local minima due to the complex loss landscape, we employ a ``curriculum learning'' strategy.
This involves training neural ODEs to make accurate and numerically stable predictions over progressively extended integration time frames.
Integration times are increased in three stages with integration windows of 15, 23.3, and 43.3 hours, with the learning rate decreased at each stage.
These values carry no particular significance apart from corresponding to the 90th, 140th, and 260th time steps at a 10-minute time step.
In each stage, the neural ODEs are trained for 2,000 epochs, and the initial model weights of each stage are selected based on the lowest training loss in the previous stage.
At the beginning of each stage, the losses are renormalized, i.e., the $A_\phi^a$ factors in the loss function are recalculated to promote a good gradient flow towards optimal model performance.
The final model weights are selected on the basis of the lowest validation loss in the final stage.
We used a total of 62 training examples to train neural ODEs, which is equivalent to 62 different initial and boundary conditions integrated forward in each epoch.
A ``full-batch gradient descent'' approach (where the loss is computed across all training simulations for each gradient descent step) is taken, as we found that performing minibatching worsens loss convergence.
Total training takes approximately $\SI{90}{hours}$ wall-clock time on a single CPU core on the Supercloud computing cluster~\cite{reuther_interactive_2018}.
We chose CPU training because of Enzyme.jl's more mature CPU support at the time of training NORi, but in future iterations it would be useful to train natively on Oceananigans.jl with GPUs.

\section{Training, validating, and comparing NORi in single-column contexts} \label{section results column}
All single-column experiments in this section use a vertical grid resolution of $\SI{8}{m}$, the grid resolution on which NORi is trained.
We chose this resolution to match that of operational global ocean models, which is $\mathcal{O}(\SI{10}{m})$ in the entrainment region in the upper ocean.

\subsection{Training and validation loss}
\begin{figure}[htbp]
\centering
\includegraphics[trim={0 8cm 0 4.5cm}, width=\textwidth]{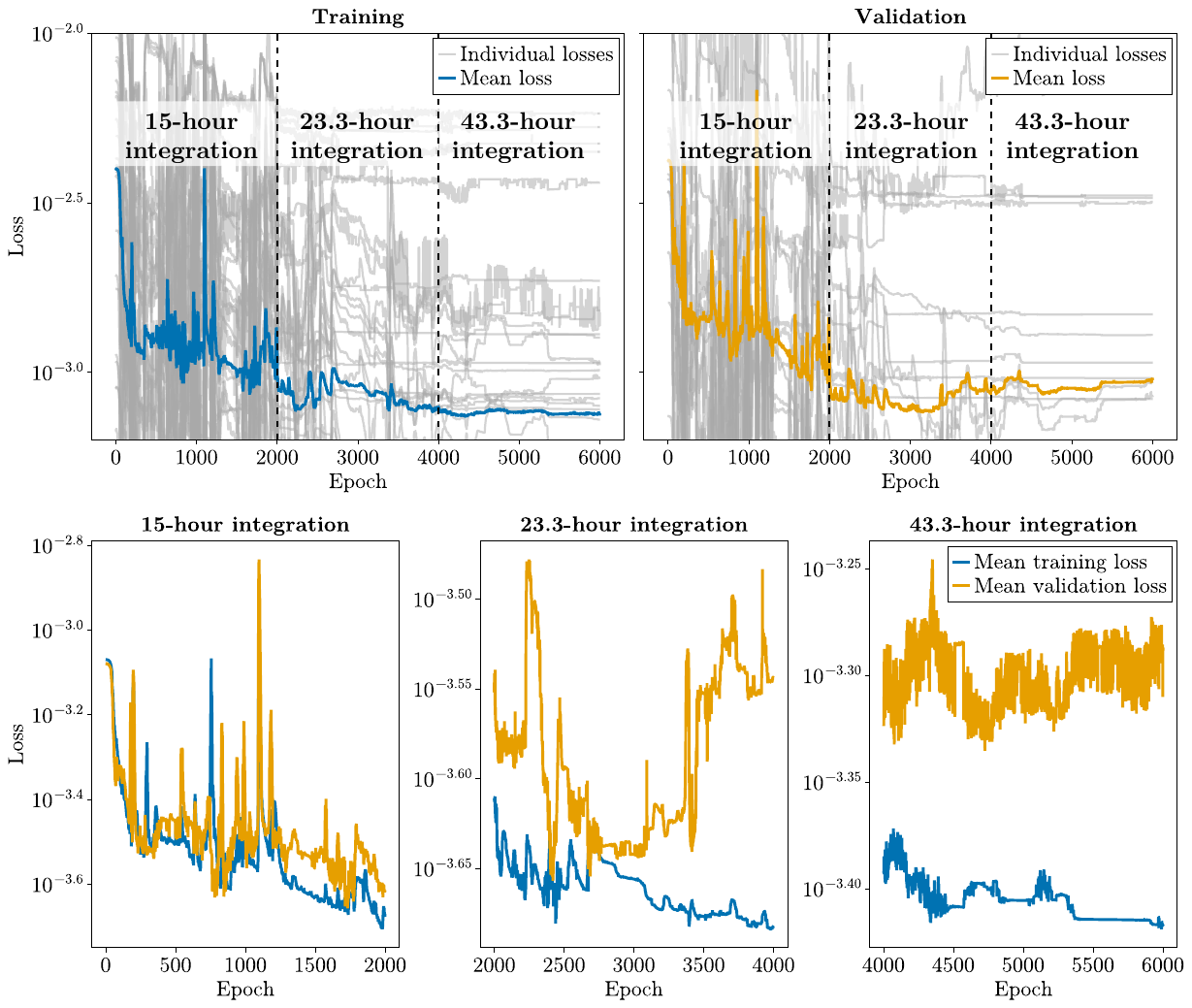}
\caption{Training and validation losses of NORi.
        Top row: mean and individual losses as a function of epoch over the final integration horizon of $\SI{43.3}{hours}$ for the training suite (Tables~\ref{table free convection LES training suite}, \ref{table wind + convection LES training suite}, and \ref{table misc LES training suite}) in the left panel and the validation suite (Table~\ref{table LES validation suite}) in the right panel.
        For the plots shown in this upper row, all losses are normalized once, at epoch 0---an epoch is defined as one iteration over the entire training suite.
        The gray lines are the individual losses for each training/validation case, while the colored lines are their means.
        Bottom row: mean training and validation losses over the three stages of ``curriculum learning,'' where each subsequent stage covers a longer integration period: $\SI{15}{hours}$, $\SI{23.3}{hours}$, and $\SI{43.3}{hours}$.
        The training loss is re-normalized at the start of each stage, as explained in \ref{section appendix loss scalings}.
        The epoch with the lowest training loss at each stage is used to initialize network weights for the next stage.
        The final model weights are selected from the model with the lowest validation loss in the final stage to avoid overfitting the model to the data.}
\label{figure NDE loss}
\end{figure}

We trained NORi on 62 cases listed in Tables~\ref{table wind driven LES suite}, \ref{table free convection LES training suite}, \ref{table wind + convection LES training suite}, and \ref{table misc LES training suite} and validated it on another 30 cases shown in Table~\ref{table LES validation suite} to select the final model weights.
The last two testing cases in Table~\ref{table wind driven LES suite} are not used to validate the components of the neural network, as they lack the nonlocal convection-driven entrainment that the neural networks are designed to capture.
Figure~\ref{figure NDE loss} shows that the neural network losses decreased during the training epochs, approaching convergence after about 4,000 epochs.
Training losses remain small in cases that begin with small losses, indicating that neural networks augment the base closure only when there are significant residual errors.
We also find that the ``curriculum'' learning approach, where neural networks are trained on progressively longer integration windows, reduces the loss over the final $\SI{43.3}{hours}$ more effectively than training directly on the full integration window.
The bulk of loss reduction occurs in the first stage of training, which covers the first $\SI{15}{hours}$, suggesting that once neural networks learn to capture entrainment physics early on, they tend to generalize reasonably well to the rest of the integration window.
The longer integration windows in the second and third stages serve to fine-tune the model weights and enhance the numerical stability of the solutions over longer time frames.
The validation losses follow a similar trend to the training losses, indicating that NORi interpolates well across the physical regimes it is trained on.

\subsection{Inference performance against large-eddy simulations}

\begin{figure}[htbp]
\centering
\includegraphics[trim={0 0 0 0}, width=\textwidth]{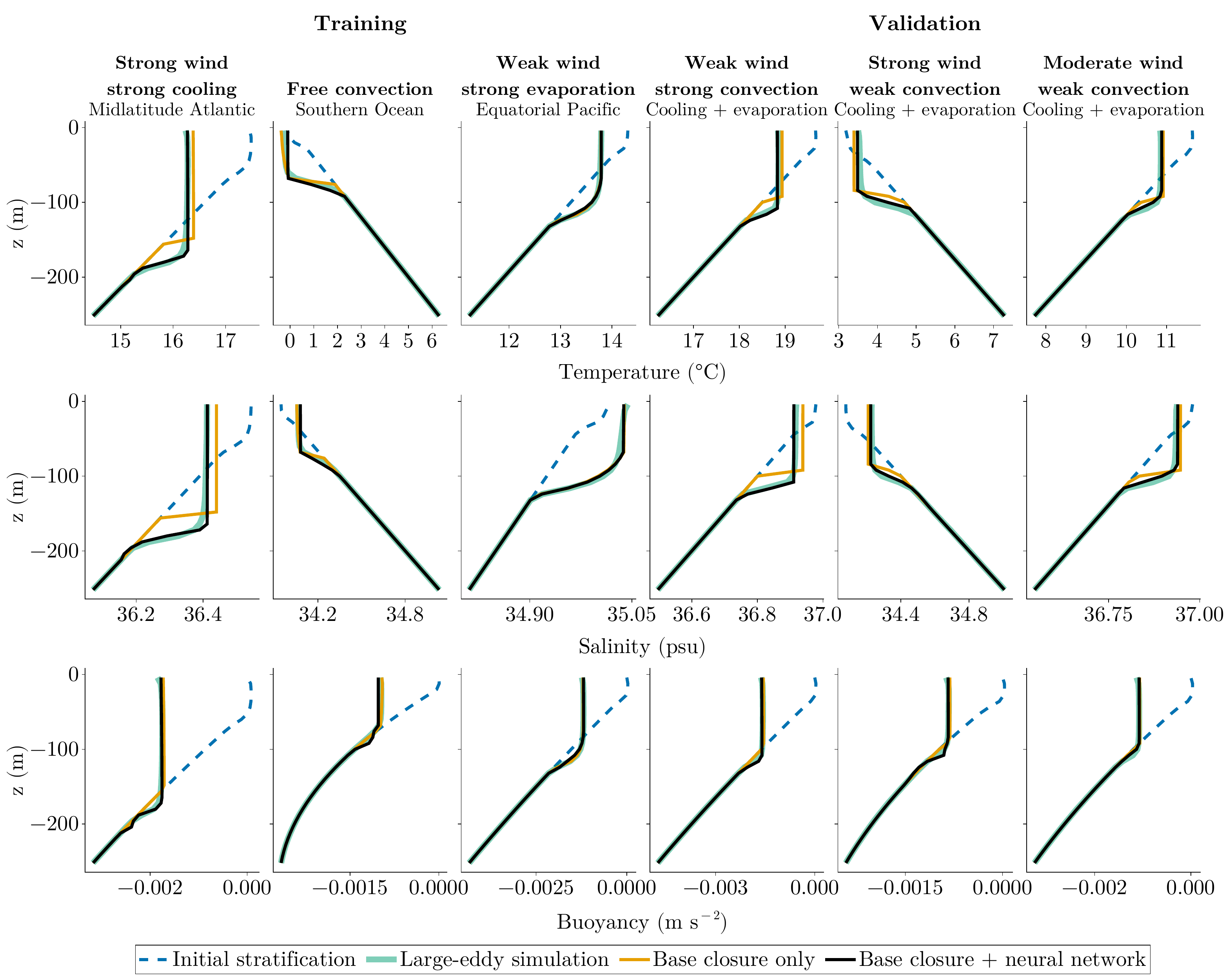}
\caption{Temperature, salinity, and buoyancy profiles for selected training (columns 1 through 3) and validation (columns 4 through 6) examples generated with NORi (black lines), the base closure (orange lines), and area-averaged large-eddy simulation (LES) solutions (green lines). 
The profiles are computed $\SI{1.75}{days}$ after initialization (dashed lines).}
\label{figure NDE training results}
\end{figure}

Once trained, NORi matches the LES solutions in all training cases and demonstrates strong predictive skill in the validation cases.
This includes the cases where the base closure was inadequate, indicating that NORi effectively captures entrainment, which is the primary deficiency of the base closure. 

We illustrate the improvement of NORi over the base closure in six representative cases: three training cases and three validation cases in Figure~\ref{figure NDE training results}. 
NORi matches the LES solutions in the training cases and approximates the validation cases well, showing substantial improvement over the base closure in simulations where the entrainment is strong. 
The first column of Figure~\ref{figure NDE training results} shows the evolution of a BL subject to strong winds and cooling.
The initial stratification is derived from a linear fit to summertime Argo profiles in the midlatitude Atlantic, where the gradient is strong~\cite{gaillard_situbased_2016}.
During the first hours, a turbulent Ekman layer driven by wind develops at the surface, with depth $L_\text{Ek} \simeq 0.25f^{-1}\sqrt{ \left| J_u^{\text{top}} \right| } \sim \SI{70}{m}$~\cite{mcwilliams_fundamentals_2011}.
But soon afterwards convective plumes start to penetrate deeper, resulting in substantial entrainment captured by NORi, but not by the base closure. 
In contrast, the third column of Figure~\ref{figure NDE training results} presents a simulation in which both NORi and the base closure match the LES solution. 
This solution is representative of the equatorial Pacific, where $f = 0$ and therefore $L_{\text{Ek}} \rightarrow \infty$, so Ekman dynamics do not apply.
Despite strong evaporation, momentum shear reaches the base of the BL and dominates over convection, resulting in minimal entrainment. 
These two examples demonstrate that NORi correctly reduces to the base closure when shear dominates turbulence and deepening scales with $Ri$, while it improves on the base closure when convection drives entrainment. 

The fourth through sixth columns of Figure~\ref{figure NDE training results} show validation examples that were not seen during training. 
NORi generalizes well to these unseen scenarios, confirming that the neural network effectively interpolates within the range of air-sea fluxes and stratification values seen in training. 
Should new cases emerge where NORi fails to match LES or observational measurements of BL evolution, its high expressivity and flexibility give us confidence that it can be retrained through fine-tuning or transfer learning to capture the new physics.

Importantly, NORi works well in the nonlinear thermodynamics regime, as demonstrated in the second and fifth columns of Figure~\ref{figure NDE training results}, where conditions similar to those of the Southern Ocean with low temperatures, destabilizing temperature stratifications, and salinity-dominated buoyancy structures are illustrated. 
The initial temperature and salinity stratifications are linear with depth, but the resulting buoyancy is not. 
NORi captures the entrainment of temperature, salinity, and buoyancy, which behave differently due to the nonlinear equation of state and are missed by the base closure.

It is worth remarking that the base closure matches the buoyancy LES profiles much better than those of temperature and salinity, whose errors compensate when combined into buoyancy.
This supports our argument that BL parameterizations must be validated with solutions and observations of both temperature and salinity profiles, rather than just buoyancy, as is often done in the literature.

\begin{figure}[htbp]
\centering
\includegraphics[trim={0 0 0 0}, width=\textwidth]{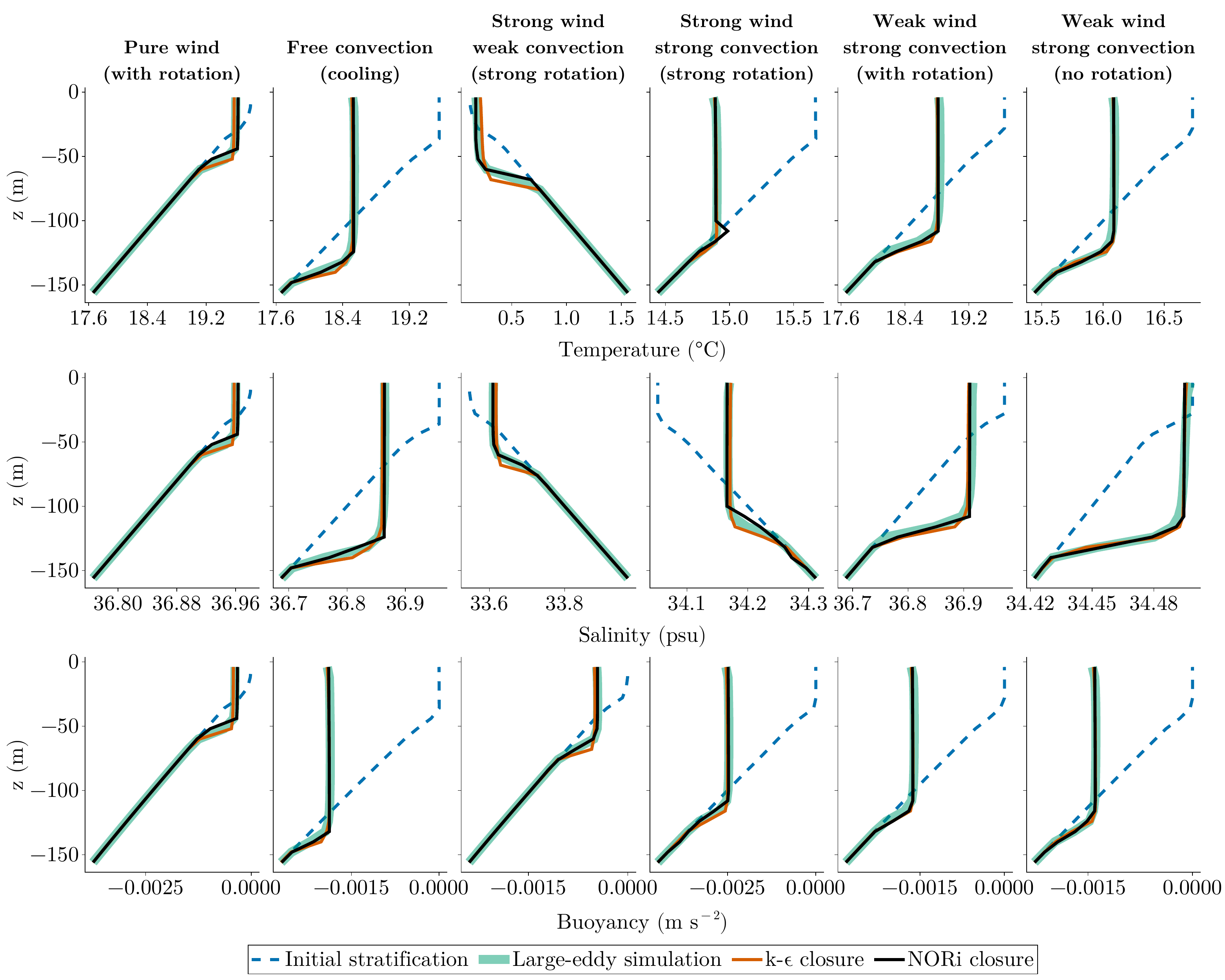}
\caption{Comparison of NORi and $k$-$\epsilon$ performance in a column-model setting. 
The model profiles of temperature, salinity, and buoyancy are plotted together with the area-averaged LES profiles. 
The examples shown are drawn from validation cases not seen by NORi during training (see Table~\ref{table LES validation suite} for a complete list). 
The profiles are computed 1.75 days after initialization (dashed lines).}
\label{figure vs k epsilon results}
\end{figure}

Figure~\ref{figure vs k epsilon results} shows that NORi compares very favorably with $k$-$\epsilon$, a state-of-the-art two-equation first-order turbulence closure that solves two additional prognostic equations for the turbulent kinetic energy $k$ and its dissipation $\epsilon$ to predict the turbulent fluxes $\overline{u^\prime w^\prime}$, $\overline{v^\prime w^\prime}$, $\overline{w^\prime T^\prime}$, and $\overline{w^\prime S^\prime}$.
The $k$-$\epsilon$ closure used in this study is the one implemented in Oceananigans.jl for oceanic applications and briefly described in \ref{section appendix k epsilon}. 
Despite its much simpler formulation, NORi, a leaner zero-equation model, produces single-column solutions that are practically indistinguishable from both the $k$-$\epsilon$ and LES solutions within the validation suite used in this work. 

NORi compares favorably also with CATKE~\cite{wagner_formulation_2025}, a recently developed 1.5-equation closure that estimates turbulent diffusivities for momentum and tracers using diagnostic equations for a mixing length and a prognostic equation for turbulent kinetic energy (TKE). 
The comparison is presented in Figure~\ref{figure vs CATKE results} in \ref{section appendix CATKE}.
Since NORi performs comparably to CATKE, which has been shown to outperform traditional parameterizations such as KPP~\cite{wagner_formulation_2025}, NORi would likewise outperform KPP for the scenarios examined in this study.
We attribute NORi's skill to a base closure that is physically simple and realistic yet expressive enough in structure for the neural networks to capture the missing physics correctly and stably, enabled by the right calibration strategy.
Therefore, we conclude that NORi is a promising candidate for coarse-resolution, long-time step simulations where few-equation closure models are preferred, though further testing in realistic configurations is needed to confirm this.

\subsection{Numerical stability} \label{subsection 1D long integration}

To assess whether NORi is numerically stable on time scales longer than the training horizon of $\SI{2}{days}$, we run NORi for $\SI{60}{days}$ in a single $\SI{768}{m}$ deep column with a constant grid resolution of $\SI{8}{m}$ and a time step of $\SI{5}{minutes}$. 
The simulation is initialized with a constant temperature gradient of $\pd{T_0}{z} = \SI{9.77d-3}{\celsius.m^{-1}}$ and a salinity gradient of $\pd{S_0}{z} = \SI{1.46d-3}{psu.m^{-1}}$, and surface values $T_0 (z = 0) = \SI{30}{\celsius}$ and $S_0 (z = 0) = \SI{37}{psu}$. 
Surface forcing is imposed through a constant momentum flux of $J_u^{\text{top}} = \SI{-1d-4}{m^2.s^{-2}}$ and time-dependent sinusoidal temperature and salinity fluxes
\begin{align}
    \label{equation temperature flux long integration 1D} J_T^{\text{top}} &= \left[ 2 \cos \left( \frac{2 \pi}{\SI{1}{day}} t \right) + 1  \right] \times 10^{-4} \, \si{\celsius.m.s^{-1}}, \\
    \label{equation salinity flux long integration 1D} J_S^{\text{top}} &= -2 \cos \left( \frac{2 \pi}{\SI{2.63158}{days}} t \right) \times 10^{-5} \, \si{psu.m.s^{-1}},
\end{align}
with a Coriolis parameter of $f = \SI{1d-4}{s^{-1}}$. 
The period $\SI{2.63158}{days}$ is chosen to be essentially incommensurate with one day so that the combined temperature and salinity fluxes generate irregular fluctuations in the surface buoyancy flux, primarily destabilizing, but occasionally stabilizing. 
Because they have different periods, the temperature and salinity fluxes are not in phase.
At certain times, they act in opposite directions (one stabilizing, the other destabilizing), while at other times they are jointly stabilizing or destabilizing.
Therefore, in addition to testing numerical stability, we test whether NORi can simulate the evolution of a BL under time-dependent air-sea fluxes despite being trained only in settings with constant air-sea fluxes. 
NORi correctly turns on the neural network to capture entrainment only when the buoyancy flux drives convection (destabilizing).

\begin{figure}[htbp]
    \centering
    \includegraphics[trim={0 0 0 0}, width=\textwidth]{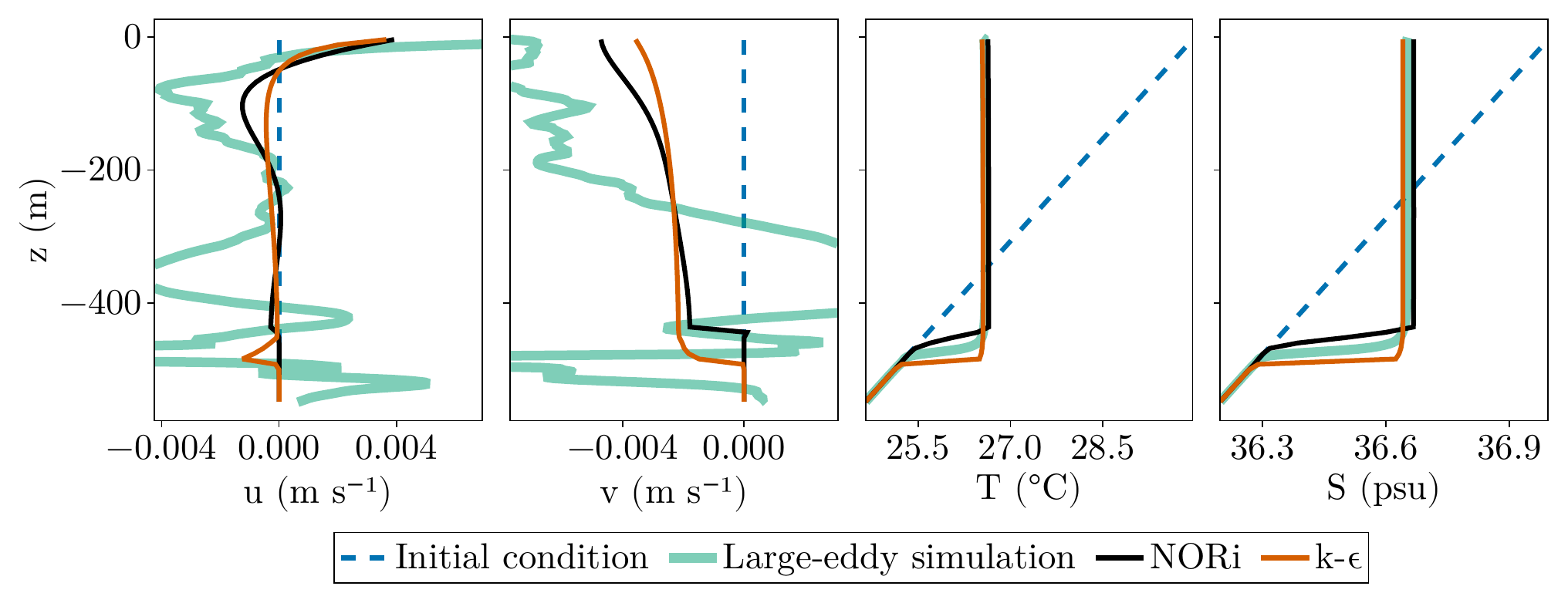}
    \caption{Velocity, temperature, and salinity profiles $\SI{60}{days}$ after initialization (dashed lines) for simulations forced with a constant wind stress ($J_u^{\text{top}} = \SI{-1d-4}{m^2.s^{-2}}$) and time-dependent heat and freshwater fluxes given by Equations~\eqref{equation temperature flux long integration 1D} and \eqref{equation salinity flux long integration 1D}. 
    The Coriolis parameter is set to $f = \SI{1d-4}{s^{-1}}$. 
    Solutions from NORi, the $k$-$\epsilon$ model, and the area-averaged LES are shown.}
\label{figure 1D model long integration}
\end{figure}

In Figure~\ref{figure 1D model long integration} we compare the NORi profiles of the velocity, temperature, and salinity fields at $\SI{60}{days}$ with the profiles of an LES and the $k$-$\epsilon$ model forced with the same time-dependent air-sea fluxes. 
We note that the momentum components $u$ and $v$ can be highly noisy in LES due to internal oscillations that do not contribute to entrainment, so the analysis of results against LES should focus on the temperature and salinity fields.
This also illuminates why the loss function used to train NORi does not include momentum terms (see our discussion in Section~\ref{section neural network}).

The profiles are quite similar, but differences do emerge: the $k$-$\epsilon$ model produces a mixed layer that is around $\SI{50}{m}$ deeper than that of NORi at the end of $\SI{60}{days}$.
Comparing both models with the LES solution, we see that NORi underestimates the penetration of mixing, while $k$-$\epsilon$ overestimates it, resulting in mixed layer biases of around $\SI{25}{m}$ in each direction. 
Further training over longer integration times may be warranted to improve NORi's representation of entrainment. 
As for $k$-$\epsilon$, there is literature confirming biases in free convection scenarios characterized by slightly stable stratification in the mixed layer and an increase in stratification at the BL base, both of which cannot be captured with purely diffusive closures~\cite{deardorff_theoretical_1972,legay_derivation_2025}.
That said, $k$-$\epsilon$ parameters have not been calibrated with the same suite of LES used to train NORi, potentially affecting its performance.
Regardless, the conclusion stands that NORi and $k$-$\epsilon$ are virtually indistinguishable over short time scales (see Figure~\ref{figure vs k epsilon results}), but small biases accumulate, leading to differences over longer time scales. 
Although these biases are minor at $\SI{60}{days}$, their impact on long global ocean simulations may be significant.

\subsection{Time step dependence}

\begin{figure}[htbp]
    \centering
    \includegraphics[trim={0 0 0 0}, width=\textwidth]{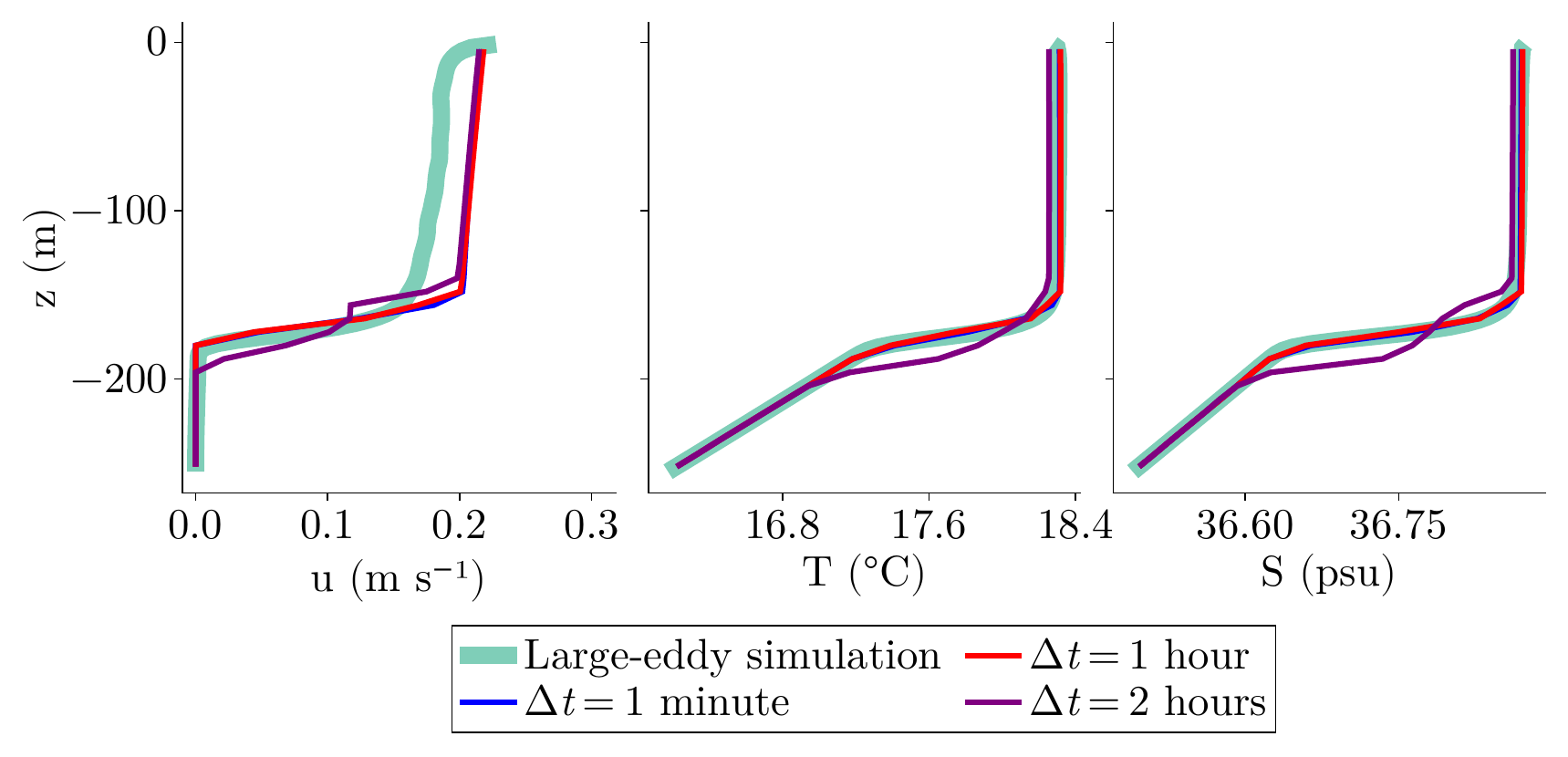}
    \caption{Velocity, temperature, and salinity profiles \SI{4}{days} after initialization generated with NORi using a range of time steps. The surface forcing parameters are given in the text. 
    The $v$-component is zero because the simulation is non-rotating and the surface momentum stress acts in the $x$-direction. 
    The LES solution is also shown for reference.}
    \label{figure NN 1D timestep}
\end{figure}

\begin{figure}[htbp]
    \centering
    \includegraphics[trim={0 0 0 0}, width=\textwidth]{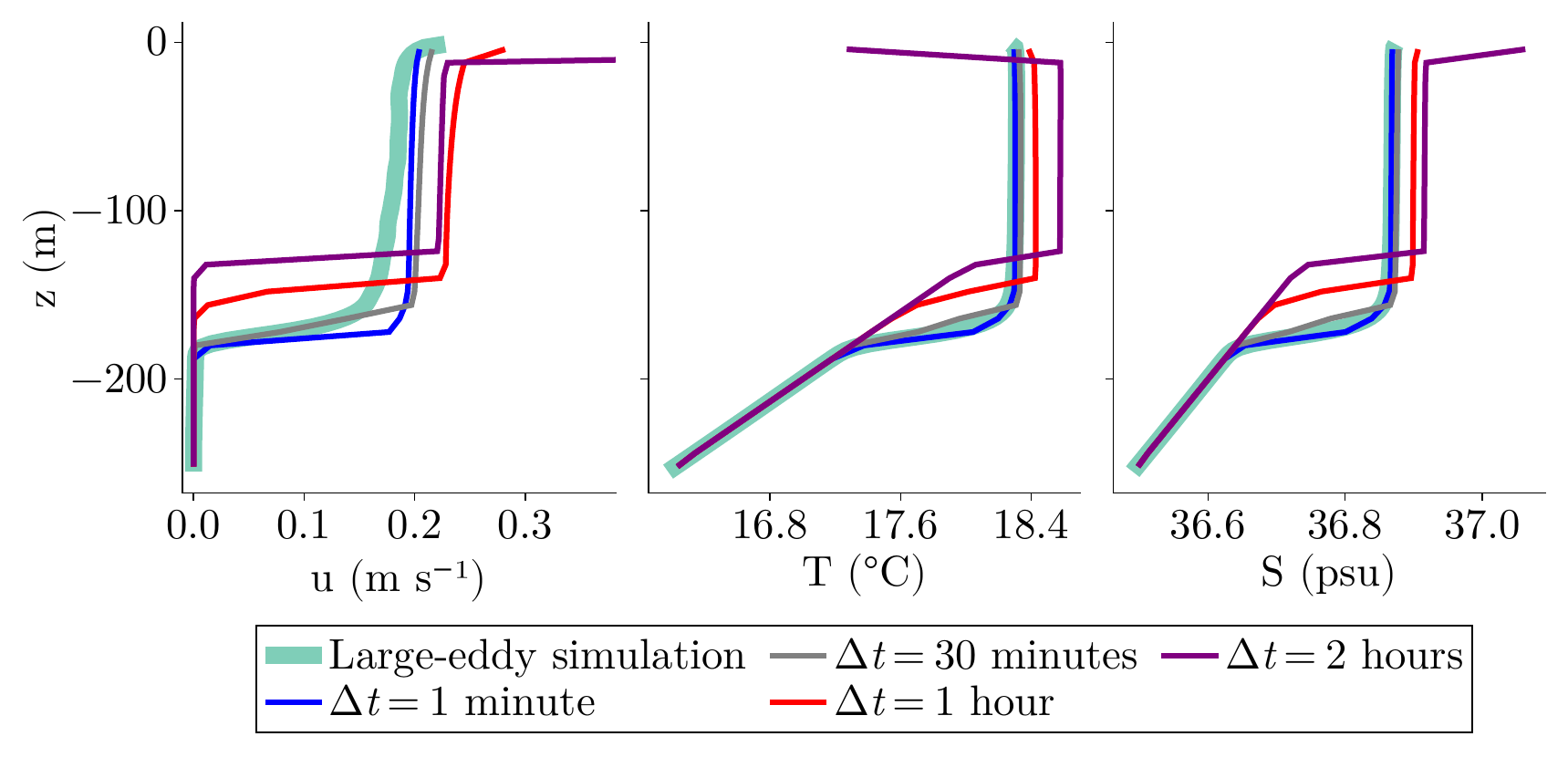}
    \caption{Same as Figure~\ref{figure NN 1D timestep} but computed with the $k$-$\epsilon$ model.}
    \label{figure kepsilon 1D timestep}
\end{figure}

An important consideration in the development of parameterizations for large-scale ocean models is the time step required to get stable and accurate solutions. 
To address this point, we run NORi and the $k$-$\epsilon$ model with increasing time steps from $\SI{1}{minute}$ to $\SI{2}{hours}$. 
To focus on the limitations of the time steps associated with the parameterizations and not with the inertial oscillations, we set $f=0$.
The simulations are initialized with a temperature gradient of $\pd{T_0}{z} = \SI{1.5d-2}{\celsius.m^{-1}}$ and a salinity gradient of $\pd{S_0}{z} = \SI{2d-3}{psu.m^{-1}}$, with surface values $T_0 (z = 0) = \SI{20}{\celsius}$ and $S_0 (z = 0) = \SI{37}{psu}$.
The surface fluxes are set to $J_u^{\text{top}} = \SI{-1d-4}{m^2.s^{-2}}$, $J_T^{\text{top}} = \SI{2d-4}{\celsius.m.s^{-1}}$, and $J_S^{\text{top}} = \SI{-2d-5}{psu.m.s^{-1}}$, respectively. 
The models are integrated for $\SI{4}{days}$.
Figures~\ref{figure NN 1D timestep} and \ref{figure kepsilon 1D timestep} show the final velocity, temperature, and salinity profiles for the different time steps computed using the NORi closure and the $k$-$\epsilon$ closure.

As shown in Figure~\ref{figure NN 1D timestep}, the NORi solutions are independent of time step for time steps shorter than $\SI{1}{hour}$. 
The solution begins to deviate from the LES solution only when the time step is increased to $\SI{2}{hours}$; the mixed layer becomes too deep because NORi begins to overestimate mixing.
This time step constraint appears to be set by the diffusive time scale associated with the diffusivity at the base of the BL rather than by the neural networks and may be further extended by using more accurate discretization schemes. 
In contrast, as seen in Figure~\ref{figure kepsilon 1D timestep}, the $k$-$\epsilon$ model achieves numerical convergence only for time steps below $\SI{30}{minutes}$, while it underestimates mixing for longer time steps. 
It is well known that two-equation BL models, such as $k$-$\epsilon$, require shorter time steps than zero-equation ones. 
Indeed, this is the main reason why they are not more commonly used in large-scale coarse-resolution ocean models that require longer time steps~\cite{reichl_simplified_2018, wagner_formulation_2025}. 
In summary, NORi can be used with time steps up to $\SI{1}{hour}$ without loss of accuracy compared to $T$ and $S$ profiles from LES solutions, making it a good candidate for use in coarse-resolution ocean simulations. 
It should also be acknowledged that, regardless of time steps, the velocity profiles produced by both NORi and $k$-$\epsilon$ show larger deviations from the LES solution than the tracer profiles do.
However, it appears that the details of the velocity profile are not key for capturing the vertical mixing of tracers.

In a highly unoptimized, naive first implementation of NORi in Oceananigans.jl where data are copied to and from the working arrays for neural network inference, NORi is around $\SI{15}{\%}$ slower than $k$-$\epsilon$ and around $\SI{45}{\%}$ slower than its unaugmented base closure on a per-time step basis.
Given that NORi remains accurate at time steps at least twice as long as those required by $k$-$\epsilon$, we conclude that there are computational advantages to using NORi over two-equation closures in large-scale simulations. 
At the same time, we have seen in Figure~\ref{figure 1D model long integration} that this set of trained NORi weights produces biases that are of the same magnitude as $k$-$\epsilon$ when compared to LES in our simulation over $\SI{60}{days}$, supporting the case for more comprehensive evaluation and testing of NORi for large-scale simulations.

\subsection{Ocean Weather Station Papa: validation against observations}

The validation of NORi by comparing with the library of LES solutions has been predicated on two assumptions. 
Firstly, the LES solutions accurately capture the physics of BL mixing. 
Secondly, the neural networks would be able to represent mixing under variable surface forcing despite being trained with LES solutions forced with constant fluxes. 
Both assumptions are now tested by running NORi as a column model under realistic settings based on observations from Ocean Weather Station Papa (OWS Papa). 
OWS Papa provides 70-year-long time series of atmospheric variables and temperature and salinity upper ocean profiles at $50.1^\circ$N, $144.9^\circ$W in the subpolar gyre of the Northeast Pacific~\cite{cronin_pmel_2023}. 
OWS Papa has been widely used to test single-column BL models as it is situated in a location where lateral fluxes are weak~\cite{burchard_comparative_2001}, making it an ideal testbed for evaluating vertical mixing parameterizations~\cite{large_oceanic_1994,kantha_improved_1994,yuan_kprofile_2024}. 

We used OWS Papa data from 2007 onward, when a new surface mooring was deployed, providing a self-consistent dataset. 
NORi was then run to simulate the fall to winter evolution of the BL for each one of eight years (we excluded years with more than 72 hours of gaps in the data). 
The simulations were initialized with observed temperature and salinity profiles from November 1 and integrated forward for $\SI{120}{days}$. 
NORi was forced with surface momentum, heat, and salinity fluxes computed from air-sea observations using the COARE3.0b algorithm~\cite{fairall_bulk_2003,cronin_assessment_2006}. 
We focused on fall to winter observations, when the BL deepens through processes parameterized in NORi.
The spring to summer evolution is instead dominated by BL restratification through mesoscale and submesoscale baroclinic instabilities that require information about lateral buoyancy gradients and are currently not included in the formulation of NORi. 

The changes in vertically integrated heat and salt content do not match the surface air-sea fluxes in the OWS Papa data. 
The mismatch is likely due to lateral advection of different water masses. 
In preliminary NORi runs, this led to large discrepancies between observed and simulated temperature and salinity profiles. 
To partially mitigate the issue, we subtracted a constant from the time-variable surface fluxes, computed as the difference between the time-integrated fluxes over the fall/winter season and the net change in temperature and salinity over the same period, integrated over the top $\SI{100}{m}$, which is the approximate depth of the mixed layer at the end of $\SI{120}{days}$. 
This constant departs from zero because the heat and freshwater budgets at the OWS Papa station are not closed. 
This correction is not closure-specific and is not tuned to improve NORi relative to the baselines. 
It is used only to remove the common budget mismatch that a one-dimensional column model cannot represent.
Although there are also temperature and salinity drifts deeper in the water column, we do not attempt to correct for those, as they are irrelevant to the testing of BL models. 

\begin{figure}[htbp]
\centering
\includegraphics[trim={0 0 0 0}, width=\textwidth]{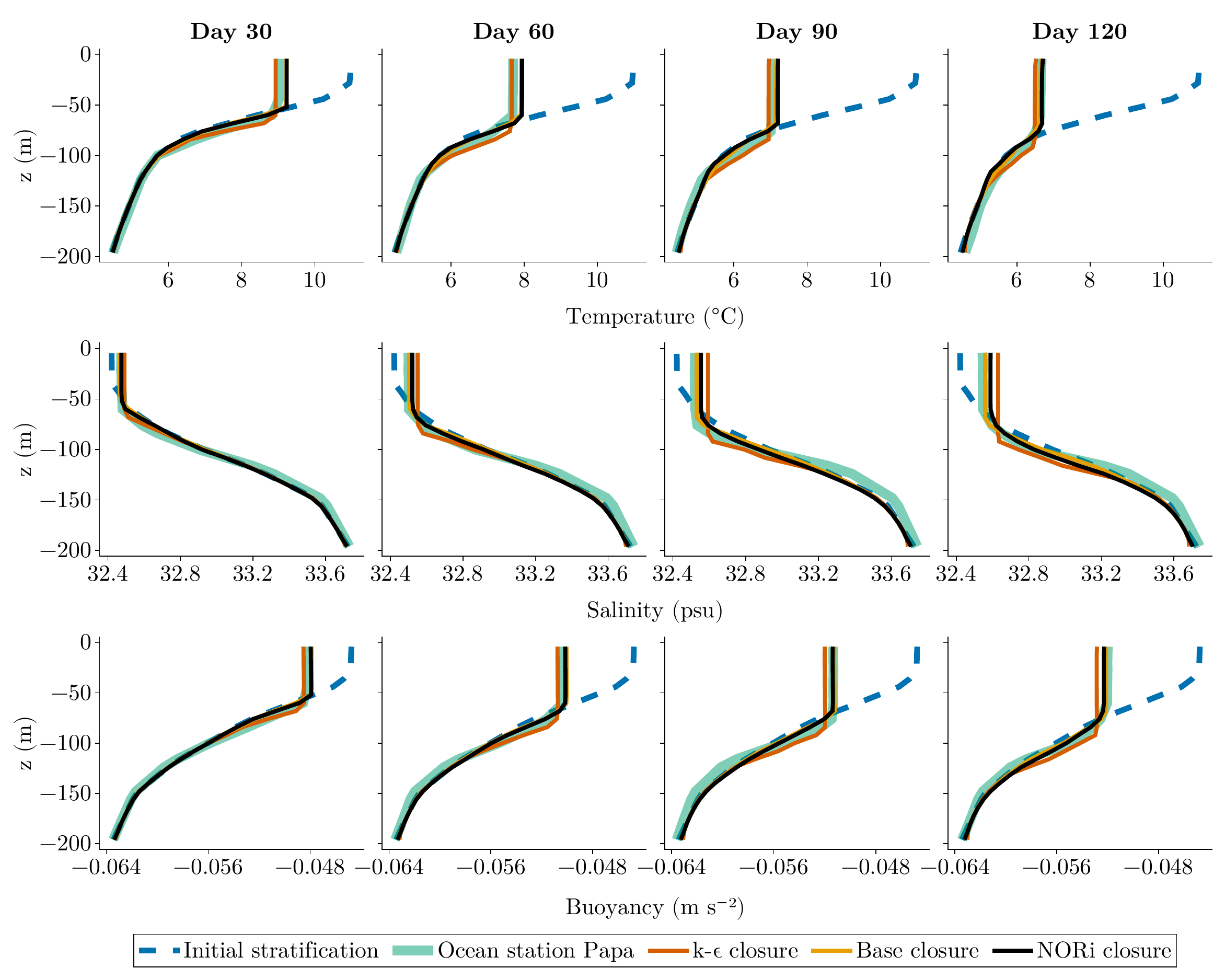}
\caption{Vertical profiles of temperature, salinity, and buoyancy from Ocean Weather Station Papa (OWS Papa) observations and column model simulations run with NORi, its base closure, and the $k$-$\epsilon$ closure with prescribed surface fluxes from OWS Papa. 
Simulations are performed for eight different years between 2007 and 2023 starting with observed profiles on November 1 and run for $\SI{120}{days}$. 
The instantaneous profiles are the mean over eight different years at thirty-day intervals. 
Some years are excluded because there are large gaps (more than \SI{72}{hours}) in the data. 
A constant flux correction is applied on top of the prescribed fluxes to mitigate the drifts in the model solution due to the discrepancies between the observed temperature and salinity budgets and the prescribed surface fluxes.}
\label{figure ows papa kepsilon}
\end{figure}

Figure~\ref{figure ows papa kepsilon} shows profiles of temperature, salinity, and buoyancy from observations and BL models (NORi, base closure, and $k$-$\epsilon$) at four times during the $\SI{120}{days}$, averaged over eight different years. 
All models produce profiles close to the observations. 
It is encouraging that NORi produces reasonable solutions under realistic, time-dependent air-sea fluxes, even though it was trained only on idealized LES forced with constant fluxes.
The similarity between the NORi and base closure predictions may naively lead to the conclusion that adding neural networks to capture entrainment was unnecessary. 
However, the explanation for this similarity is that convection is very weak at OWS Papa, as evidenced by the BL depth changing by less than $\SI{50}{m}$ over $\SI{120}{days}$. 
In such a scenario, entrainment is expected to be weak. 
It is actually encouraging that, even though NORi was not explicitly trained on such weak surface forcings, its neural networks do not produce spurious entrainment but instead remain essentially inactive. 
We note that $k$-$\epsilon$ slightly overestimates the mixed layer depth, confirming the same bias reported in the validation test cases shown in Figures~\ref{figure vs k epsilon results} and \ref{figure 1D model long integration}. 
More details about the model-data comparison are given in \ref{section appendix ows papa} where we show the evolution of stratification and model-data differences in Figure~\ref{figure ows papa stratification} and additional comparisons with the CATKE model in Figure~\ref{figure ows papa CATKE}.

\subsection{Summary of single-column comparisons} \label{subsection single column summary}

Taken together, the validation against LES and OWS Papa observations paints a consistent picture of how NORi compares with the state-of-the-art closures $k$-$\epsilon$ and CATKE.
In regimes where turbulence is driven primarily by shear and entrainment is weak, the base closure alone is already sufficient to reproduce the LES and observed profiles, and NORi, $k$-$\epsilon$, and CATKE are essentially indistinguishable.
However, under convective regimes, the base closure systematically underestimates entrainment and the addition of the neural networks is the key for NORi to achieve the same skill as $k$-$\epsilon$ and CATKE while retaining the longer time steps afforded by a zero-equation closure.

\section{Testing NORi on a large-scale, centennial double-gyre simulation} \label{section results double gyre}

It is difficult to make statements about the numerical stability of neural network-based parameterizations based only on single-column simulations, since the lateral effects of advection and feedbacks between different columns are missing. 
To test NORi's numerical stability in large-scale simulations, an idealized double-gyre simulation is set up that includes many key components of realistic ocean simulations. 
This test is not intended to be an evaluation of NORi's predictive skill in realistic ocean simulations, since there is no ``ground truth'' in a double-gyre setup. 
Evaluating NORi's skill in large-scale settings would require global/regional ocean simulations and reliable observations to compare against, which is beyond the scope of this work.

Implementing NORi in Oceananigans.jl was straightforward; it is only necessary to set up input and output fields for neural network inference, as the neural network package Lux.jl easily integrates into Oceananigans.jl without the need for intermediate wrappers.
NORi works flexibly with both CPU and GPU backends in Oceananigans.jl, allowing us to take advantage of hardware acceleration for large-scale simulations.

\begin{figure}[htbp]
\centering
\includegraphics[trim={0 12cm 0 8cm}, width=\textwidth]{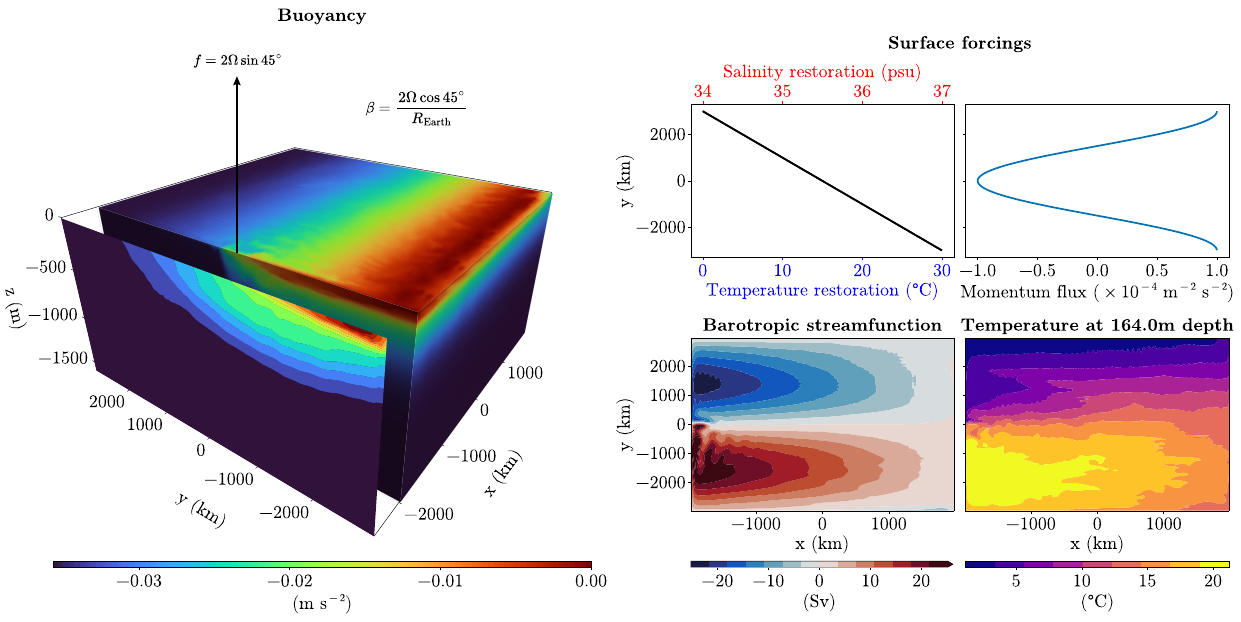}
\caption{The configuration of the double-gyre simulation.
        Results from the NORi closure (implemented in Oceananigans.jl and run on GPUs) at $t = \SI{100}{years}$.
        The subfigure on the left shows a 3D snapshot of the buoyancy field as well as the zonally averaged buoyancy of the simulation.
        The subfigures on the right show the temperature and salinity surface-restoring profiles (top left), the surface wind stress (top right), the barotropic stream function averaged over the last 10 years (bottom left), and a horizontal temperature slice at a depth of $\SI{164}{m}$ (bottom right). 
        The barotropic stream function reaches a maximum of around $\SI{40}{Sv}$ in the subtropical gyre.}
\label{figure double gyre setup}
\end{figure}

The double-gyre has dimensions of $(L_x, L_y, L_z) = (4000, 6000, 1.6) \, \si{km}$ with a grid size of $(N_x, N_y, N_z) = (100, 100, 200)$, which fits on a single NVIDIA V100 GPU. 
We do not include any other parameterizations to isolate and observe only the effects of BL schemes.
The setup uses a $\beta$-plane approximation with $f = 2\Omega \sin 45^\circ \approx \SI{1.0d-4}{s^{-1}}$ at the center of the domain and $\beta = \frac{2 \Omega \cos 45^\circ}{R_\text{Earth}} \approx \SI{1.6d-11}{m^{-1}.s^{-1}}$, where $\Omega$ is the rotation rate of the Earth and $R_\text{Earth}$ is the radius of the Earth. 
The meridional extent of the gyre is $\SI{6000}{km}$ and corresponds approximately to $\SI{54}{\degree}$ in latitude. 
Forcing consists of restoring temperature and salinity on an 8-day time scale to linear profiles decreasing with latitude from $\SI{30}{\celsius}$ to $\SI{0}{\celsius}$ and from $\SI{37}{psu}$ to $\SI{34}{psu}$, respectively (see Figure~\ref{figure double gyre setup}). 
Momentum forcing is through a sinusoidal wind stress of magnitude $\SI{1d-4}{m^2.s^{-2}}\approx\SI{0.1}{N.m^{-2}}$ driving easterlies north and south and westerlies at the center latitude, as shown in Figure~\ref{figure double gyre setup}. 
A linear drag with a damping time scale of $\SI{30}{days}$ is imposed at the deepest grid level.
The double-gyre simulation is run for $\SI{100}{years}$ with a constant time step of $\SI{5}{minutes}$, chosen short enough that the NORi and $k$-$\epsilon$ BL model results can be directly compared. 
The simulation is initialized with a uniform temperature field in the horizontal direction and with a linear vertical gradient decreasing from $\SI{30}{\celsius}$ at the top to $\SI{10}{\celsius}$ at the bottom. 
The salinity field is initialized without vertical gradient, but with a linear meridional gradient decreasing from $\SI{37}{psu}$ to $\SI{34}{psu}$ with latitude. 
Figure~\ref{figure double gyre setup} shows the equilibrated 3D buoyancy field, a horizontal temperature slice at a depth of $\SI{164}{m}$, and the vertically integrated barotropic stream function $\Psi$ averaged over $\SI{10}{years}$, defined as
\begin{align}
    \pd{\Psi}{y} &= -\int_{-L_z}^0 u \dd z, \, \pd{\Psi}{x} = \int_{-L_z}^0 v \dd z, \\
    \overline{\Psi} &= \frac{1}{\SI{10}{years}} \int_{\text{year } 90}^{\text{year } 100} \Psi \dd t.
\end{align}
The barotropic stream function is characterized by two intensified gyres on the western side of the basin. 
A thermocline develops in the southern half of the domain consistent with the ventilated thermocline theory~\cite{pedlosky_ocean_1996, vallis_atmospheric_2017}. 
Deep convection generates a well-mixed water column at the northern edge of the domain.

We run simulations using NORi, the base closure, and the $k$-$\epsilon$ model. 
In this naive first implementation of NORi in Oceananigans.jl, we allocated additional working arrays to store the inputs and outputs of the neural networks; thus, a neural network inference is a forward pass over the input working array.
Admittedly, this is a highly inefficient approach as it incurs severe memory overhead as well as data duplication.
However, this simple implementation has already yielded a performance that is only $\SI{15}{\%}$ slower than $k$-$\epsilon$ on a per-time step basis.
Combined with the larger time steps that NORi supports (see Figures~\ref{figure NN 1D timestep} and \ref{figure kepsilon 1D timestep}), a more efficient implementation has the potential to be significantly faster than two-equation closures when deployed in large-scale simulations.

\begin{figure}[htbp]
\centering
\includegraphics[width=\textwidth]{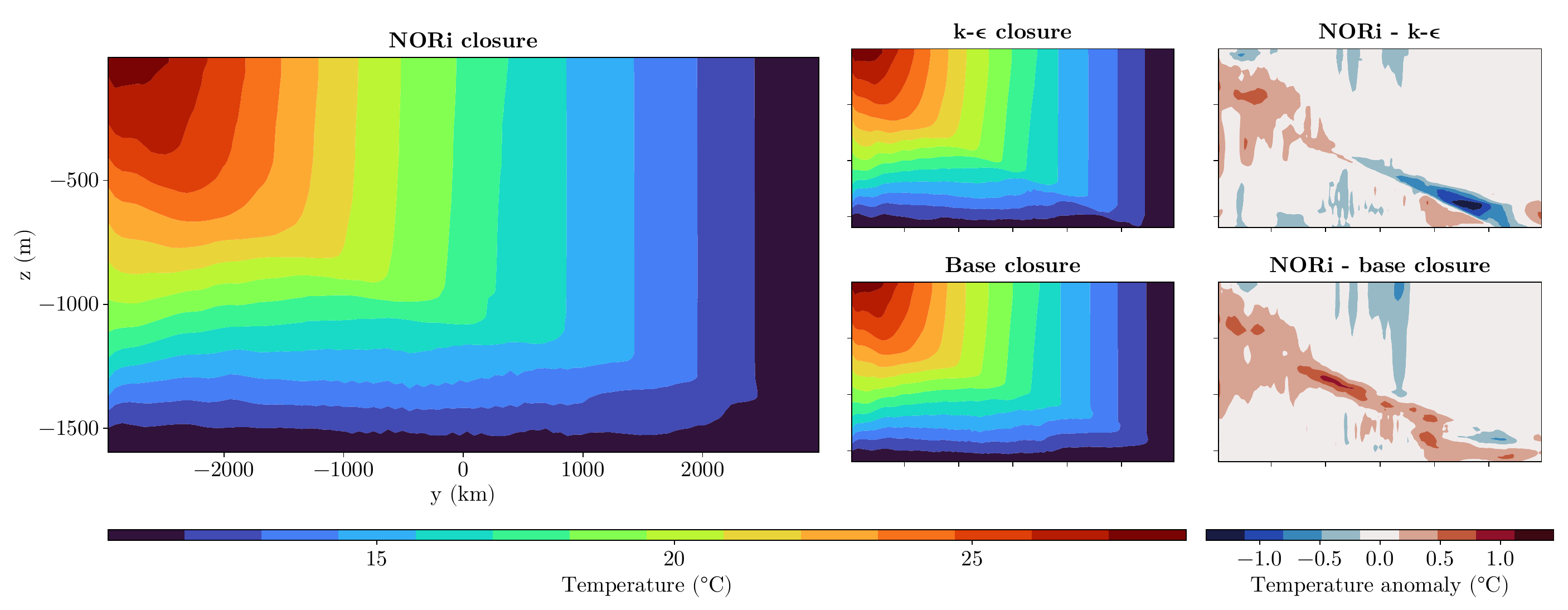}
\caption{Zonally averaged temperature slice from the three BL models in the double-gyre setup: NORi, $k$-$\epsilon$, and the base closure.
The snapshots are taken at $t = \SI{2.5}{years}$.
The rightmost column shows the temperature differences between NORi and the other two models.}
\label{figure double gyre yz stratification}
\end{figure}

NORi is stable for the full 100 years of simulation despite being trained on an ensemble of LES with integration times of less than $\SI{2}{days}$.
In the many different setups we have run during experimentation, NORi has never crashed due to numerical instabilities.
This is significant as numerical stability is one of the biggest challenges facing hybrid modeling of PDEs~\cite{kochkov_neural_2024,frezat_posteriori_2022}, where equations are strongly coupled spatially and temporally such that they can be unstable against noisy neural network outputs.
Figure~\ref{figure double gyre yz stratification} compares the zonally averaged temperature at $t = \SI{2.5}{years}$ from double-gyre simulations using the three BL models. 
NORi simulates deeper mixed layers than the base closure as expected, because the base closure lacks the representation of entrainment. 
In contrast, NORi produces shallower mixed layers than $k$-$\epsilon$, especially in the northern part of the domain.
Interestingly, this difference pattern is the same as what we observed in the single-column comparison at $\SI{60}{days}$ with LES in Figure~\ref{figure 1D model long integration}.
Although NORi and $k$-$\epsilon$ produce very similar solutions during the 2-day training periods, important differences can emerge over longer time periods, as also discussed in Section~\ref{subsection 1D long integration}. 
However, the discrepancies in this case may also arise from interactions between the parameterizations and the large-scale ocean dynamics.
Figure~\ref{figure double gyre yz stratification k epsilon 100 years} in \ref{section appendix double gyre 100 years} confirms that these discrepancies persist in time by showing the same zonally averaged temperature after $\SI{100}{years}$. 
In \ref{section appendix CATKE} we further show that NORi generates deeper mixed layers than CATKE~\cite{wagner_formulation_2025} at $\SI{100}{years}$, suggesting that more studies where objective comparisons against observationally constrained data are available, such as regional/global ocean simulations, are needed to determine which model is more accurate.

The results of the double-gyre solutions highlight the importance of recalibrating BL parameterizations in global simulations with climatological observations, because the small residuals in the calibration with LES can build up over longer times and larger scales. 
However, calibration with short-term, high-resolution simulations or observations provides essential initial estimates for the model parameters, reducing the amount of data required for calibration in the global context.

\section{Conclusions} \label{section conclusion}

We introduced NORi, a parameterization of oceanic surface BL turbulence formulated with neural ODEs where neural networks augment a simple physical closure.
To overcome the triple conundrum of high numerical stability, fast model inference, and relatively cheap training, the neural networks, as expressive as they are, must be constrained by our physical understanding of the vertical mixing process.
Incorporating physical knowledge into model design drastically reduces the amount of data required and the size of the model~\cite{goyal_inductive_2022}, allowing us to arrive at a physical closure that is ``simple but not any simpler''.
Layered on top of this closure are highly expressive neural networks that capture more complicated processes, i.e., the nonlocal entrainment fluxes that are missing from the local eddy-diffusivity closure and whose accurate representations have eluded traditional parameterizations for decades.
With this approach, NORi outperforms physics-based closures while preserving physical principles that are well understood, suggesting that it is worthwhile to augment traditional parameterizations with neural networks rather than replace them with neural networks altogether.

During the NORi design process, we paid special attention to ensuring that every choice is physically motivated.
Since neural networks predict fluxes instead of tendencies, NORi satisfies conservation laws to machine precision by construction.
The nonlinear thermodynamics of seawater are explicitly incorporated, so NORi can be readily deployed in realistic ocean simulations.
By sharing the same neural network weights across all vertical levels, we ensure that NORi's representation of entrainment is independent of mixed layer depth.
However, this also means that NORi cannot capture deep convection dynamics (see our discussion in Section~\ref{subsection neural network architecture}), but a simple extension of including the mixed layer depth in the neural network inputs can address this limitation.
To reduce inference cost, the neural networks in NORi are relatively lean, with only three hidden layers and 128 neurons per hidden layer, and further work on distillation and pruning could make them even smaller without a significant loss in accuracy.
It would also be worth investigating whether evaluating the neural networks in half precision increases inference speed while retaining sufficient accuracy near the sharp gradients at the BL base.
NORi generalizes well across a wide range of wind, cooling, evaporation, and rotational strengths under various stratifications and thermodynamic regimes, correctly weakens its neural network components to produce solutions at OWS Papa that are comparable to, if not better than, those obtained with the $k$-$\epsilon$ model, and is time step-independent below $\SI{1}{hour}$, which is appropriate for large-scale, coarse-resolution ocean simulations.

NORi was designed to be calibrated with \textit{a posteriori} training, where the loss function depends on the full time history of coarse-grained variables rather than on instantaneous turbulent fluxes alone, ensuring that we optimize the model directly to our desired target.
By using the neural ODE approach, we also explicitly optimized for numerical stability, since numerically unstable models will incur a high loss and are thus avoided.
As a result, NORi is numerically stable for at least $\SI{100}{years}$ despite being trained on an ensemble of LES with timespans of roughly $\SI{2}{days}$, and in the many double-gyre configurations we have explored during experimentation, NORi has never crashed due to numerical instabilities.
However, in pursuit of numerical stability, we have made sacrifices in the scalability of the calibration process because of the serial-in-time nature of neural ODEs.
Despite these drawbacks, we have found that \textit{a posteriori} training is essential for building a model that is usable in practice, as training neural networks directly on instantaneous turbulent fluxes leads to models that are highly numerically unstable and exhibit finite-time blowup.
Ultimately, providing more structure and physical inductive biases in the model design is essential for improving training outcomes.

We have designed NORi to be relatively straightforward to implement and deploy in both CPU- and GPU-based ocean models.
Importantly, automatic differentiability is only required during training; at inference time, NORi only requires a standard forward pass through the neural networks, making it compatible with any ocean model that supports neural network evaluation.
Using the Julia software stack, we integrated NORi directly into Oceananigans.jl without intermediate wrappers, and took advantage of its GPU-oriented design to avoid CPU--GPU data transfers during inference.
In our first naive implementation, additional working arrays were allocated for neural network inputs and outputs, leading to memory-bound performance due to large transfers of values between working arrays and native Oceananigans.jl fields.
Despite such inefficiencies, NORi's performance is only $\SI{15}{\%}$ slower than the state-of-the-art $k$-$\epsilon$ closure on a per-time step basis.
A more efficient implementation that writes custom forward-pass kernels accessing native Oceananigans.jl arrays directly would eliminate the need for these working arrays and is a performance engineering task for future work.

A few important challenges remain to be tackled in future work. 
NORi has been trained and deployed with a fixed grid resolution of \SI{8}{m}, which is comparable to that of state-of-the-art global ocean models (around $\SI{10}{m}$ near the BL base). 
However, NORi cannot yet be used flexibly with variable vertical grid sizes. 
A straightforward extension may be to interpolate between the dynamics grid and the physics grid, running NORi inference at a fixed $\SI{8}{m}$ resolution and then interpolating the fluxes back to the dynamics grid, similar to what is done in \citeA{molod_running_2009}. 
A first attempt at this approach using simple linear interpolation yielded unsatisfactory results, as entrainment fluxes weaken with coarser vertical resolutions that fail to represent the sharp gradients at the BL base. 
More sophisticated reconstruction techniques, like making resolution an input to the neural networks, may be potential solutions. 
Alternatively, one could implement a resolution-independent representation of the entrainment fluxes---\citeA{vanzanten_entrainment_1999} infers the fluxes from prognostic equations for the inversion depth and minimum buoyancy flux---though this would require a reformulation of the neural network architecture.
Failing all of the above, one may simply retrain the neural networks on the desired grid, even though this would limit the utility of the neural networks except in operational applications where the grid resolution is fixed.

Once variable vertical grid support is implemented and validated, NORi could be run in a regional and/or global ocean simulation with realistic forcings---setups that can be verified against observations. 
This would allow us to assess whether NORi alleviates known mixed layer depth biases in large-scale climate simulations. 
Ultimately, we plan to calibrate NORi jointly with other ocean parameterizations, such as the Gent--McWilliams~\cite{gent_isopycnal_1990} and Redi~\cite{redi_oceanic_1982} parameterizations as well as interior mixing parameterizations which also enhance vertical mixing, \textit{a posteriori} within both regional and global ocean simulations, where the model is initialized, forced, and verified with observations.
This would allow us to fine-tune NORi and reduce biases found in large-scale simulations, but not present in idealized column setups.

Joint calibration of different parameterizations is useful to eliminate compensating errors. 
A first step in that direction would be to jointly calibrate NORi's base closure and neural networks to ensure that the neural networks do not compensate for biases other than nonlocal entrainment. 
The neural network weights and base closure parameters learned from the separate calibration would be used as priors to reduce the parameter space to be explored for \textit{a posteriori} calibration given the complexity of the forward model. 
The next step would be to compare NORi with current state-of-the-art BL parameterizations calibrated against a common suite of ocean scenarios to rigorously assess their respective biases. 
Should the comparison identify structural deficiencies, one may consider adopting a more sophisticated base closure, like $k$-$\epsilon$~\cite{umlauf_generic_2003} or CATKE~\cite{wagner_formulation_2025}, or increasing the depth of the neural networks.

Beyond free-running comparisons, \citeA{wei_direct_2026} also demonstrated an assessment approach in which mean fields in single-column models are strongly nudged toward LES values so that flux skill can be diagnosed independently of the coupled mean-field evolution, since the errors in mean fields and turbulent fluxes are entangled. 
This may serve as a useful tool for evaluating NORi alongside other parameterizations in future work, revealing systematic biases NORi might possess.

Another important consideration is that in its current implementation, NORi is active for five grid points below the BL base, resulting in a $\SI{40}{m}$ layer below the BL base, which is thicker than the plume penetration depth in all the LES cases we have run, including the testing case where the BL deepened to around $\SI{500}{m}$.
In that scenario, there is a nonzero wind stress, thus the thickness of the entrainment layer is also influenced by shear.
For free convection scenarios, the $1/5$ rule applies~\cite{carson_development_1973}, so it is possible that our current implementation holds only for BLs up to $\SI{1}{km}$.
To address such regimes, we can run LES with much longer integration times and deeper BLs and train NORi with additional grid points below the BL base.
Even though this extension is straightforward, we have opted not to pursue this as this exercise is purely academic, since global ocean models have vertical resolutions that coarsen with depth.
Under such circumstances, five grid points may remain sufficient.

The suite of LES should also be extended to include missing BL physics.
Surface wave effects, i.e., Langmuir turbulence, are not accounted for in the LES generated in this work but are known to substantially affect the evolution of the BL under very strong winds.
Different LES numerical implementations, such as explicit rather than implicit LES closure or addition of wall-models, may provide additional validation of the dataset used for calibration. 
Retraining or fine-tuning NORi with additional LES would be straightforward, since neither the architecture nor the training procedure would need to change.

We have focused on the deepening phases of the BL in the absence of lateral density gradients, when the dynamics is controlled by destabilizing surface fluxes. 
However, submesoscale instabilities at upper ocean buoyancy fronts affect the deepening of the BL~\cite{taylor_buoyancy_2010,thomas_friction_2008} and dominate restratification when surface fluxes are stabilizing~\cite{fox-kemper_parameterization_2008-1,taylor_submesoscale_2023}.
In the future, we aim to extend NORi to include the impact of submesoscale processes within a similar framework, thereby allowing for a more complete representation of the BL dynamics across the entire seasonal cycle.

\section*{Open Research Section}
The LES data used for the training and validation of NORi are generated from open source software \href{https://github.com/CliMA/Oceananigans.jl}{Oceananigans.jl}~\cite{ramadhan_oceananigansjl_2020,wagner_high-level_2025}.
These LES data are packaged into a publicly available dataset \href{https://doi.org/10.5281/zenodo.16278000}{SOBLLES - A Salty Ocean Boundary Layer Large-Eddy Simulation Dataset} hosted on Zenodo~\cite{lee_soblles_2025}.
The base closure in NORi is calibrated using open source software \href{https://github.com/CliMA/EnsembleKalmanProcesses.jl}{EnsembleKalmanProcesses.jl}~\cite{dunbar_ensemblekalmanprocessesjl_2022}.
The neural networks in NORi are implemented using open source software \href{https://github.com/LuxDL/Lux.jl}{Lux.jl}~\cite{pal_lux_2023} and calibrated using \href{https://github.com/EnzymeAD/Enzyme.jl}{Enzyme.jl}~\cite{moses_instead_2020}.
Scripts for the training and validation of NORi, as well as its implementation on Oceananigans.jl, together with scripts to reproduce all figures in this document, are available in the GitHub repository \href{https://github.com/xkykai/NORiOceanParameterization.jl}{NORiOceanParameterization.jl}~\cite{lee_xkykainorioceanparameterizationjl_2025}.
The data required to reproduce all figures in this paper are available through a data companion to \href{https://github.com/xkykai/NORiOceanParameterization.jl}{NORiOceanParameterization.jl} hosted on \href{https://zenodo.org/doi/10.5281/zenodo.17605194}{Zenodo}~\cite{lee_data_2026}.

\section*{Conflict of Interest Disclosure}
The authors declare there are no conflicts of interest for this manuscript.

\acknowledgments
The authors thank Aakash Sane and the two other anonymous reviewers for their insightful comments and constructive suggestions that helped strengthen this work.
This project is supported by Schmidt Sciences, LLC.
X. K. L. was also partly supported by the Norman C. Rasmussen Fellowship and the Callahan-Dee Fellowship.
The work was initiated while X. K. L. was an undergraduate student in the Department of Physics at Imperial College London under the Imperial-MIT student exchange program, working with A. R. and J. M. Preliminary results are reported in A. R.'s PhD thesis~\cite{ramadhan_data-driven_2023}.
Artificial intelligence tools, including large language models and code-completion tools offered by OpenAI, Anthropic, and GitHub Copilot, were used to assist with code development, manuscript editing, and proofreading.
All AI-generated suggestions were thoroughly reviewed and verified by the authors.

\newpage

\appendix
\section{Column model derivation of the ocean surface boundary layer} \label{section appendix column model}
For easier reference, we first rewrite the equations \eqref{equation LES u}, \eqref{equation LES T}, \eqref{equation LES S}, and \eqref{equation LES incompressibility} in their component form:
\begin{align}
    \label{equation LES u appendix} \pd{u}{t} &= -\nabla \cdot (\boldsymbol{u} u) + fv - \partial_x p, \\
    \label{equation LES v appendix} \pd{v}{t} &= -\nabla \cdot (\boldsymbol{u} v) - fu - \partial_y p,\\
    \label{equation LES w appendix} \pd{w}{t} &= -\nabla \cdot (\boldsymbol{u} w) - \partial_z p + b,\\
    \label{equation LES T appendix} \pd{T}{t} &= -\nabla \cdot (\boldsymbol{u} T),\\
    \label{equation LES S appendix} \pd{S}{t} &= -\nabla \cdot (\boldsymbol{u} S), \\
    \label{equation LES incompressibility appendix} \nabla \cdot \boldsymbol{u} &= 0.
\end{align}
We again note that by decomposing a field into its horizontally averaged component and perturbations away from the average $\phi = \overline{\phi}(z, t) + \phi^\prime(x, y, z, t)$, in a laterally doubly periodic domain, the horizontal averaging operator $\overline{(\cdot)}$ has the property that
\begin{align}
    \label{equation averaging rule x average} \overline{\pd{\phi}{x}} = \pd{\overline{\phi}}{x} &= 0, \\
    \label{equation averaging rule y average} \overline{\pd{\phi}{y}} = \pd{\overline{\phi}}{y} &= 0, \\
    \label{equation averaging rule deviation} \overline{\phi^\prime} &\equiv 0.
\end{align}

Applying the horizontal averaging to Equation~\eqref{equation LES incompressibility appendix} and noting properties \eqref{equation averaging rule x average} and \eqref{equation averaging rule y average}, we obtain the following:
\begin{align}
    \cancelto{0}{\overline{\pd{u}{x}}} + \cancelto{0}{\overline{\pd{v}{y}}} + \overline{\pd{w}{z}} &= 0, \\
    \pd{\overline{w}}{z} &= 0, \\
    \therefore \overline{w} &= C
\end{align}
where $C$ is a constant. 
We now apply the horizontal averaging operator to the impenetrability boundary condition on the surface $z=0$, i.e., Equation~\eqref{equation LES w no penetration}, which yields
\begin{align}
    \overline{w (z=0)} &= 0, \\
    \overline{w} &= C = 0,
\end{align}
since $\overline{w}$ is a constant everywhere in the column.

Expanding the terms in Equation~\eqref{equation LES u appendix} and applying the horizontal averaging operator, we have (in Cartesian coordinates)
\begin{align}
    \pd{\overline{u}}{t} &= -\cancelto{0}{\partial_x \overline{u u}} - \cancelto{0}{\partial_y \overline{v u}} - \partial_z \overline{w u} + f \overline{v} - \cancelto{0}{\partial_x \overline{p}}, \\
    &= -\pd{}{z} \overline{\left( \overline{u} \, \cancelto{0}{\overline{w}} + u^\prime \cancelto{0}{\overline{w}} + \overline{u} w^\prime + u^\prime w^\prime \right)} + f \overline{v}, \\
    &= -\pd{}{z} \left( \overline{u} \cancelto{0}{\overline{w^\prime}} + \overline{u^\prime w^\prime}  \right) + f \overline{v}, \\
    &= -\pd{\overline{u^\prime w^\prime}}{z} + f \overline{v},
\end{align}
where $\overline{u^\prime w^\prime}$ is the vertical flux of horizontal momentum in the $x$-direction, yielding Equation~\eqref{equation u 1D}. 
Similarly, we can apply the same procedure to Equations~\eqref{equation LES v appendix}, \eqref{equation LES T appendix}, and \eqref{equation LES S appendix} to arrive at Equations~\eqref{equation v 1D}, \eqref{equation T 1D}, and \eqref{equation S 1D}, respectively.

\section{The implementation of the $k$-$\epsilon$ boundary layer closure} \label{section appendix k epsilon}

The $k$-$\epsilon$ implementation in Oceananigans.jl follows the generic length-scale (GLS) framework of \citeA{umlauf_generic_2003} tailored to oceanic applications, using the Canuto-A stability functions of \citeA{canuto_ocean_2001} together with the realizability constraints of \citeA{umlauf_second-order_2005} on the stratification and shear numbers $\alpha^N = N^2(k/\epsilon)^2$ and $\alpha^M = |S|^2(k/\epsilon)^2$ where $N^2$ is the buoyancy frequency and $S$ is the shear.
Because the momentum and tracer stability functions $\mathbb{S}_u(\alpha^N,\alpha^M)$ and $\mathbb{S}_c(\alpha^N,\alpha^M)$ are rational functions of the local stratification and shear that share a common denominator but differ in their numerators, the turbulent Prandtl number $\mathrm{Pr}_t = \mathbb{S}_u/\mathbb{S}_c$ is itself a function of $\alpha^N$ and $\alpha^M$ rather than a fixed constant.
The eddy viscosity and diffusivities are reconstructed as $\nu_t = \mathbb{S}_u\, k^2/\epsilon$ and $\kappa_t = \mathbb{S}_c\, k^2/\epsilon$, with separate Schmidt-number scalings for the $k$ and $\epsilon$ diffusivities.
The turbulent kinetic energy ($k$) equation balances vertical turbulent diffusion, shear production, buoyancy production, and viscous dissipation with TKE Schmidt number $\sigma_k = 1$~\cite{launder_numerical_1974}.

The dissipation ($\epsilon$) equation uses $C_{\epsilon 1} = 1.44$ for shear production and $C_{\epsilon 2} = 1.92$ for dissipation~\cite{launder_numerical_1974}, and Schmidt number $\sigma_\epsilon = 1.2$ fixed by consistency with the law of the wall in a neutral logarithmic boundary layer~\cite{umlauf_generic_2003}. 
The buoyancy-production coefficient is implemented with the standard sign-of-$N^2$ split into stable and unstable branches $C_{\epsilon 3}^{+}$ and $C_{\epsilon 3}^{-}$. 
Although the standard GLS practice uses regime-dependent values, here we set $C_{\epsilon 3}^{+} = C_{\epsilon 3}^{-} = -0.65$ uniformly. 
To preserve positivity of $k$ on long integrations, dissipation is treated implicitly via $\omega_e = \epsilon/k$, and the negative part of the buoyancy production term is moved to the implicit side following a Patankar-style splitting~\cite{burchard_high-order_2003}.
Vertical diffusion of $u$, $v$, $T$, $S$, $k$, and $\epsilon$ is handled with a vertically implicit time discretization.
Two limiters keep the closure well-posed in strongly stratified regions.
Firstly, the turbulent length scale is capped by a Galperin-type bound and by the domain depth, $\ell_{\max} = \min\!\bigl(L_z,\, 0.75\sqrt{k/N^2}\bigr)$~\cite{galperin_quasi-equilibrium_1988}.
Secondly, since the GLS closure relates dissipation to length scale via $\epsilon = (c_\mu^0)^3\, k^{3/2}/\ell$, this length-scale cap induces a corresponding lower bound on dissipation, $\epsilon \ge (c_\mu^0)^3\, k^{3/2}/\ell_{\max}$, which we enforce together with a hard floor of $\SI{1d-12}{m^{2}.s^{-3}}$.
A TKE floor of $k_{\min} = \SI{1d-6}{m^{2}.s^{-2}}$ and a one-minute Euler damping of any spurious negative TKE provide additional numerical robustness.

At the surface, we impose a zero-flux TKE boundary condition, omitting the wave-breaking TKE injection of~\citeA{craig_modeling_1994}.
Surface forcing of turbulence enters only through the dissipation boundary condition, which is set using a Charnock roughness length $\ell_r = 0.11\, u_\star^2/g$~\cite{charnock_wind_1955,umlauf_generic_2003}.

\section{Training and testing suite} \label{section appendix LES table}
In this section, we provide the complete training and testing suite used to train, validate, and test NORi. Table~\ref{table wind driven LES suite} shows the training cases driven by shear instability (wind-driven mixing) and two testing cases used to evaluate the base closure, Table~\ref{table free convection LES training suite} shows the training free convection cases driven by buoyancy instability (no winds), Table~\ref{table wind + convection LES training suite} shows the training cases driven by both shear and convective instabilities, while Table~\ref{table misc LES training suite} shows the remaining miscellaneous training cases that include cases with the same forcings and stratifications as some of those before but in different thermodynamic regimes (lower temperatures) as well as cases with no temperature or salinity gradient.
Finally, Table~\ref{table LES validation suite} shows the validation cases used to select the final weights of the neural network after training.
Note that for all tables we adopt this sign convention: positive $J_T^\text{top}$ denotes upward heat flux (cooling); negative $J_S^\text{top}$ denotes downward salt flux (evaporation); negative $J_u^\text{top}$ denotes downward momentum flux (i.e., into the ocean).
Also, $J_v^\text{top} = 0$ in all simulations since imposing nonzero $J_v^\text{top}$ is dynamically equivalent to imposing nonzero $J_u^\text{top}$ during training and inference since momentum is only implicitly provided to NORi via the Richardson number.
Initial profiles $\phi$ are linear with vertical gradient $\partial_z \phi$ and surface value $\phi^\text{top}$.

\begin{table}[htbp]
\centering
\caption{Cases used to train and evaluate the base closure.
Rows 1--6 are shear-driven training cases; rows 7--8 are held-out testing cases for the base closure and are not used to validate the neural networks.}
\label{table wind driven LES suite}
\begin{tabular}{@{}cccccccc@{}}
\toprule
$J_T^\text{top}$                        & $J_S^\text{top}$                         & $J_u^\text{top}$                   & $\pd{T}{z}$                & $\pd{S}{z}$                 & $f$               & $T^\text{top}$                 & $S^\text{top}$                  \\ \midrule
$(\unit{\celsius.m.s^{-1}})$ & $(\unit{psu.m.s^{-1}})$ & $(\unit{m^{2}.s^{-2}})$ & $(\unit{\celsius.m^{-1}})$ & $(\unit{psu.m^{-1}})$ & $(\unit{s^{-1}})$ & $(\unit{\celsius})$ & $(\unit{psu})$ \\ \midrule
0                            & 0                             & $\num{-5d-4}$           & $\num{1.4d-2}$             & $\num{2.1d-3}$              & $\num{8d-5}$      & 18                  & 36.6                 \\
0                            & 0                             & $\num{-2d-4}$           & $\num{1.4d-2}$             & $\num{2.1d-3}$              & $\num{8d-5}$      & 18                  & 36.6                 \\
0                            & 0                             & $\num{-5d-4}$           & $\num{1.3d-2}$             & $\num{7.5d-4}$              & 0                 & 14.5                & 35                   \\
0                            & 0                             & $\num{-2d-4}$           & $\num{1.3d-2}$             & $\num{7.5d-4}$              & 0                 & 14.5                & 35                   \\
0                            & 0                             & $\num{-5d-4}$           & $\num{-2.5d-2}$            & $\num{-4.5d-3}$             & $\num{-1.25d-4}$   & 0                   & 33.9                 \\
0                            & 0                             & $\num{-2d-4}$           & $\num{-2.5d-2}$            & $\num{-4.5d-3}$             & $\num{-1.25d-4}$   & 0                   & 33.9                 \\
$\num{-3d-4}$                & $\num{3d-5}$                 & $\num{-5d-4}$           & $\num{1.5d-2}$             & $\num{1.5d-3}$              & $\num{1d-4}$      & 20                  & 37                   \\
0                            & 0                             & $\num{-3.5d-4}$         & $\num{1.5d-2}$             & $\num{1.5d-3}$              & $\num{1d-4}$      & 20                  & 37                   \\
\bottomrule
\end{tabular}
\end{table}

\begin{table}[htbp]
\centering
\caption{Free convection cases used to train NORi.}
\label{table free convection LES training suite}
\begin{tabular}{@{}cccccccc@{}}
\toprule
$J_T^\text{top}$                        & $J_S^\text{top}$                         & $J_u^\text{top}$                   & $\pd{T}{z}$                & $\pd{S}{z}$                 & $f$               & $T^\text{top}$                 & $S^\text{top}$                  \\ \midrule
$(\unit{\celsius.m.s^{-1}})$ & $(\unit{psu.m.s^{-1}})$ & $(\unit{m^{2}.s^{-2}})$ & $(\unit{\celsius.m^{-1}})$ & $(\unit{psu.m^{-1}})$ & $(\unit{s^{-1}})$ & $(\unit{\celsius})$ & $(\unit{psu})$ \\ \midrule
$\num{5d-4}$                 & 0                             & 0                       & $\num{1.4d-2}$             & $\num{2.1d-3}$              & $\num{8d-5}$      & $18$                & $36.6$               \\
$\num{1d-4}$                 & 0                             & 0                       & $\num{1.4d-2}$             & $\num{2.1d-3}$              & $\num{8d-5}$      & $18$                & $36.6$               \\
0                            & $\num{-5d-5}$                 & 0                       & $\num{1.4d-2}$             & $\num{2.1d-3}$              & $\num{8d-5}$      & $18$                & $36.6$               \\
0                            & $\num{-1d-5}$                 & 0                       & $\num{1.4d-2}$             & $\num{2.1d-3}$              & $\num{8d-5}$      & $18$                & $36.6$               \\
$\num{3d-4}$                 & $\num{-3d-5}$                 & 0                       & $\num{1.4d-2}$             & $\num{2.1d-3}$              & $\num{8d-5}$      & $18$                & $36.6$               \\
$\num{3d-4}$                 & $\num{-3d-5}$                 & 0                       & $\num{1.4d-2}$             & $\num{-2.1d-3}$             & $\num{8d-5}$      & $18$                & $36.6$               \\
$\num{1d-4}$                 & $\num{-1d-5}$                 & 0                       & $\num{1.4d-2}$             & $\num{2.1d-3}$              & $\num{8d-5}$      & $10$                & $36.6$               \\
$\num{5d-4}$                 & 0                             & 0                       & $\num{1.3d-2}$             & $\num{7.5d-4}$              & 0                 & $14.5$              & $35$                 \\
$\num{1d-4}$                 & 0                             & 0                       & $\num{1.3d-2}$             & $\num{7.5d-4}$              & 0                 & $14.5$              & $35$                 \\
0                            & $\num{-5d-5}$                 & 0                       & $\num{1.3d-2}$             & $\num{7.5d-4}$              & 0                 & $14.5$              & $35$                 \\
0                            & $\num{-1d-5}$                 & 0                       & $\num{1.3d-2}$             & $\num{7.5d-4}$              & 0                 & $14.5$              & $35$                 \\
$\num{3d-4}$                 & $\num{-3d-5}$                 & 0                       & $\num{1.3d-2}$             & $\num{7.5d-4}$              & 0                 & $14.5$              & $35$                 \\
$\num{3d-4}$                 & $\num{-3d-5}$                 & 0                       & $\num{1.3d-2}$             & $\num{-7.5d-4}$             & 0                 & $14.5$              & $35$                 \\
$\num{1d-4}$                 & $\num{-1d-5}$                 & 0                       & $\num{1.3d-2}$             & $\num{7.5d-4}$              & 0                 & $10$                & $35$                 \\
$\num{5d-4}$                 & 0                             & 0                       & $\num{-2.5d-2}$            & $\num{-4.5d-3}$             & $\num{-1.25d-4}$  & 0                   & $33.9$               \\
$\num{1d-4}$                 & 0                             & 0                       & $\num{-2.5d-2}$            & $\num{-4.5d-3}$             & $\num{-1.25d-4}$  & 0                   & $33.9$               \\
0                            & $\num{-5d-5}$                 & 0                       & $\num{-2.5d-2}$            & $\num{-4.5d-3}$             & $\num{-1.25d-4}$  & 0                   & $33.9$               \\
0                            & $\num{-1d-5}$                 & 0                       & $\num{-2.5d-2}$            & $\num{-4.5d-3}$             & $\num{-1.25d-4}$  & 0                   & $33.9$               \\
$\num{3d-4}$                 & $\num{-3d-5}$                 & 0                       & $\num{-2.5d-2}$            & $\num{-4.5d-3}$             & $\num{-1.25d-4}$  & 0                   & $33.9$               \\
\bottomrule
\end{tabular}
\end{table}

\begin{table}[htbp]
\centering
\caption{Wind $+$ convection cases used to train NORi.}
\label{table wind + convection LES training suite}
\begin{tabular}{@{}cccccccc@{}}
\toprule
$J_T^\text{top}$                        & $J_S^\text{top}$                         & $J_u^\text{top}$                   & $\pd{T}{z}$                & $\pd{S}{z}$                 & $f$               & $T^\text{top}$                 & $S^\text{top}$                  \\ \midrule
$(\unit{\celsius.m.s^{-1}})$ & $(\unit{psu.m.s^{-1}})$ & $(\unit{m^{2}.s^{-2}})$ & $(\unit{\celsius.m^{-1}})$ & $(\unit{psu.m^{-1}})$ & $(\unit{s^{-1}})$ & $(\unit{\celsius})$ & $(\unit{psu})$ \\ \midrule
$\num{5d-4}$                 & 0                             & $\num{-2d-4}$           & $\num{1.4d-2}$             & $\num{2.1d-3}$              & $\num{8d-5}$      & $18$                & $36.6$               \\
$\num{1d-4}$                 & 0                             & $\num{-2d-4}$           & $\num{1.4d-2}$             & $\num{2.1d-3}$              & $\num{8d-5}$      & $18$                & $36.6$               \\
$\num{5d-4}$                 & 0                             & $\num{-5d-4}$           & $\num{1.4d-2}$             & $\num{2.1d-3}$              & $\num{8d-5}$      & $18$                & $36.6$               \\
$\num{1d-4}$                 & 0                             & $\num{-5d-4}$           & $\num{1.4d-2}$             & $\num{2.1d-3}$              & $\num{8d-5}$      & $18$                & $36.6$               \\
0                            & $\num{-5d-5}$                 & $\num{-2d-4}$           & $\num{1.4d-2}$             & $\num{2.1d-3}$              & $\num{8d-5}$      & $18$                & $36.6$               \\
0                            & $\num{-1d-5}$                 & $\num{-2d-4}$           & $\num{1.4d-2}$             & $\num{2.1d-3}$              & $\num{8d-5}$      & $18$                & $36.6$               \\
0                            & $\num{-5d-5}$                 & $\num{-5d-4}$           & $\num{1.4d-2}$             & $\num{2.1d-3}$              & $\num{8d-5}$      & $18$                & $36.6$               \\
0                            & $\num{-1d-5}$                 & $\num{-5d-4}$           & $\num{1.4d-2}$             & $\num{2.1d-3}$              & $\num{8d-5}$      & $18$                & $36.6$               \\
$\num{3d-4}$                 & $\num{-3d-5}$                 & $\num{-2d-4}$           & $\num{1.4d-2}$             & $\num{2.1d-3}$              & $\num{8d-5}$      & $18$                & $36.6$               \\
$\num{3d-4}$                 & $\num{-3d-5}$                 & $\num{-5d-4}$           & $\num{1.4d-2}$             & $\num{2.1d-3}$              & $\num{8d-5}$      & $18$                & $36.6$               \\
$\num{5d-4}$                 & 0                             & $\num{-2d-4}$           & $\num{1.3d-2}$             & $\num{7.5d-4}$              & 0                 & $14.5$              & $35$                 \\
$\num{1d-4}$                 & 0                             & $\num{-2d-4}$           & $\num{1.3d-2}$             & $\num{7.5d-4}$              & 0                 & $14.5$              & $35$                 \\
$\num{5d-4}$                 & 0                             & $\num{-5d-4}$           & $\num{1.3d-2}$             & $\num{7.5d-4}$              & 0                 & $14.5$              & $35$                 \\
$\num{1d-4}$                 & 0                             & $\num{-5d-4}$           & $\num{1.3d-2}$             & $\num{7.5d-4}$              & 0                 & $14.5$              & $35$                 \\
0                            & $\num{-5d-5}$                 & $\num{-2d-4}$           & $\num{1.3d-2}$             & $\num{7.5d-4}$              & 0                 & $14.5$              & $35$                 \\
0                            & $\num{-1d-5}$                 & $\num{-2d-4}$           & $\num{1.3d-2}$             & $\num{7.5d-4}$              & 0                 & $14.5$              & $35$                 \\
0                            & $\num{-5d-5}$                 & $\num{-5d-4}$           & $\num{1.3d-2}$             & $\num{7.5d-4}$              & 0                 & $14.5$              & $35$                 \\
0                            & $\num{-1d-5}$                 & $\num{-5d-4}$           & $\num{1.3d-2}$             & $\num{7.5d-4}$              & 0                 & $14.5$              & $35$                 \\
$\num{3d-4}$                 & $\num{-3d-5}$                 & $\num{-2d-4}$           & $\num{1.3d-2}$             & $\num{7.5d-4}$              & 0                 & $14.5$              & $35$                 \\
$\num{3d-4}$                 & $\num{-3d-5}$                 & $\num{-5d-4}$           & $\num{1.3d-2}$             & $\num{7.5d-4}$              & 0                 & $14.5$              & $35$                 \\
$\num{5d-4}$                 & 0                             & $\num{-2d-4}$           & $\num{-2.5d-2}$            & $\num{-4.5d-3}$             & $\num{-1.25d-4}$  & 0                   & $33.9$               \\
$\num{1d-4}$                 & 0                             & $\num{-2d-4}$           & $\num{-2.5d-2}$            & $\num{-4.5d-3}$             & $\num{-1.25d-4}$  & 0                   & $33.9$               \\
$\num{5d-4}$                 & 0                             & $\num{-5d-4}$           & $\num{-2.5d-2}$            & $\num{-4.5d-3}$             & $\num{-1.25d-4}$  & 0                   & $33.9$               \\
$\num{1d-4}$                 & 0                             & $\num{-5d-4}$           & $\num{-2.5d-2}$            & $\num{-4.5d-3}$             & $\num{-1.25d-4}$  & 0                   & $33.9$               \\
0                            & $\num{-5d-5}$                 & $\num{-2d-4}$           & $\num{-2.5d-2}$            & $\num{-4.5d-3}$             & $\num{-1.25d-4}$  & 0                   & $33.9$               \\
0                            & $\num{-1d-5}$                 & $\num{-2d-4}$           & $\num{-2.5d-2}$            & $\num{-4.5d-3}$             & $\num{-1.25d-4}$  & 0                   & $33.9$               \\
0                            & $\num{-5d-5}$                 & $\num{-5d-4}$           & $\num{-2.5d-2}$            & $\num{-4.5d-3}$             & $\num{-1.25d-4}$  & 0                   & $33.9$               \\
0                            & $\num{-1d-5}$                 & $\num{-5d-4}$           & $\num{-2.5d-2}$            & $\num{-4.5d-3}$             & $\num{-1.25d-4}$  & 0                   & $33.9$               \\
$\num{5d-4}$                 & $\num{-5d-5}$                 & $\num{-2d-4}$           & $\num{-2.5d-2}$            & $\num{-4.5d-3}$             & $\num{-1.25d-4}$  & 0                   & $33.9$               \\
$\num{5d-4}$                 & $\num{-5d-5}$                 & $\num{-5d-4}$           & $\num{-2.5d-2}$            & $\num{-4.5d-3}$             & $\num{-1.25d-4}$  & 0                   & $33.9$               \\
\bottomrule
\end{tabular}
\end{table}

\begin{table}[htbp]
\centering
\caption{Miscellaneous training cases used to train NORi, including cases with no temperature or salinity gradients and cases with very strong stratification at warm temperatures.}
\label{table misc LES training suite}
\begin{tabular}{@{}cccccccc@{}}
\toprule
$J_T^\text{top}$                        & $J_S^\text{top}$                         & $J_u^\text{top}$                   & $\pd{T}{z}$                & $\pd{S}{z}$                 & $f$               & $T^\text{top}$                 & $S^\text{top}$                  \\ \midrule
$(\unit{\celsius.m.s^{-1}})$ & $(\unit{psu.m.s^{-1}})$ & $(\unit{m^{2}.s^{-2}})$ & $(\unit{\celsius.m^{-1}})$ & $(\unit{psu.m^{-1}})$ & $(\unit{s^{-1}})$ & $(\unit{\celsius})$ & $(\unit{psu})$ \\ \midrule
$\num{3d-4}$                 & 0                             & $\num{-1d-4}$           & $\num{1d-2}$               & 0                           & $\num{1d-4}$      & $10$                & $35$                 \\
$\num{5d-4}$                 & 0                             & $\num{-1d-4}$           & $\num{1.4d-2}$             & 0                           & $\num{1d-4}$      & $10$                & $37$                 \\
0                            & $\num{-5d-5}$                 & $\num{-1d-4}$           & 0                          & $\num{-5d-3}$               & $\num{1.5d-4}$    & 0                   & $34$                 \\
0                            & $\num{-5d-5}$                 & $\num{-1d-4}$           & 0                          & $\num{-1d-3}$               & $\num{1d-4}$      & $10$                & $35$                 \\
$\num{5d-4}$                 & $\num{-5d-5}$                 & $\num{-1d-4}$           & $\num{6d-2}$               & $\num{8d-3}$                & $\num{1d-4}$      & $30$                & $37$                 \\
$\num{3d-4}$                 & $\num{-3d-5}$                 & $\num{-1d-4}$           & $\num{6d-2}$               & $\num{8d-3}$                & $\num{1d-4}$      & $30$                & $37$                 \\
$\num{1d-4}$                 & $\num{-1d-5}$                 & $\num{-1d-4}$           & $\num{6d-2}$               & $\num{8d-3}$                & $\num{1d-4}$      & $30$                & $37$                 \\
\bottomrule
\end{tabular}
\end{table}

\begin{table}[htbp]
\centering
\caption{Validation cases to evaluate performance and select final trained weights of NORi.}
\label{table LES validation suite}
\begin{tabular}{@{}cccccccc@{}}
\toprule
$J_T^\text{top}$                        & $J_S^\text{top}$                         & $J_u^\text{top}$                   & $\pd{T}{z}$                & $\pd{S}{z}$                 & $f$               & $T^\text{top}$                 & $S^\text{top}$                  \\ \midrule
$(\unit{\celsius.m.s^{-1}})$ & $(\unit{psu.m.s^{-1}})$ & $(\unit{m^{2}.s^{-2}})$ & $(\unit{\celsius.m^{-1}})$ & $(\unit{psu.m^{-1}})$ & $(\unit{s^{-1}})$ & $(\unit{\celsius})$ & $(\unit{psu})$ \\ \midrule
$\num{3.5d-4}$               & 0                             & 0                       & $\num{1.5d-2}$             & $\num{2d-3}$                & $\num{1d-4}$      & 20                  & 37                   \\
0                            & $\num{-3.5d-5}$               & 0                       & $\num{1.5d-2}$             & $\num{2d-3}$                & $\num{1d-4}$      & 20                  & 37                   \\
$\num{2d-4}$                 & $\num{-2d-5}$                 & $\num{-1d-4}$           & $\num{1.5d-2}$             & $\num{2d-3}$                & $\num{1d-4}$      & 20                  & 37                   \\
$\num{2d-4}$                 & $\num{-1.5d-5}$               & $\num{-3.5d-4}$         & $\num{1d-2}$               & $\num{1.5d-3}$              & $\num{1d-4}$      & 17                  & 36                   \\
$\num{1.5d-4}$               & $\num{-2d-5}$                 & $\num{-2.5d-4}$         & $\num{1.7d-2}$             & $\num{1.8d-3}$              & $\num{1d-4}$      & 12                  & 37                   \\
$\num{5d-5}$                 & $\num{-5d-6}$                 & $\num{-4d-4}$           & $\num{1.2d-2}$             & $\num{1.2d-3}$              & $\num{1d-4}$      & 10                  & 37                   \\
$\num{3.5d-4}$               & $\num{-3.5d-5}$               & $\num{-3d-4}$           & $\num{1d-2}$               & $\num{-2d-3}$               & $\num{1.5d-4}$    & 16                  & 34                   \\
$\num{4.5d-4}$               & $\num{-4d-5}$                 & $\num{-4.5d-4}$         & $\num{1.3d-2}$             & $\num{-1d-3}$               & $\num{7d-5}$      & 13                  & 35                   \\
$\num{3.5d-4}$               & 0                             & 0                       & $\num{1d-2}$               & $\num{5d-4}$                & 0                 & 17                  & 34.5                 \\
0                            & $\num{-3.5d-5}$               & 0                       & $\num{1d-2}$               & $\num{5d-4}$                & 0                 & 17                  & 34.5                 \\
$\num{2d-4}$                 & $\num{-2d-5}$                 & $\num{-1d-4}$           & $\num{1d-2}$               & $\num{5d-4}$                & 0                 & 17                  & 34.5                 \\
$\num{2d-4}$                 & $\num{-1.5d-5}$               & $\num{-3.5d-4}$         & $\num{1.2d-2}$             & $\num{7d-4}$                & $\num{-1d-5}$     & 13                  & 36                   \\
$\num{1.5d-4}$               & $\num{-2d-5}$                 & $\num{-2.5d-4}$         & $\num{1.6d-2}$             & $\num{6d-4}$                & $\num{-2d-5}$     & 16                  & 34                   \\
$\num{5d-5}$                 & $\num{-5d-6}$                 & $\num{-4d-4}$           & $\num{1.1d-2}$             & $\num{3d-4}$                & $\num{3d-5}$      & 10                  & 36.5                 \\
$\num{3.5d-4}$               & $\num{-3.5d-5}$               & $\num{-3d-4}$           & $\num{1.5d-2}$             & $\num{-5d-4}$               & $\num{-5d-5}$     & 20                  & 37                   \\
$\num{4.5d-4}$               & $\num{-4d-5}$                 & $\num{-4.5d-4}$         & $\num{1.7d-2}$             & $\num{-8d-4}$               & $\num{-1d-4}$     & 12                  & 35                   \\
$\num{3.5d-4}$               & 0                             & 0                       & $\num{-2d-2}$              & $\num{-4.7d-3}$             & $\num{-1.5d-4}$   & 0                   & 34.5                 \\
0                            & $\num{-3.5d-5}$               & 0                       & $\num{-2d-2}$              & $\num{-4.7d-3}$             & $\num{-1.5d-4}$   & 0                   & 34.5                 \\
$\num{2d-4}$                 & $\num{-2d-5}$                 & $\num{-1d-4}$           & $\num{-2d-2}$              & $\num{-4.7d-3}$             & $\num{-1.5d-4}$   & 0                   & 34.5                 \\
$\num{2d-4}$                 & $\num{-1.5d-5}$               & $\num{-3.5d-4}$         & $\num{-1.7d-2}$            & $\num{-4d-3}$               & $\num{1.5d-4}$    & 3                   & 34                   \\
$\num{4d-4}$                 & $\num{-4d-5}$                 & $\num{-2.5d-4}$         & $\num{-1.5d-2}$            & $\num{-4d-3}$               & $\num{1d-4}$      & -1                  & 36                   \\
$\num{5d-5}$                 & $\num{-5d-6}$                 & $\num{-4d-4}$           & $\num{-1d-2}$              & $\num{-3d-3}$               & $\num{1.25d-4}$   & 0                   & 33.5                 \\
$\num{3.5d-4}$               & $\num{-3.5d-5}$               & $\num{-3d-4}$           & $\num{-2d-2}$              & $\num{-4d-3}$               & $\num{-1.4d-4}$   & 1                   & 35.5                 \\
$\num{4.5d-4}$               & $\num{-4d-5}$                 & $\num{-4.5d-4}$         & $\num{-2d-2}$              & $\num{-4.5d-3}$             & $\num{-1d-4}$     & 2                   & 34                   \\
$\num{4.5d-4}$               & 0                             & $\num{-2d-4}$           & $\num{3d-2}$               & 0                           & $\num{1d-4}$      & 20                  & 37                   \\
$\num{1.5d-4}$               & 0                             & $\num{-3d-4}$           & $\num{1d-2}$               & 0                           & $\num{1d-4}$      & 15                  & 35                   \\
0                            & $\num{-4d-5}$                 & $\num{-3d-4}$           & 0                          & $\num{-4.5d-3}$             & $\num{-1d-4}$     & 17                  & 36                   \\
0                            & $\num{-2d-5}$                 & $\num{-1.5d-4}$         & 0                          & $\num{-2.5d-3}$             & $\num{-1d-4}$     & 16                  & 34                   \\
$\num{4d-4}$                 & $\num{-4d-5}$                 & $\num{-4d-4}$           & $\num{5d-2}$               & $\num{-7d-3}$               & $\num{1d-4}$      & 25                  & 36                   \\
$\num{3d-4}$                 & $\num{-2d-5}$                 & $\num{-2d-4}$           & $\num{3d-2}$               & $\num{-7.5d-3}$             & $\num{1d-4}$      & 27                  & 35                   \\
\bottomrule
\end{tabular}
\end{table}

\section{Loss scalings} \label{section appendix loss scalings}
As seen in Equations~\eqref{loss function base closure} and \eqref{eq:lossfunction}, during training of both the base closure and the neural networks, we use a weighted loss function to balance the contributions of each variable. 
The weights are set using normalization factors $A_\phi^a$ for the field $\phi$ in simulation $a$ within the training suite.
The difference between the loss functions for the base closure and neural network training is that velocity components $u$ and $v$ are not included in the neural network training loss as these fields are dominated by inertial oscillations which do not affect vertical mixing.
The values of $A_\phi^a$ are recalculated at the beginning of each training stage to renormalize the training loss. 
$A_\phi^a$ is dependent on the relative contribution of temperature and salinity to the potential density in the initial condition of the neural ODEs, as well as the relative contribution of each loss component to the total loss. 
$A_T^a$ and $A_S^a$ are calculated as follows:
\begin{align}
    A_T^a &= \frac{\alpha \left\Vert \Delta T_0^a \right\Vert_\infty + \beta \left\Vert \Delta S_0^a \right\Vert_\infty}{\alpha \left\Vert \Delta T_0^a \right\Vert_\infty},\\
    A_S^a &= \frac{\alpha \left\Vert \Delta T_0^a \right\Vert_\infty + \beta \left\Vert \Delta S_0^a \right\Vert_\infty}{\beta \left\Vert \Delta S_0^a \right\Vert_\infty}
\end{align}
where $T_0^a$ and $S_0^a$ denote the initial temperature and salinity profiles in training case $a$, with $\left\Vert \Delta \phi_0^a \right\Vert_\infty = \max \overline{\phi}_0^a - \min \overline{\phi}_0^a$ giving their vertical range. 
Here, $\alpha$ and $\beta$ are the thermal expansion coefficient and haline contraction coefficient of seawater, respectively.
The remaining normalization factors are chosen such that
\begin{align}
    \frac{\mathcal{L}_\sigma^a}{\mathcal{L}_T^a + \mathcal{L}_S^a} &= \frac{1}{9}, \\
    \mathcal{L}_\text{profile}^a &= \mathcal{L}_\text{gradient}^a, \\
    \mathcal{L}_\text{gradient}^a &= A_{\partial_z T}^a \delta \left( \partial_z \overline{T} \right)^a + A_{\partial_z S}^a \delta \left( \partial_z \overline{S} \right)^a + A_{\partial_z \sigma}^a \delta \left( \partial_z \overline{\sigma} \right)^a \\
    &= A_\text{gradient}^a \left( A_{T}^a \delta \left( \partial_z \overline{T} \right)^a + A_{S}^a \delta \left( \partial_z \overline{S} \right)^a + A_{\sigma}^a \delta  \left( \partial_z \overline{\sigma} \right)^a \right), \\
    \delta \overline{\phi} &\equiv \frac{1}{N_tN_z} \sum_{i=1}^{N_z} \sum_{j=1}^{N_t} \left| \overline{\phi}^{i,j} - \overline{\phi}^{i,j}_{LES} \right|^2
\end{align}
where $\overline{\phi}$ and $\overline{\phi}_{LES}$ are one of the vertical gradient profiles (temperature, salinity, potential density) as a function of vertical level $i$, time $j$, and scenario $a$---the overbars are a reminder that we are only parameterizing the area-averaged profiles.

\section{Neural network architecture} \label{section appendix neural network architecture}
In NORi, neural networks are used to predict temperature and salinity entrainment fluxes, thereby augmenting the base physical closure.
This is reflected in the design of neural networks, where they are only active within the zone where entrainment occurs and are inactive everywhere else in the vertical column.
For cell indices centered on the interior face $i = 2, 3, \dots, N$, where $i=1$ indicates the location of the face at the ocean surface and $i = N+1$ indicates the grid point at the bottom of the domain, 
\begin{align}
    J_{\mathbb{NN}_\phi}^i &= 
      \begin{cases}
    \mathbb{NN}_\phi (\boldsymbol{h}^i; \boldsymbol{\theta}_\phi), & \text{for } i = i_\text{min}, i_\text{min} + 1, \dots , i_\text{max}, \\
    0, & \text{otherwise},\\
  \end{cases}\\
    \boldsymbol{h}^i &= \left[ \left(\partial_z \boldsymbol{ \overline{T}}\right)^{i}_\text{zone}, \left(\partial_z \boldsymbol{ \overline{S}}\right)^{i}_\text{zone},  \left(\partial_z \boldsymbol{ \overline{\sigma}}\right)^{i}_\text{zone}, \arctan(\boldsymbol{Ri}^{i}_\text{zone}), J_b^\text{top} \right]^\top, \\
    \boldsymbol{\phi}^i_\text{zone} &\equiv \bigg[ \phi^{\min \{ i+2, N \}}, \phi^{\min \{ i+1, N \}}, \phi^{i}, \phi^{\max \{ i-1, 2 \}}, \phi^{\max \{ i-2, 2 \}} \bigg]
\end{align}
where $\mathbb{NN}_\phi$ is a neural network to predict the turbulent flux of tracer $\phi$, $\boldsymbol{h}^i$ is the input vector to the neural network at index $i$, $\boldsymbol{\theta}_\phi$ are the trainable weights for the neural network $\mathbb{NN}_\phi$, $i_\text{min}$ and $i_\text{max}$ are the first and last grid indices, between which (inclusive) the neural networks are active, $\boldsymbol{\phi}^i_\text{zone}$ denotes the 5-element vector of variable $\phi$ around the local neighborhood of index $i$, $\partial_z$ is the discrete gradient operator in $z$, and $J_b^\text{top}$ is the surface buoyancy flux.
For any face-centered grid point within the ``entrainment zone'', the values of the temperature and salinity fluxes are predicted by $\mathbb{NN}_T$ and $\mathbb{NN}_S$, respectively, with each network outputting a scalar that is the predicted value of temperature or salinity flux at the grid point.

The inputs of $\mathbb{NN}_T$ and $\mathbb{NN}_S$ are the temperature gradient, salinity gradient, potential density gradient, and gradient Richardson number of five grid points within the neighborhood of the grid point intended for inference, as well as the surface buoyancy flux.
The inputs and outputs of the neural networks are normalized using a Z-score normalization against the entire training suite to ensure that for each variable, the distribution of their values within the training suite has zero mean and unit variance.
More details on feature normalization can be found in \ref{section appendix normalization}.

\begin{figure}[htbp]
\centering
\includegraphics[trim={0 3cm 0 3cm}, width=\textwidth]{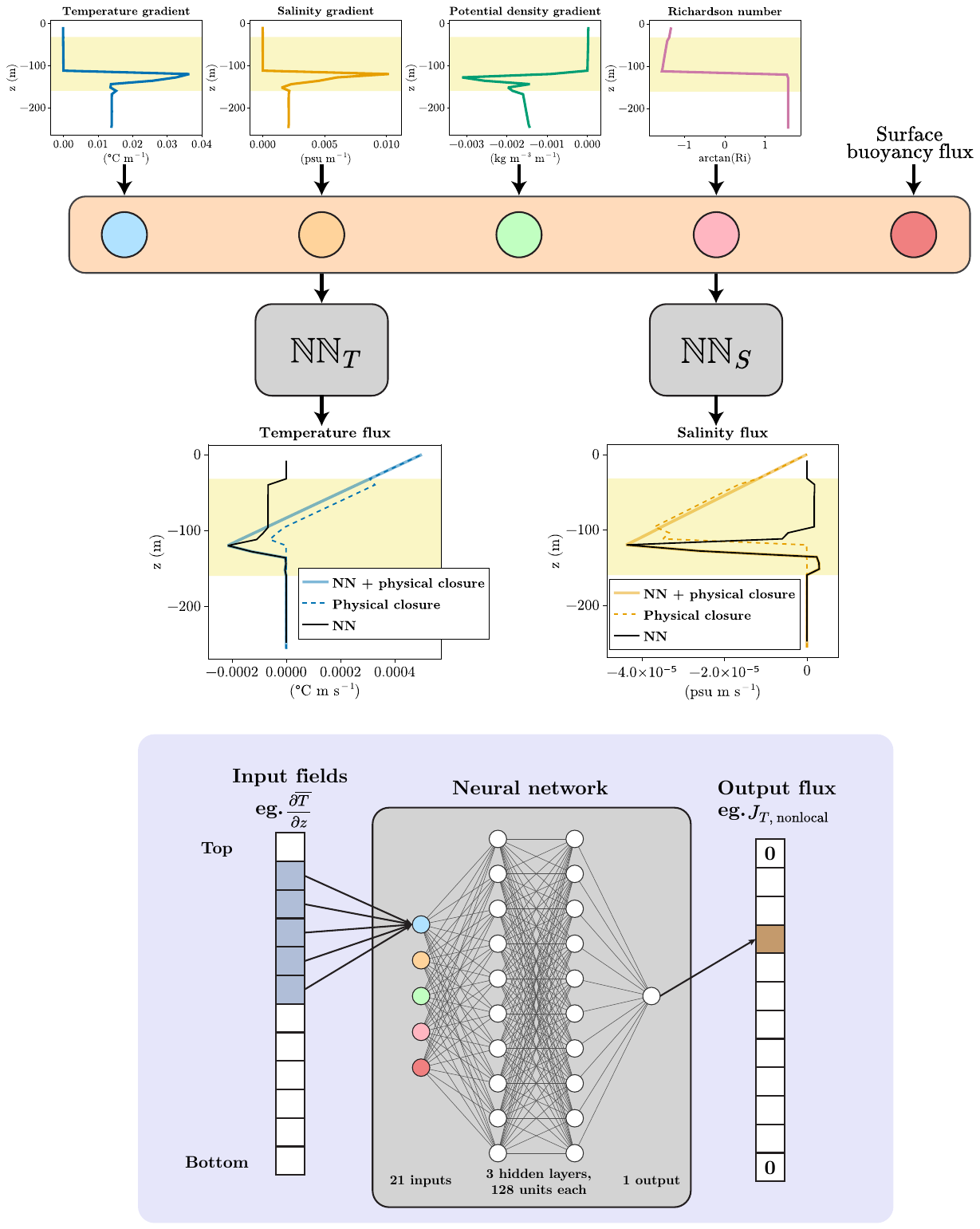}
\caption{Schematic of NORi's neural network architecture, augmenting the base closure to capture missing entrainment fluxes.
Top panel: the inputs of the neural networks $\mathbb{NN}_T: \mathbb{R}^{21} \rightarrow \mathbb{R}$ and $\mathbb{NN}_S: \mathbb{R}^{21} \rightarrow \mathbb{R}$, as well as their outputs, which are the entrainment temperature and salinity fluxes $J_{T, \text{ nonlocal}}^i$, $J_{S, \text{ nonlocal}}^i$.
The light yellow zones indicate the depth range where the neural networks are active.
Beyond these zones, neural network fluxes are zero.
The temperature and salinity output fluxes show the neural networks augmenting the base closure to capture entrainment.
Bottom panel: neural network architecture.
The neural networks each produce a single output: the residual flux at grid point $i$ given input $\boldsymbol{h}^i \in \mathbb{R}^{21}$, consisting of tracer gradients at five grid points in the local neighborhood of grid point $i$ and the surface buoyancy flux (see \ref{section appendix neural network architecture}).
The neural networks do not predict fluxes at the top and bottom grid points.}
\label{figure NN schematic}
\end{figure}

We adopt the principle of ``simple, but not any simpler'' in selecting the appropriate input variables to be included in the feature vector $\boldsymbol{h}^i$, and our rationale for each is as follows:
\begin{enumerate}
    \item The rate of nonlocally forced entrainment, which is the process we want to model using neural networks, depends primarily on the strength of the buoyancy flux $J_b^\text{top}$ at the ocean surface.
This is because the surface buoyancy flux is the source of entrainment mixing through the generation of vertically coherent plumes.
The buoyancy flux is a function of the temperature and salinity fluxes, but we provide only the buoyancy flux to the neural networks to eliminate the need for neural networks to learn the implicit correlation between temperature, salinity, and buoyancy fluxes.
This is because the entrainment penetration depth is a function of buoyancy, not temperature or salinity separately.
    \item The strength of entrainment mixing depends strongly on the ratio between local shear and background stratification at the base of the mixed layer, which is encoded by $1/Ri$.
A strong stratification in the thermocline would reduce the extent of penetrative convection, in which convective plumes overshoot into the interior due to inertia, while stronger local shear leads to stronger plume ``rollup'', which also reduces the plume penetration.
Information about local shear and stratification is encoded in the local Richardson number $Ri$.
    \item The mixing rate due to entrainment depends on local tracer gradients regardless of their absolute values.
To enforce this tracer invariance, the temperature, salinity, and potential density gradients $\pd{\overline{T}}{z}$, $\pd{\overline{S}}{z}$, and $\pd{\overline{\sigma}}{z}$, as well as the Richardson number $Ri$, are used as inputs to the neural networks instead of $\overline{u}$, $\overline{v}$, $\overline{T}$, $\overline{S}$, and $\overline{\sigma}$, since their actual values do not play a direct role in mixing dynamics.
    \item NORi is designed for the nonlinear equation of state TEOS-10~\cite{roquet_accurate_2015}.
Therefore, the relationship between the temperature and salinity gradients on the one hand and the stratification on the other depends on the actual values of temperature and salinity.
To encode information on nonlinearities in the equation of state explicitly to the neural networks, the potential density gradient $\pd{\overline{\sigma}}{z}$ is also provided.
    \item Entrainment can be thought of in terms of the interaction between nonlocal buoyancy fluxes and local tracer gradients.
The mixing occurs only locally due to a nonlocal flux source.
Therefore, tracer gradients are provided locally, while the surface buoyancy flux is provided nonlocally to the neural networks.
\end{enumerate}
The weights of $\mathbb{NN}_T$ and $\mathbb{NN}_S$ are shared across all grid points.
The local gradient Richardson number is provided as input to the neural networks in the form of $\arctan (Ri)$ so that it is bounded from above and below by $\pm \frac{\pi}{2}$, with appreciable gradient in the $|Ri| \lesssim 1$ regime, where turbulent dynamics are most sensitive to shear.
The ``entrainment zone'' where the neural network is active is diagnosed from the diffusivity field $\kappa$ by
\begin{align}
    i_\kappa &= \min \{i \in \{1, 2, \dots, N+1\} : \kappa[i] = \kappa_0 \}, \\
    i_\text{min} &= \max \{ i_\kappa - 10, 2 \}, \\
    i_\text{max} &= \min \{ i_\kappa + 5, N \},
\end{align}
where $i_\kappa$ is the grid point closest to the ocean surface where the base closure diffusivity $\kappa$ is equal to the background diffusivity $\kappa_0$.
This is equivalent to the nearest point from the surface where $Ri \geq Ri^c$, with $Ri^c$ defined by the base closure.
The neural networks do not predict turbulent fluxes at the grid point at the ocean surface ($i = 1$) and at the bottom ($i = N+1$).
At these boundary points, the temperature and salinity fluxes are given by the boundary conditions.
This restriction of the prediction zone of the neural networks reduces inference time, but implicitly assumes that convective plumes cannot penetrate deeper than $\SI{40}{m}$ below the BL base, which has been the case in all the LES generated.
Figure~\ref{figure NN schematic} illustrates a schematic of the neural network architecture.

This design enforces some important physical principles:
\begin{enumerate}
    \item The top and bottom fluxes are prescribed to satisfy the boundary conditions.
This enforces tracer conservation as a hard constraint; NORi can only redistribute tracers in the interior.
    \item In NORi, the local neighborhood visible to the neural networks as inputs spans $\SI{48}{m}$, as we have found this to be a vertical extent that is sufficient for the neural networks to characterize the location of inference relative to the BL base while ensuring generalizability across different locations within the vertical column.
    \item Entrainment has local effects: it enhances mixing in a small neighborhood near the base of the BL as it brings denser fluid from the ocean interior into the mixed layer.
    Therefore, the neural networks only need to know the information around their neighborhood to determine if entrainment occurs and produce a local flux to represent any mixing.
\end{enumerate}

\section{PDE discretization of the 1D column model} \label{section appendix discretization}
To solve the partial differential equations \eqref{equation u 1D}, \eqref{equation v 1D}, \eqref{equation T 1D}, and \eqref{equation S 1D} numerically, we discretize the equations spatially and temporally.
The spatial dimensions are discretized using the finite volume method, while we use a split implicit-explicit (IMEX) method for the temporal discretization.
The diffusion terms are time stepped with the implicit backward Euler method, while the advective terms ($f \overline{u}$ and $f \overline{v}$) and neural network fluxes are time stepped with the explicit forward Euler method.
The discretized equations are given by
\begin{align}
    \label{u IMEX} \frac{\boldsymbol{\overline{u}}^{n+1} - \boldsymbol{\overline{u}}^{n}}{\Delta t}  &= \partial_z \left( \boldsymbol{\nu} \partial_z \boldsymbol{\overline{u}}^{n+1} \right)  + f \boldsymbol{\overline{v}}^n, \\
    \label{v IMEX} \frac{\boldsymbol{\overline{v}}^{n+1} - \boldsymbol{\overline{v}}^{n}}{\Delta t} &= \partial_z \left( \boldsymbol{\nu} \partial_z \boldsymbol{\overline{v}}^{n+1} \right) - f \boldsymbol{\overline{u}}^n,\\
    \label{T IMEX} \frac{\boldsymbol{\overline{T}}^{n+1} - \boldsymbol{\overline{T}}^{n}}{\Delta t} &= \partial_z \left( \boldsymbol{\kappa} \partial_z \boldsymbol{\overline{T}}^{n+1} \right) - \partial_z \boldsymbol{J}_{\mathbb{NN}_T}^n, \\
    \label{S IMEX} \frac{\boldsymbol{\overline{S}}^{n+1} - \boldsymbol{\overline{S}}^{n}}{\Delta t} &= \partial_z \left( \boldsymbol{\kappa} \partial_z \boldsymbol{\overline{S}}^{n+1} \right) - \partial_z \boldsymbol{J}_{\mathbb{NN}_S}^n,
\end{align}
where $n = 0, 1, \cdots$ indexes discrete time steps with $t^{n+1} - t^{n} = \Delta t$ and $\partial_z$ is the discrete gradient operator in $z$.

\section{Variable normalization and model nondimensionalization} \label{section appendix normalization}
To promote better training behavior of neural networks and to allow the loss function to combine quantities with different units, we normalize the field and flux variables using a Z-score normalization such that the variables in the training suite have a zero mean and unit variance.
This approach ensures that all features are treated equally during neural network training.
The Z-score normalization is given by
\begin{align}
    \overline{u}^\star &= \frac{\overline{u} - \mu_{\overline{u}}}{\sigma_{\overline{u}}}, \\
    \overline{v}^\star &= \frac{\overline{v} - \mu_{\overline{v}}}{\sigma_{\overline{v}}}, \\
    \overline{T}^\star &= \frac{\overline{T} - \mu_{\overline{T}}}{\sigma_{\overline{T}}}, \\
    \overline{S}^\star &= \frac{\overline{S} - \mu_{\overline{S}}}{\sigma_{\overline{S}}}, \\
    \overline{u^\prime w^\prime}^\star &= \frac{\overline{u^\prime w^\prime} - \mu_{\overline{u^\prime w^\prime}}}{\sigma_{\overline{u^\prime w^\prime}}}, \\
    \overline{v^\prime w^\prime}^\star &= \frac{\overline{v^\prime w^\prime} - \mu_{\overline{v^\prime w^\prime}}}{\sigma_{\overline{v^\prime w^\prime}}}, \\
    \overline{w^\prime T^\prime}^\star &= \frac{\overline{w^\prime T^\prime} - \mu_{\overline{w^\prime T^\prime}}}{\sigma_{\overline{w^\prime T^\prime}}}, \\
    \overline{w^\prime S^\prime}^\star &= \frac{\overline{w^\prime S^\prime} - \mu_{\overline{w^\prime S^\prime}}}{\sigma_{\overline{w^\prime S^\prime}}},
\end{align}
where $\mu_\phi$ and $\sigma_\phi$ are the mean and standard deviation of the field $\phi$ computed across all training simulations and in both space and time.
Using the normalization above, starred quantities $\phi^\star$ are nondimensional.
We also nondimensionalize time and space using
\begin{align}
    t^\star &= \frac{t}{\tau}, \\
    z^\star &= \frac{z}{H},
\end{align}
where $\tau$ is the training horizon for each simulation, while $H$ is the total depth of the domain.
Substituting the nondimensional variables above and decomposing the fluxes into local (base-closure) and nonlocal (neural-network) components, the equations solved during training of the NORi closure are
\begin{align}
    \pd{\overline{u}^\star}{t^\star} &= \frac{\tau}{H^2} \pd{}{z^\star} \left( \nu \pd{\overline{u}^\star}{z^\star} \right) + \frac{f \tau}{\sigma_{\overline{u}}} (\sigma_{\overline{v}} \overline{v}^\star + \mu_{\overline{v}}), \\
    \pd{\overline{v}^\star}{t^\star} &= \frac{\tau}{H^2} \pd{}{z^\star} \left( \nu \pd{\overline{v}^\star}{z^\star} \right) - \frac{f \tau}{\sigma_{\overline{v}}} (\sigma_{\overline{u}} \overline{u}^\star + \mu_{\overline{u}}), \\
    \pd{\overline{T}^\star}{t^\star} &= \frac{\tau}{H^2} \pd{}{z^\star} \left( \kappa \pd{\overline{T}^\star}{z^\star} \right) -\frac{\tau}{H} \frac{\sigma_{\overline{w^\prime T^\prime}}}{\sigma_{\overline{T}}} \pd{}{z^\star} J_{\mathbb{NN}_T}^\star, \\
    \pd{\overline{S}^\star}{t^\star} &= \frac{\tau}{H^2} \pd{}{z^\star} \left( \kappa \pd{\overline{S}^\star}{z^\star} \right) -\frac{\tau}{H} \frac{\sigma_{\overline{w^\prime S^\prime}}}{\sigma_{\overline{S}}} \pd{}{z^\star} J_{\mathbb{NN}_S}^\star.
\end{align}
Here, $J_{\mathbb{NN}_T}^\star$ and $J_{\mathbb{NN}_S}^\star$ denote the normalized neural-network flux outputs.

\section{Stratification comparison with Ocean Weather Station Papa} \label{section appendix ows papa}

\begin{figure}[htbp]
\centering
\includegraphics[width=\textwidth]{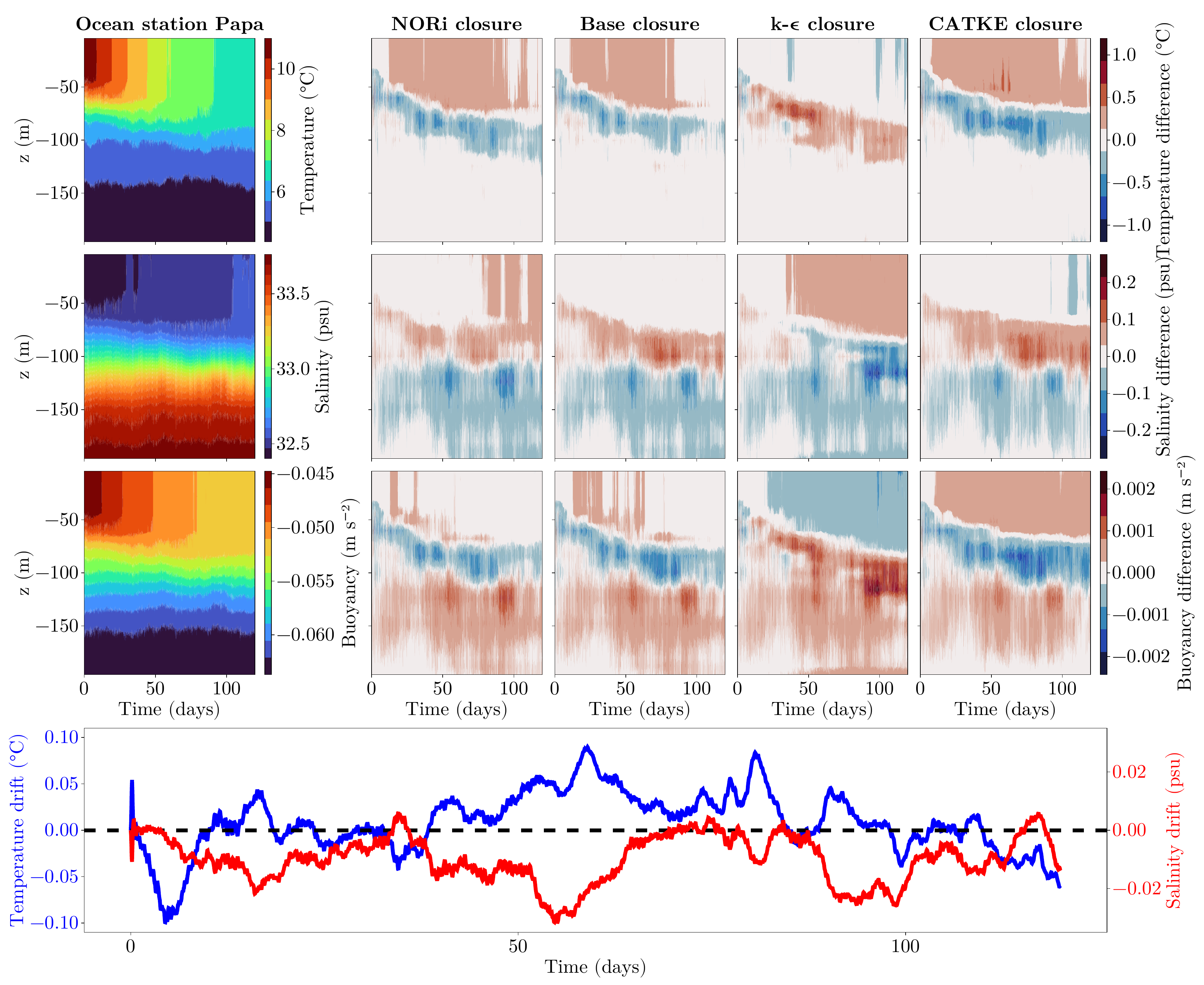}
\caption{Left column, from top to bottom: Evolution of temperature, salinity, and buoyancy in OWS Papa observations. 
Right columns, from left to right: The temperature, salinity, and buoyancy differences between solutions of NORi, its unaugmented base closure, the $k$-$\epsilon$ closure of \citeA{umlauf_generic_2003}, and the CATKE closure of \citeA{wagner_formulation_2025} in the OWS Papa simulation.
The differences are computed as model solutions minus observations.
All contour plots are eight-year composites, taken from years with no large gaps in the OWS Papa data.
Bottom row: The composite vertically averaged temperature and salinity drift between the models and observations.
Since all models are forced with the same surface fluxes, their drifts are identical.
The drifts shown are averaged over a running 24-hour window to smooth out high-frequency variability.
}
\label{figure ows papa stratification}
\end{figure}

In our comparison of the column model solutions forced with realistic conditions, we find that NORi, its unaugmented base closure, the $k$-$\epsilon$ closure of \citeA{umlauf_generic_2003}, and the CATKE closure of \citeA{wagner_formulation_2025} all produce a comparable stratification evolution when compared against OWS Papa observations.
Figure~\ref{figure ows papa stratification} shows the evolution of temperature, salinity, and buoyancy in OWS Papa observations, the model-observation differences in these fields, as well as the vertically averaged temperature and salinity drifts between the models and observations.
In general, all models mentioned above capture the evolution of the depth of the mixed layer against observations, as the temperature differences in the mixed layer are on the order of $\SI{0.25}{\celsius}$, while the salinity differences in the mixed layer are on the order of $\SI{0.05}{psu}$.
NORi and its base closure produce fields that are highly similar, indicating that nonlocal entrainment is weak at OWS Papa.
NORi, its base closure, and CATKE generally underestimate the mixed layer depth slightly, while $k$-$\epsilon$ produces a deeper mixed layer.
However, these differences are small and the overall skill of all models is high.

The constant flux adjustment imposed on the ocean surface in addition to the prescribed surface fluxes given by the OWS Papa observations minimizes the temperature and salinity drifts, which can be seen from the small values of the average drifts in the bottom row of Figure~\ref{figure ows papa stratification}.
The temperature has a maximum drift of around $\SI{0.1}{\celsius}$, while the salinity has a maximum drift of around $\SI{0.02}{psu}$.
Furthermore, most of the drift originates in the deep ocean, where lateral advection---which the 1D column model cannot represent---modifies the temperature and salinity budgets.

\section{Double-gyre solution at 100 years} \label{section appendix double gyre 100 years}

\begin{figure}[htbp]
\centering
\includegraphics[width=\textwidth]{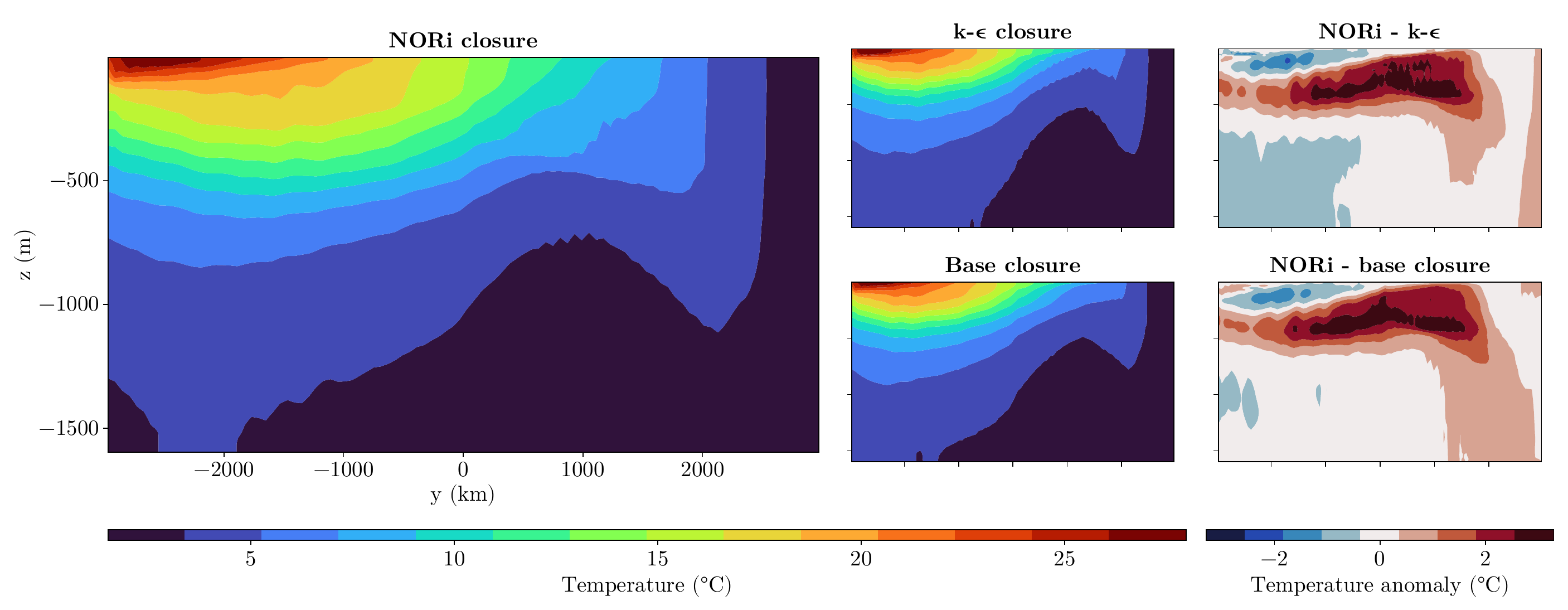}
\caption{Zonally averaged temperature stratification comparison between NORi, the $k$-$\epsilon$ closure of \citeA{umlauf_generic_2003}, and the unaugmented base closure in the double-gyre simulation.
The snapshots are taken at $t = \SI{100}{years}$.
At this time, the simulation is not yet fully equilibrated.
The rightmost column shows the temperature differences in stratification between NORi and the other two closures ($k$-$\epsilon$ and the unaugmented base closure).}
\label{figure double gyre yz stratification k epsilon 100 years}
\end{figure}

Here we show a snapshot of the zonal average of the double-gyre simulation at $t = \SI{100}{years}$ to complement Figure~\ref{figure double gyre yz stratification} in the main text.
At $t = \SI{100}{years}$, the double-gyre simulations begin to show significant differences between NORi, its unaugmented base closure, and $k$-$\epsilon$ closure.
Figure~\ref{figure double gyre yz stratification k epsilon 100 years} shows that, compared to $k$-$\epsilon$ and the base closure, NORi produces deeper mixed layers along the entire meridional extent of the basin, except in the northernmost regions where the deep mixed layers reach the bottom of the domain.
However, we do not have a ``ground truth'' solution as a reference since this is an idealized setup.
Therefore, further quantification of each model's skill requires running realistic ocean simulations, where observational data are available for comparison.
However, this is beyond the present scope, and we aim to provide a thorough and systematic evaluation of various closures in realistic ocean simulations in the future.

\section{NORi solution comparison with CATKE} \label{section appendix CATKE}
In this section, we show results comparing NORi and CATKE~\cite{wagner_formulation_2025}, a physics-based BL parameterization that uses a diagnostic mixing length with a convective adjustment component and a prognostic turbulent kinetic energy (TKE).
CATKE demonstrated superior performance compared to commonly used BL parameterizations such as K-Profile Parameterization (KPP)~\cite{large_oceanic_1994} and the SMC-LT closure of \citeA{harcourt_improved_2015} when benchmarked against idealized and realistic LES~\cite{wagner_formulation_2025}.

Before making any comparisons, one important caveat is that CATKE is trained with an LES suite that takes into account the effects of additional mixing due to surface waves under shear-driven mixing scenarios, i.e., Langmuir turbulence.
In addition, CATKE and NORi are calibrated using different training suites with LES of different background stratification, surface forcings, equations of state, and rotation rates.
Under the same surface wind stress and initial conditions, LES that take into account wave effects generally produce deeper mixed layers.
We expect CATKE to produce deeper mixed layers than NORi in strongly wind-driven scenarios, since NORi is trained on simulations that do not take into account wave effects.
This is because our goal in developing NORi is to illustrate a new design and training paradigm for data-driven ocean modeling, and we have chosen to demonstrate it using a reduced set of physics that capture the leading-order effects of vertical mixing due to winds and convection.
Although the specific physical details of the training data are crucial, the NORi framework can be flexibly extended to incorporate other microturbulence physics through fine-tuning and/or transfer learning.
Using this approach of augmenting a robust and simple physical closure with neural network components trained end-to-end, future versions of NORi can account for additional physical processes such as wave effects and rotation-dependent deep convection.
They can also solve additional prognostic equations for turbulent quantities.

\begin{figure}[htbp]
\centering
\includegraphics[trim={0 0 0 0}, width=\textwidth]{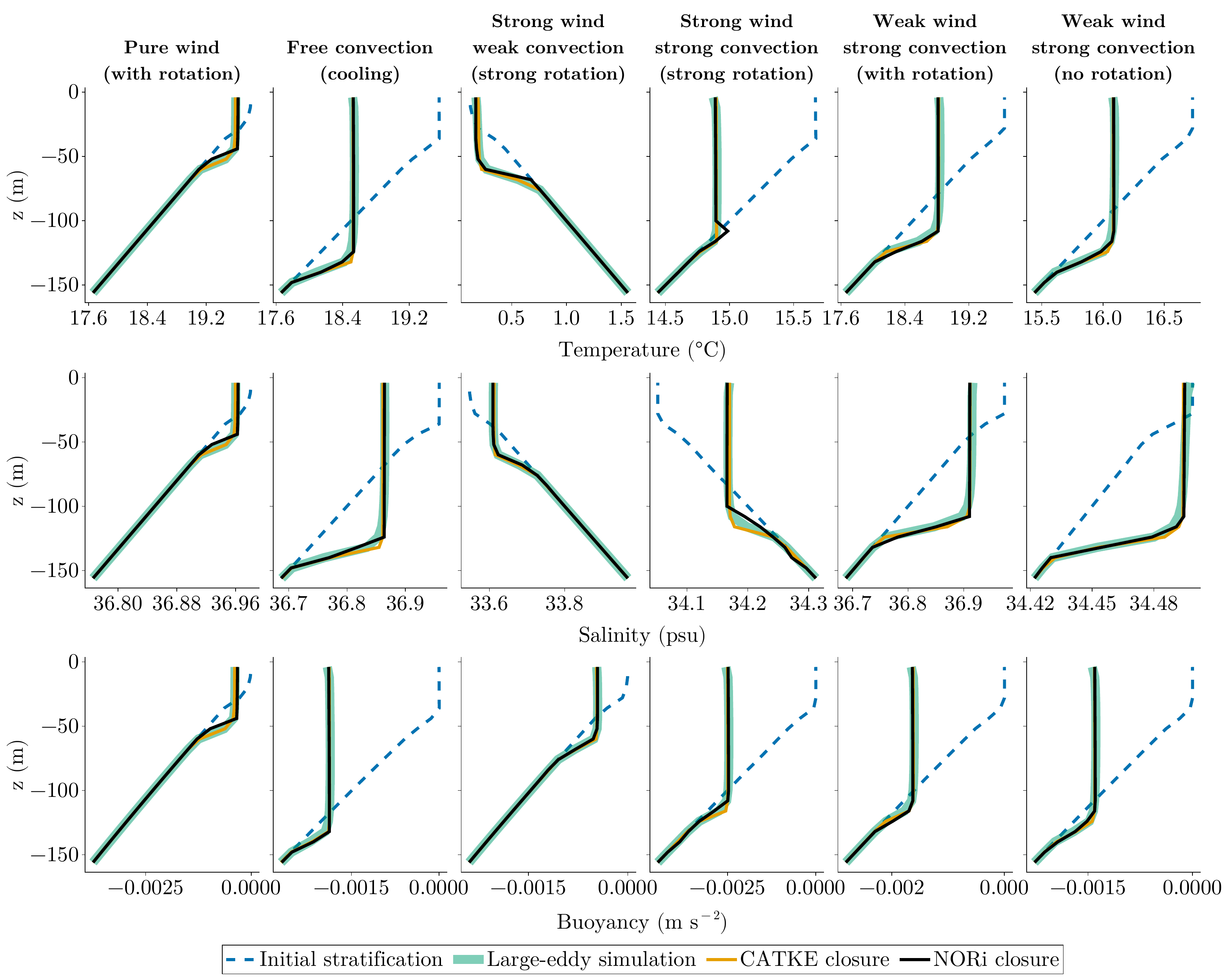}
\caption{Comparison of NORi and CATKE performance in the column-model setting. CATKE~\cite{wagner_formulation_2025} is a physics-based BL parameterization that uses a diagnostic mixing length with a convective adjustment component and a prognostic turbulent kinetic energy (TKE).
We compare temperature, salinity, and buoyancy fields against LES solutions.
All examples shown are validation cases which were not seen by NORi during training.}
\label{figure vs CATKE results}
\end{figure}

As expected, CATKE produces deeper mixed layers in single-column training and validation cases driven by strong winds.
In Figure~\ref{figure vs CATKE results}, the first and fourth columns are cases of strong wind, where CATKE produces slightly deeper mixed layers than NORi.
As noted earlier, the LES solutions shown in Figure~\ref{figure vs CATKE results} do not take into account wave effects; therefore, it is not surprising that they are generally shallower than CATKE's predictions.
Discrepancies between CATKE and NORi are expected given the differences in their training data.

\begin{figure}[htbp]
\centering
\includegraphics[width=\textwidth]{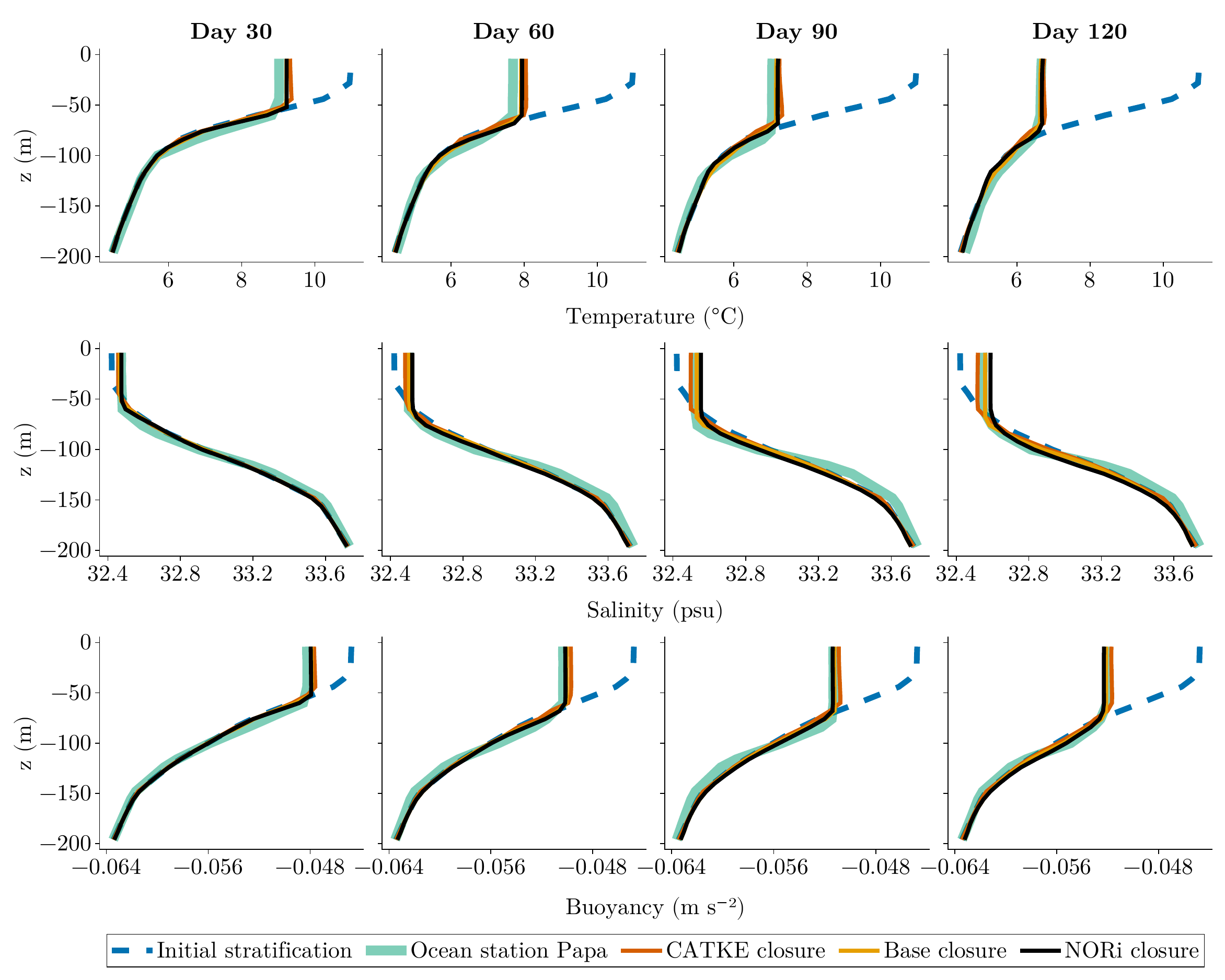}
\caption{Monthly comparison of column-model simulations from NORi, its base closure, and CATKE against OWS Papa observations, run with prescribed surface fluxes given by OWS Papa observations.
The simulations are initialized using the observed temperature and salinity profiles on November 1 and run for $\SI{120}{days}$.
The profiles shown are composites averaged over eight different years between 2007 and 2023.
Some years are excluded because there are large gaps (more than \SI{72}{hours}) within the data.
A constant flux correction is applied on top of the prescribed fluxes to mitigate the drifts in the model solution due to the discrepancies between the observed temperature and salinity budgets and those given by the prescribed fluxes.
All models are forced with identical surface fluxes.}
\label{figure ows papa CATKE}
\end{figure}

When tested in realistic scenarios based on OWS Papa observations, NORi and CATKE produce very similar results as shown in Figure~\ref{figure ows papa CATKE}.
CATKE produces slightly weaker mixing, as indicated by its shallower mixed layer depth.
However, these differences are very small as both CATKE and NORi agree well with the observed profiles.

\begin{figure}[htbp]
\centering
\includegraphics[width=\textwidth]{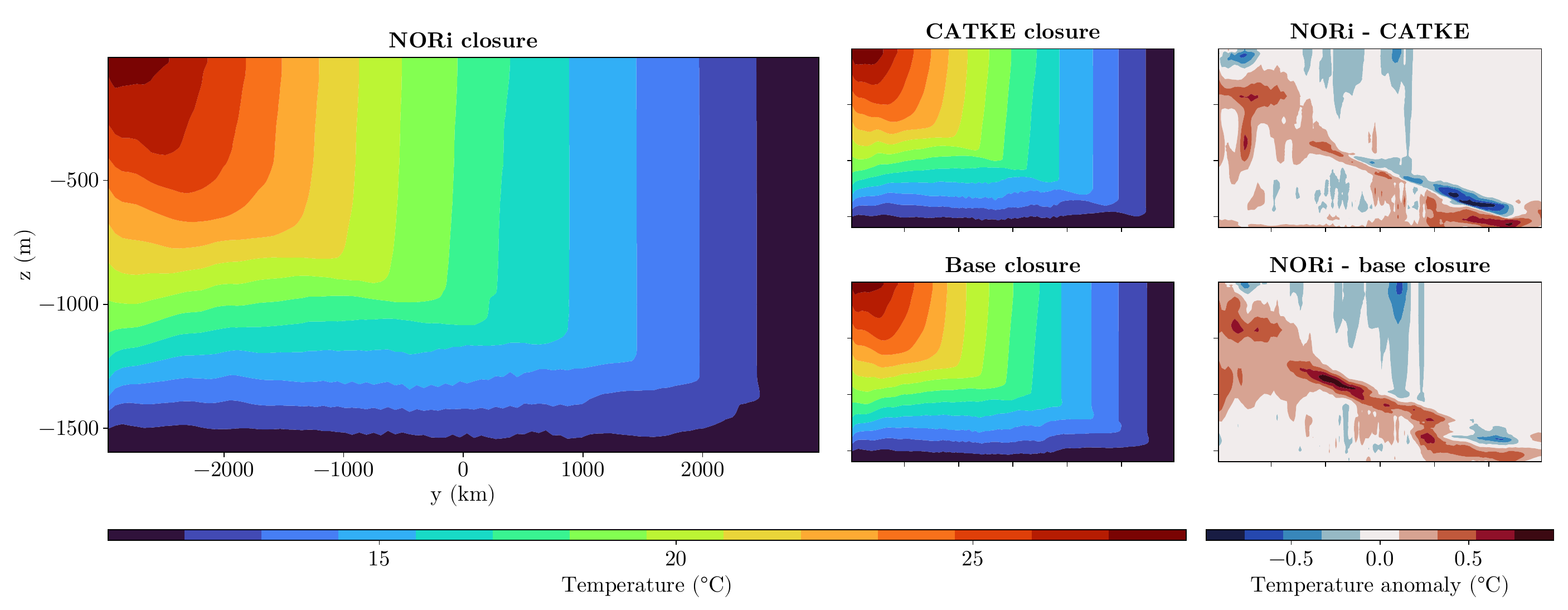}
\caption{Zonally averaged temperature stratification comparison between NORi, CATKE~\cite{wagner_formulation_2025}, and the unaugmented base closure in the double-gyre simulation.
The snapshots are taken at $t = \SI{2.5}{years}$, when the mixed layer extends to almost the entire water column in the north.
The rightmost column shows the temperature differences in stratification between NORi and the other two closures (CATKE and the unaugmented base closure).}
\label{figure double gyre yz stratification CATKE 2.5 years}
\end{figure}

\begin{figure}[htbp]
\centering
\includegraphics[width=\textwidth]{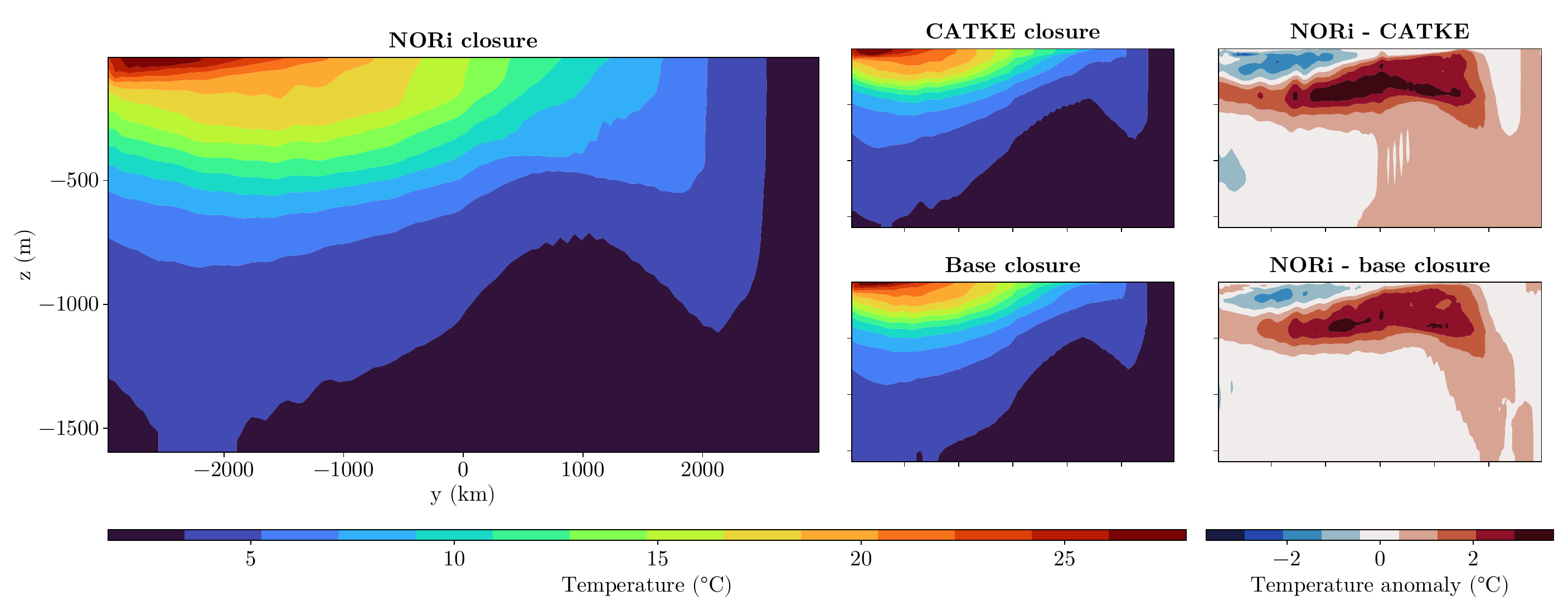}
\caption{Same as Figure~\ref{figure double gyre yz stratification CATKE 2.5 years}, but at $t = \SI{100}{years}$.}
\label{figure double gyre yz stratification CATKE 100 years}
\end{figure}

In the double-gyre simulation, the differences between CATKE and NORi are similar to those between $k$-$\epsilon$ and NORi (see Figures~\ref{figure double gyre yz stratification} and \ref{figure double gyre yz stratification k epsilon 100 years}).
At $\SI{2.5}{years}$, as shown by the zonally averaged temperature in Figure~\ref{figure double gyre yz stratification CATKE 2.5 years}, NORi produces shallower mixed layers than CATKE in the northern regions of the simulation.
Interestingly, NORi also produces slightly deeper mixed layers in the southern part of the domain.
At $t = \SI{100}{years}$, we can see from Figure~\ref{figure double gyre yz stratification CATKE 100 years} that NORi generates mixed layers that are much deeper than CATKE in most regions, except in the northern part of the basin, where deep mixed layers reach the bottom, similar to the comparison with $k$-$\epsilon$.
As we do not have a ground truth solution for this idealized setup, a thorough evaluation of NORi and CATKE would require running realistic regional/global ocean simulations where observational data are available for comparison.
However, it is not surprising that CATKE and $k$-$\epsilon$ behave similarly, as CATKE is heavily inspired by two-equation closures such as $k$-$\epsilon$.

\bibliography{references}

\end{document}